\documentclass[fleqn,usenatbib]{mnras}
\usepackage{newtxtext,newtxmath}

\usepackage[T1]{fontenc}

\DeclareRobustCommand{\VAN}[3]{#2}
\let\VANthebibliography\thebibliography
\def\thebibliography{\DeclareRobustCommand{\VAN}[3]{##3}\VANthebibliography}

\usepackage{graphicx}
\usepackage{amsmath}
\usepackage[dvipsnames]{xcolor}
\usepackage{tabularx}
\usepackage{multirow}

\usepackage{subcaption}

\newcommand\chimp{h^{-1}{\rm Mpc}}
\newcommand\chikp{h^{-1}{\rm kpc}}

\newcommand\hisun{h^{-1}{\rm M}_\odot}
\newcommand\hiimsun{h^{-2}{\rm M}_\odot}
\newcommand\himsun{h^{-1}{\rm M}_\odot}

\newcommand\zs{z_{\rm s}}
\newcommand\zl{z_{\rm l}}
\newcommand\ws{w_{\rm s}}
\newcommand\wl{w_{\rm l}}
\newcommand\wls{w_{\rm ls}}

\newcommand\fsat{f_{\rm sat}}
\newcommand\Mprog{M_{\rm prog}}
\newcommand\Vpeak{V_{\rm peak}}

\newcommand{\avg}[1]{{\langle #1 \rangle}}
\newcommand{\ncen}{N_{\rm c}}
\newcommand{\nsat}{N_{\rm s}}
\newcommand{\Mmin}{M_{\rm min}}
\newcommand{\Mcen}{M_{\rm cen}}
\newcommand{\avMcen}{\avg{M_{\rm cen}}}
\newcommand{\logMmin}{\log M_{\rm min}}
\newcommand{\sigmalogM}{\sigma_{\log M}}
\newcommand{\rmd}{{\rm d}}
\newcommand{\ngal}{n_{\rm gal}}

\newcolumntype{C}{>{\centering\arraybackslash}p{1.4cm}}

\title[Galaxy-dark matter connection in HSC]{Galaxy-dark matter connection of photometric galaxies from the HSC-SSP Survey: Galaxy-galaxy lensing and the halo model}

\author[Chaurasiya et al.]{
Navin Chaurasiya,$^{1}$\thanks{E-mail: navin@iucaa.in }
Surhud More,$^{1,2}$\thanks{E-mail: surhud@iucaa.in}
Shogo Ishikawa,$^{3}$
Shogo Masaki,$^{4,5}$
Daichi Kashino,$^{5,6,7}$
\newauthor{Teppei Okumura$^{2,8}$}
\\
$^{1}$Inter-University Centre for Astronomy and Astrophysics, Ganeshkhind, Pune 411007, IN\\
$^{2}$Kavli Institute for the Physics and Mathematics of the Universe (WPI), University of Tokyo, 5-1-5, Kashiwanoha, 2778583, JP\\
$^{3}$Center for Gravitational Physics and Quantum Information, Yukawa Institute for Theoretical Physics, Kyoto University, Sakyo-ku, Kyoto 606-8502, Japan\\
$^{4}$National Institute of Technology, Suzuka College, Suzuka, Mie 510-0294, Japan\\
$^{5}$Department of Physics, Nagoya University, Nagoya, Aichi 464-8601, Japan\\
$^{6}$Institute for Advanced Research, Nagoya University, Nagoya, Aichi 464-8601, Japan\\
$^{7}$Division of Science, National Astronomical Observatory of Japan, 2-21-1 Osawa, Mitaka, Tokyo 181-8588, Japan\\
$^{8}$Academia Sinica Institute of Astronomy and Astrophysics, AS/NTU Astronomy-Mathematics Building, No.1, Sec. 4, Roosevelt Rd, Taipei 10617, Taiwan, R.O.C. \\
}

\date{Accepted XXX. Received YYY; in original form ZZZ}

\pubyear{2023}

\begin{document}
\label{firstpage}
\pagerange{\pageref{firstpage}--\pageref{lastpage}}
\maketitle

\begin{abstract}
We infer the connection between the stellar mass of galaxies from the Subaru Hyper Suprime-Cam (HSC) survey, and their dark matter halo masses and its evolution in two bins of redshifts between $[0.3, 0.8]$. We use the measurements of the weak gravitational lensing signal of the galaxies using background galaxies from the Year 1 catalog of galaxy shapes from the HSC survey. We bin the galaxies in stellar mass with varying thresholds ranging from $8.6 \leq \log \left[ M_*/(h^{-2}\rm{M_\odot})\right] \leq 11.2$ and use stringent cuts in the selection of source galaxies to measure the weak lensing signal. We present systematic and null tests to demonstrate the robustness of our measurements. We model these measurements of the weak lensing signal together with the abundance of galaxies in the halo occupation distribution framework. For each stellar mass threshold bin, we obtain constraints on the halo occupation parameters of central galaxies $M_{\rm min}$ and $\sigma_{\log M}$, which correspond to the halo mass at which central galaxies in the threshold sample reach half occupancy, and its scatter, respectively, along with parameters that describe the occupation of the satellite galaxies. The measurements of abundance and weak lensing individually constrain different degeneracy directions in the $M_{\rm min}$ and $\sigma_{\log M}$ plane, thus breaking the degeneracy in these parameters. We demonstrate that the weak lensing measurements are best able to constrain the average central halo masses, $\avg{\Mcen}$. We compare our measurements to those obtained using the abundance and clustering of these galaxies as well as the subhalo abundance matching measurements and demonstrate qualitative agreement. We find that the galaxy-dark matter connection does not vary significantly between redshift bins we explore in this study. Uncertainties in the photometric redshift of the lens galaxies imply that more efforts are required to understand the true underlying stellar mass-halo mass relation of galaxies and its evolution over cosmic epoch.
\end{abstract}

\begin{keywords}
galaxies: evolution – galaxies: haloes – (cosmology:) large-scale structure of Universe - gravitational lensing: weak - cosmology: observations
\end{keywords}

\section{Introduction} \label{intro}
In the standard cosmological model, structure formation in the Universe is governed by the interplay between dark matter, which enhances overdensities of matter distribution, and dark energy, which acts to hinder such growth. Dark matter halos form the basic unit of the large scale structure, and their abundance is highly sensitive to this interplay between the cosmological parameters \citep[see e.g.,][]{2008MNRAS.391.1940M, 2011ApJ...732..122B, 2011MNRAS.410.1911C, 2012MNRAS.424..993C, 2013MNRAS.434L..61M}. The formation and evolution of galaxies in dark matter halos is a result of complex astrophysical processes related to the formation and evolution of stars, its effect on the gas, the feedback from supermassive black holes at their centers, as well as, the mergers of galaxies \citep[see e.g.,][]{2005MNRAS.363....2K, 2006MNRAS.370..645B, 2014MNRAS.444.1518V, 2014MNRAS.445..175G, 2014ARA&A..52..291C, 2020NatRP...2...42V}. Direct inference of the connection between dark matter halos and galaxies is thus important to understand these astrophysical processes \citep[see e.g.,][]{2018MNRAS.475..676S, 2019MNRAS.490.5693B}. In turn, an accurate determination of this connection can help in the inference of cosmological parameters \citep[see e.g.,][]{2005ApJ...631...41T, 2012ApJ...745...16T, 10.1093/mnras/sts525, 2014ApJ...783..118R, 2015ApJ...806....2M, 2022arXiv221003110D}.

The stellar mass contained within galaxies reflects the integrated star formation efficiency of dark matter halos of various masses. It is now well established that the star formation efficiency of halos peaks around intermediate mass halos of around $10^{12} \hisun$ \citep[see e.g.,][]{2010ApJ...717..379B, 2012ApJ...744..159L, 2016MNRAS.457.4360Z} and halos on either side of this are less efficient due to various forms of feedback associated with star formation at the low mass end and supermassive black holes at the high mass end. The evolution of the stellar mass-halo mass relation can thus provide insights into how this star formation efficiency changes with time \citep[see e.g.,][]{2013ApJ...770...57B, 2019MNRAS.488.3143B, 2017MNRAS.470..651R}.

Various observational techniques have been used to probe the dark matter halos of galaxies. One of the techniques that directly probes the halo masses beyond few tens of $\chikp$ is the inference of masses using the kinematics of satellite galaxies in dark matter halos \citep[see e.g.,][]{2009MNRAS.392..917M, 2011MNRAS.410..210M}. Satellite kinematics however has to rely on the assumption of virial equilibrium, the anisotropy of the dispersion in the orbits of satellite galaxies in dark matter halos, velocity bias which could arise from the differences in the distribution of matter compared to satellite galaxies, and accurate determination of the interloper galaxies which could masquerade as satellites. Indirect techniques such as subhalo abundance matching \citep[see e.g.,][]{2006ApJ...647..201C} instead rely on the ansatz of a monotonous relation between the stellar mass and halo masses of galaxies, along with a scatter in addition to a fixed set of cosmological parameters which determines the (sub)halo abundances. The technique of matching these abundances to the abundance of galaxies measured as the stellar mass function, allows an inference of the stellar mass-halo mass relation \citep[see e.g.,][]{2004ApJ...609...35K, 2010MNRAS.404.1111G, 2010ApJ...717..379B, 2013ApJ...771...30R}. The clustering of galaxies on large scales can also indirectly provide information about this relation \citep[see e.g.,][]{2005ApJ...631...41T, 2007ApJ...667..760Z, 2011ApJ...736...59Z} by utilizing the dependence of the large scale bias of halos on the mass of halos \citep[see e.g.,][]{2010ApJ...724..878T}.

The weak gravitational lensing signal \citep[see e.g.,][]{2006MNRAS.368..715M, 2009MNRAS.394..929C, 10.1093/mnras/sts525} of galaxies provides another direct method to constrain the galaxy-dark matter connection. According to general theory of relativity, an overdensity of matter warps spacetime in its vicinity in a manner that distorts light bundles from distant background sources traveling toward us. In its weak form, gravitational lensing causes a coherent tangential distortions in the shapes of such background galaxies. The distortion in the shape of a single galaxy due to weak lensing is quite small and difficult to disentangle from the intrinsic elliptical shape of its isophotes. A statistical averaging of the shapes of many such background galaxies gets rid of the uncorrelated intrinsic shapes of galaxies and allows the measurement of the coherent shear imprinted on the background galaxies due to weak lensing. Measurements of shapes of galaxies from ground based imaging data is challenging (see e.g., \citet{2014ApJS..212....5M, 2015MNRAS.450.2963M, 2018MNRAS.481.3170M}), as atmospheric light propagation and the telescope optics can also corrupt the measurements of shapes of galaxies. A number of tests need to be conducted for residual systematics in weak lensing measurements, but once modelled, the weak lensing signal can also provide constraints on the stellar mass-halo mass relation of galaxies \citep[see. e.g.,][]{2005ApJ...635...73H, 2006MNRAS.368..715M, 2011ApJ...738...45L, 2012ApJ...744..159L, 2011A&A...534A..14V, 2015A&A...579A..26V, 2016MNRAS.459.3251V, 10.1093/mnras/sts525, 2014MNRAS.437.2111V, 2015ApJ...806....2M, 2015MNRAS.454.1161Z, 2016MNRAS.457.4360Z, 2015MNRAS.449.1352C, 2015MNRAS.452.3529V, 2020MNRAS.492.3685H, 2020A&A...642A..83D, 2022arXiv221003110D, 2020MNRAS.499.2896T, 2022MNRAS.510.5408R, 2023arXiv230104664M}.

A number of ongoing weak lensing surveys cover large areas of sky with excellent quality imaging in order to map out the dark matter distribution in the Universe. The Dark Energy Survey (DES)\footnote{\url{http://darkenergysurvey.org}}, the Kilo Degree Survey (KiDS)\footnote{\url{http://kids.strw.leidenuniv.nl}}, and the Subaru Hyper Suprime-Cam survey (HSC)\footnote{\url{http://hsc.mtk.nao.ac.jp/ssp}} have covered areas that range from 1000 to 5000 sq. degree in this pursuit. Amongst these, the HSC is the deepest and thus allows us to carry out studies of evolution of the connection between galaxies and their dark matter halos that extend over a wide range of stellar masses. In this paper, we use galaxies from the HSC survey along with their stellar mass and photometric redshift estimates from their photometry in order to infer the stellar mass-halo mass relation in two redshift bins, $[0.30-0.55]$ and $[0.55-0.80]$.

In recent works, \citet[][hereafter, I20]{2020ApJ...904..128I} and \citet[hereafter, M22]{2022arXiv221011713M}, the clustering and abundance of galaxies have been used to constrain the galaxy-dark matter connection of the same sample of galaxies. The former amongst these studies, model their measurements of the clustering signal using an analytical halo occupation distribution (HOD) framework, while the latter use a modification to the traditional subhalo abundance matching method in order to explain the same observables. These different methodologies can explain the measurements equally well, even though they may not agree on the prescription of how galaxies occupy their dark matter halos and thus predict a different weak lensing signal. Our weak lensing signal (hereafter, WLS) measurement can thus be used as a discriminant for such theoretical models and the assumptions that they rely on.

This paper is organised as follows: We describe the lens and source data in section~\ref{data}. Sec.~\ref{abundance} describes the abundance data we use to constrain our HOD model and to study the impact of abundances on scaling relations. The formalism of stacked weak lensing signal computations and tests of survey systematics have been detailed in sec.~\ref{methodology}. We summarise our theoretical HOD modelling formalism and model fitting details in sec.~\ref{theory}. Results and inferences are discussed in sec.~\ref{results} and previous studies employing the same datasets have been compared in sec.~\ref{comparison}. We finally discuss the issues and challenges associated with photometric datasets in inferring galaxy-halo connections and possible future directions of improvements in sec.~\ref{challenges} and present the summary of the results from this paper in sec.~\ref{summary}.     

In this paper, we assume a standard 6-parameter flat $\Lambda$CDM cosmology with cosmological parameters set by cosmic microwave background observations \citep{2016A&A...594A..13P}. We use $(\Omega_{m} \, ,\Omega_{\Lambda} \, ,\Omega_{b} \, ,\sigma_{8} \, , n_s \, ,h ) = ( 0.309, 0.691, 0.049, 0.816, 0.967, 0.677)$, where, $\Omega_{m}, \Omega_{\Lambda}, \Omega_{b}$ denote the matter, dark energy and baryonic density with respect to the critical density of the Universe, $\sigma_8$ is related to the variance of density fluctuations on scale of $8\chimp$, $n_{\rm s}$ is the power law index of the power spectrum of density fluctuations on large scales, and $h$ is the dimensionless Hubble parameter given by $h = H_0/ 100\, {\rm kms^{-1} Mpc^{-1}}$. All the distances are measured in comoving units of $h^{-1}{\rm Mpc}$ and stellar, halo masses are expressed in units of $h^{-2}{\rm M_\odot}$ and $h^{-1}{\rm M_\odot}$ respectively. Throughout the paper, we use $\log$ to denote 10-based logarithms.

\section{Data} \label{data}
\subsection{HSC-SSP survey}
The Hyper Suprime-Cam instrument \citep[HSC;][]{2018PASJ...70S...1M, 2018PASJ...70S...2K} is a wide field imaging camera ($1.5\degr$ FoV diameter) mounted on the prime focus of the 8.2m Subaru Telescope located at the summit of Mauna kea in Hawaii. The Hyper Suprime-Cam survey, a Subaru Strategic Program \citep[HSC-SSP;][]{2018PASJ...70S...4A, 2018PASJ...70S...8A, 2019PASJ...71..114A, 2022PASJ...74..247A} is a three-layered (wide, deep and ultra-deep), multi-band ($grizy$ plus 4 narrow-band filters) imaging survey. The HSC survey has efficiently imaged $\sim 1200$ sq.~deg. of the sky in its wide layer, utilizing the excellent seeing conditions at the summit and the large FoV of the camera. The data is processed using a fork of the Rubin LSST science pipelines \citep{2017ASPC..512..279J}. The processed data from the survey has been released publicly at regular intervals. The measurement of the weak lensing signal requires well calibrated measurements of the shapes of galaxies. In our work, we use the first year shape catalog made public by the HSC survey collaboration to measure the weak lensing signal.

\subsection{First year HSC shape catalog}
The first year HSC shape catalog is based on an internal data release of the HSC survey (S16A). It consists of wide layer data observed over a period of 90 nights between March 2014 - April 2016. It covers an area of ${\rm \sim 140\ deg^2}$ spread over six disjoint fields - HECTOMAP, VVDS, WIDE12H, GAMA15H, GAMA09H, and XMM. The shape measurements are performed in the $i$-band. Therefore, the imaging in the $i$-band was carried out when the full width at half maximum (FWHM) for the seeing was better than $\sim 0.8\arcsec$. This results in the median $i$-band seeing FWHM of $0.58\arcsec$. The corresponding $5\sigma$ point-source depth of the survey is $i\sim26$ averaged over the area covered by S16A. 

The resulting data was processed with the HSC pipeline \citep[\texttt{hscPipe} \textsc{v4.0.2},][]{2018PASJ...70S...5B} and the shape catalog was curated by applying a number of quality flags and several selection criteria as described in \citet[][]{2018PASJ...70S..25M}. The resultant catalog covers an area of $\sim {\rm 136.9\ deg^2}$. The shapes of galaxies were measured using a moments based method which corrects for the effects of the PSF using the re-Gaussianization technique \citep[][]{2003MNRAS.343..459H}. The two components of the ellipticities are given by,
\begin{align}
    (e_1, e_2) = \frac{1-r^2}{1+r^2}(\cos 2\psi, \sin 2\psi)
\end{align}
where $r$ denotes the minor-to-major axis ratio and $\psi$ the angle made by the major axis with respect to the equatorial coordinate system.

The final shape catalog consists of galaxies selected from the full depth-full color region in all five filters. Apart from some basic quality cuts related to pixel level information, the catalog includes extended objects with an extinction corrected cmodel magnitude $i<24.5$, $i$-band SNR$\ge 10$, resolution factor $R_2\ge0.3$, $>5\sigma$ detection in at least two bands other than $i$ and a cut on the blendedness of the galaxy in the $i$-band. This conservative selection of galaxies results in an unweighted (raw) source number density of $24.6\ {\rm arcmin^{-2}}$. When lensing related weights are taken in to consideration, the effective number density of sources is $\sim 21.8\ {\rm arcmin^{-2}}$, with a sample of galaxies with median redshift of $\sim 0.8$. The additive ($c_1, c_2$) and multiplicative biases ($m$) in the shape measurements, as well as the RMS intrinsic distortion of shapes ($e_{\rm rms}$) and the photon noise component ($\sigma_{\rm e}$) were calibrated using detailed image simulations \citep{2018MNRAS.481.3170M} with the software {\sc GALSIM} \citep[][v1.4.2]{2015A&C....10..121R}. These image simulations account for the survey characteristics such as the variation in depth and seeing. The shape catalog is accompanied with inverse variance weights $w_{\rm s}$ for each galaxy which is given by
\begin{align}
    w_{\rm s} = \frac{1}{\sigma_{\rm e}^2 + e^2_{\rm rms}}\, .
    \label{eq:shape_weight}
\end{align}
The shape catalog satisfies a number of systematics and null tests, with residual systematics at the level of $0.01$, sufficient to carry out cosmological analyses with the data.

The shape catalog is also supplemented with six different catalogs of photometric redshifts of galaxies as inferred by a number of methods, some relying on fitting the photometry using templates, while others use machine learning \citep{2018PASJ...70S...9T}. In our analysis we use the estimates of the redshifts provided by \textsc{MIZUKI} code \citep{2015ApJ...801...20T}, which uses templates of galaxy spectral energy distributions (SEDs) and priors to fit the observed photometry of galaxies. It assumes an exponentially decaying star formation history with a variable decay time scale, along with a solar metallicity for the SED templates. It also assumes that the initial mass function is Chabrier \citep{2003PASP..115..763C} and that the dust attenuation is given by \citet{2000ApJ...533..682C}. Finally nebular emission lines are also added to the SEDs. In addition to various point estimates (e.g. mean, median, mode, best) and the posterior distribution functions (PDFs) of the redshift for individual galaxies, the code also outputs several physical properties, such as stellar mass and specific star formation rate of these galaxies. We will use galaxies with reliable photometric redshifts and thus restrict our source galaxy sample to those galaxies which have {\sc photoz\_risk\_best} $< 0.5$.

\begin{figure}
    \centering
    \includegraphics[width=0.99\columnwidth]{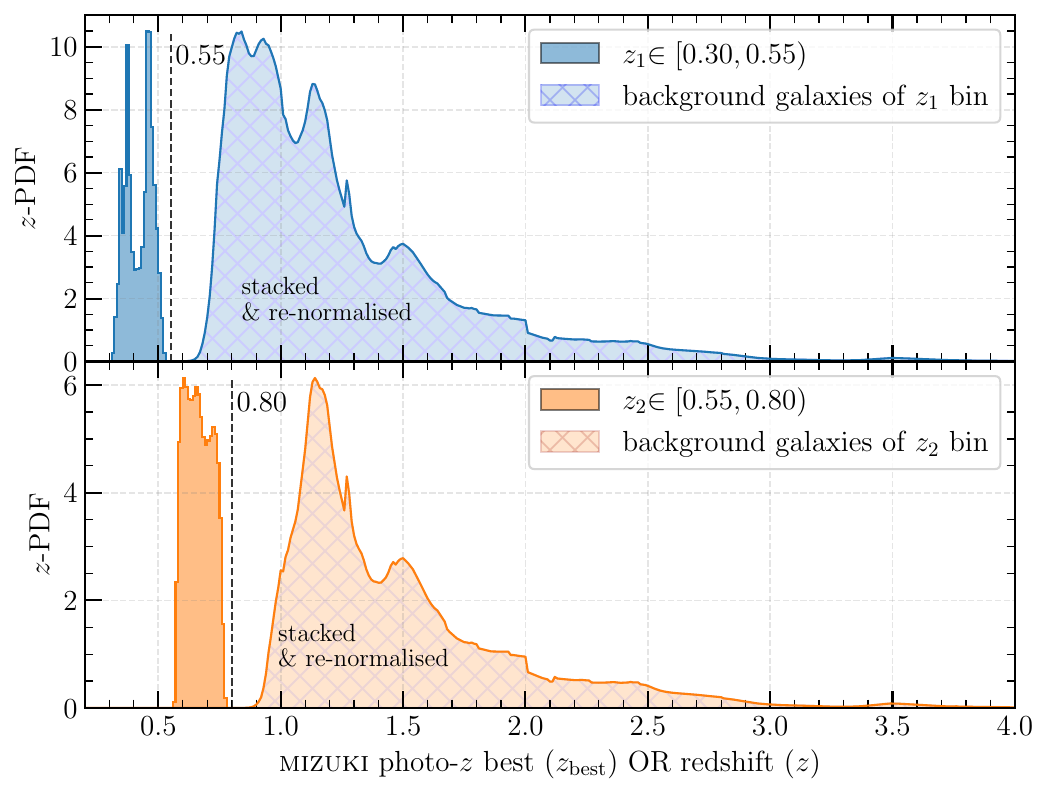}
    \caption{The distribution of lens redshifts in the {\sc mizuki} sample of galaxies after applying the full depth-full color and the star mask for each redshift bin is shown with solid shaded region. This can be compared to the stacked distribution of redshift for the source galaxies in each redshift bin, shown with hatched region.}
    \label{fig:galaxy_count_zbins}
\end{figure}

\subsection{Lens galaxies} \label{lens_sampling}
As our lens galaxies, we will use the galaxy samples presented by \citet[][]{2020ApJ...904..128I} in their HOD analysis of the clustering of these galaxies. In brief, our sample excludes galaxies centered on pixels at the edge of photometric images, or affected by cosmic rays, or have saturated pixels using the following flags: {\sc flags\_pixel\_edge, flags\_pixel\_interpolated\_center, flags\_pixel\_saturated\_center, flags\_pixel\_cr\_center}, and {\sc flags\_pixel\_bad}. We also avoid galaxies with bad fits to the SED models and remove those with $\chi^2/{\rm dof} \geq 3$ or {\sc photoz\_risk\_best} $\geq 0.1$ to use as our lens galaxies. In addition to the above cuts already mentioned in I20, we also apply the full-depth full color mask to the lens galaxy sample, to avoid selecting lenses from regions which were not observed in all bands to the nominal depth of the HSC survey. Finally, we also apply the same star mask \citep{2018PASJ...70S...7C} as that applied to the weak lensing shape catalog (S16A) which ensures full overlap of the lens galaxies spanning 125.7 ${\rm deg^2}$ on the sky with the source catalog.

We will focus on the first two redshift bins presented in I20 and use galaxy samples with $0.30 \leq z_{\rm best} < 0.55$ (Bin $z_1$) and $0.55\leq z_{\rm best} < 0.80$ (Bin $z_2$). These subsamples have redshifts that are smaller than the median of the redshifts of the source galaxies we use for the weak lensing signals. This allows us to get better signal-to-noise ratios in our measurements. In order to select lens galaxies that reliably lie in the redshift bins of our interest, we follow \citet[see Eq. 3,][]{2020ApJ...904..128I} and exclude galaxies which are within one standard deviation error (as reported by {\sc MIZUKI}) from the bin edges that define the galaxy samples. The redshift distribution of the samples can be seen in Fig.~2 of I20 and Fig.~\ref{fig:galaxy_count_zbins} after applying additional quality masks as mentioned above.

We will further divide the galaxy samples in each redshift bin using $M_*$ - the median estimate of the stellar mass posterior distribution as provided by {\sc MIZUKI}. We note that \citet{2018PASJ...70S...9T} uses $h=0.7$ to convert h-factors in $M_*$ and we also use the same to change stellar mass units from $\hiimsun$ to $\himsun$ whenever required. We construct stellar mass threshold subsamples within each of the redshift bins. Given the flux limit of HSC, we do not use galaxies with stellar masses below $10^{8.6}\hiimsun$ and $10^{9}\hiimsun$ for the redshift bins $z_1$ and $z_2$, respectively. For bin $z_1$ ($z_2$) we make 13 (12) stellar mass threshold subsamples, whose statistics are listed in Table~\ref{tab:z1z2data}.
\begin{table}
    \centering 
    \begin{tabular}{llcclc}
        \hline\hline \\[-1.5ex]
        \multicolumn{1}{c}{} & \multicolumn{2}{c}{$ z_1\, \in [0.30, 0.55)$} & & \multicolumn{2}{c}{$z_2\, \in [0.55, 0.80)$} \\[1ex]
        \cline{2-3} \cline{5-6}\\[-2ex]
        \multicolumn{1}{c}{ $\log \left[ \frac{M_{*,\rm lim}}{\hiimsun} \right]$ } & $N$ & $z_{\rm med}$ & & $N$ & $z_{\rm med}$ \\[1ex]
        \hline\\[-2.5ex] 
        8.6  & 892206 & 0.45 & & ---     & ---  \\
        8.8  & 727385 & 0.45 & & ---     & ---  \\
        9.0  & 592648 & 0.46 & & 1038903 & 0.67 \\
        9.2  & 479783 & 0.46 & & 876600  & 0.68 \\
        9.4  & 384732 & 0.46 & & 723725  & 0.68  \\
        9.6  & 305332 & 0.45 & & 590583  & 0.68 \\
        9.8  & 235427 & 0.45 & & 470783  & 0.68 \\
        10.0 & 173907 & 0.45 & & 361233  & 0.69 \\
        10.2 & 121214 & 0.45 & & 257323  & 0.69 \\
        10.4 & 76394  & 0.45 & & 166123  & 0.70 \\
        10.6 & 41422  & 0.45 & & 93603   & 0.70 \\
        10.8 & 17827  & 0.46 & & 43658   & 0.71 \\
        11.0 & 4967   & 0.47 & & 15944   & 0.71 \\ 
        11.2 & ---    & ---  & & 4356    & 0.72 \\
        \hline
    \end{tabular}
    \caption{The number of galaxies and the median redshifts of stellar-mass threshold sub-samples within redshift bins $z_1$ and $z_2$ that we use in our study. For sample construction details, please refer to Section~\ref{lens_sampling}.}
    \label{tab:z1z2data}
\end{table}

\section{Abundance of galaxies} \label{abundance}

We adopt the measurements of the abundance of galaxies as reported in I20 in order to adopt consistent abundances while comparing the results of the clustering analysis with those obtained from weak lensing. In their work, I20 compare their estimates of the SMF of photometric MIZUKI-HSC galaxies in bins of MIZUKI stellar masses and redshifts with those obtained using a multi-band, multi-survey data available in COSMOS/UltraVISTA field over 1.62 $\rm deg^2$ sky area with a $\rm K_s$-band limit of 23.4 mag (90\% complete). This allows I20 to infer the completeness of the photometric HSC galaxy sample. They also computed the abundances of MIZUKI galaxies in stellar mass threshold bins\footnote{I20 and M13 abundances are available at: \url{https://github.com/0Navin0/galaxy_halo_connection_in_HSC/tree/main/abundances}}.

Abundances of galaxies derived from photometric galaxy catalogs are prone to errors and systematics due to modelling uncertainties in their redshift and stellar mass estimates. These uncertainties in photometric redshifts are also expected to be correlated, a systematic error which results in a higher (lower) redshift for the galaxy, will also end up in systematic error which assigns a higher (lower) stellar mass to the galaxy. Errors in photometric redshifts also potentially translate into errors in the abundance. To reduce the systematics related to photometric redshifts on the abundance estimates, I20 carry out a `trimming' procedure in their section 2.3.2, which removes galaxies at the redshift bin edges with uncertain redshifts. This results in a loss of volume, but can improve the reliability of the lensing measurements, by keeping galaxies which have a higher probability of being in a given redshift bin. As the photometric measurement errors and the associated photometric redshift errors are expected to increase for fainter galaxies, this trimming method is nevertheless expected to systematically affect the abundances of fainter galaxies. The comparison with COSMOS/UltraVISTA in I20 is designed to keep a tab on such effects.

\begin{figure}
        \includegraphics[width=0.99\columnwidth]{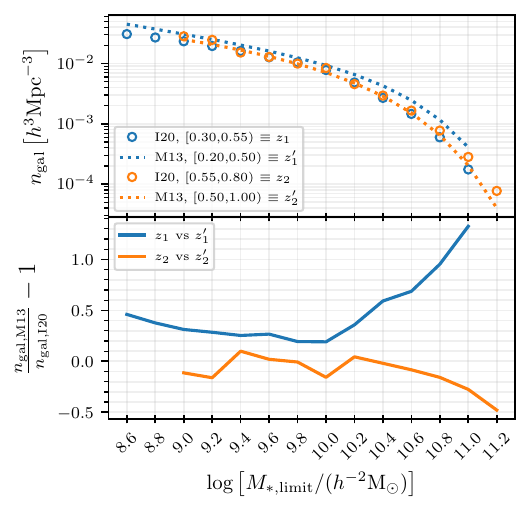} 
        \caption{ {Top panel}: Abundance measurements from literature. The measurements of the abundances from I20 for the two redshift bins are shown with blue and orange open circles, respectively. The measurements of abundances from M13 for redshift bins with significant overlap are shown as blue and orange dashed lines. These are derived using the best fit Schechter function models to the observed SMFs of galaxies from UltraVISTA. {Bottom panel:} The fractional difference between abundances from M13 versus I20 as function of stellar mass threshold in two redshift bins denoted by $z_1$ (with $z^{\prime}_1$) and $z_2$ (with $z^{\prime}_2$). We adopt a 15\% statistical error on each of these abundances in our joint HOD modelling analysis.}
        \label{fig:abundances}
\end{figure}

In order to study the impact of varying the abundances of galaxies, we will also carry out our analysis using the abundances that we compute from the best fit Schechter function models to the observed SMFs of galaxies from UltraVISTA in \citet[hereafter, M13]{2013ApJ...777...18M} and label them as M13 abundances\footnotemark[4] . 
In their study, M13 provide single and double Schechter fitting functions for SMFs of galaxies in two redshift bins [0.20,0.50)$\coloneqq z^{\prime}_1$ and [0.50,1.00)$\coloneqq z^{\prime}_2$ which are closer to our original redshift bins $z_1$ and $z_2$, respectively. We plot and compare the I20 and M13 abundances as a function of the stellar mass thresholds in Fig.~\ref{fig:abundances}.
The abundance of central galaxies is related to the abundance of dark matter halos via their halo occupation distribution. In general, galaxies in a catalog do not necessarily come with a label of central or satellite. Although algorithms to group galaxies together exist, the large errors in photometric redshifts imply that it is increasingly difficult to do so in photometric surveys. Therefore, we use a relatively large 15\% error on the abundance of galaxies for the I20 and M13 abundance measurements, so that they do not excessively drive the constraints on the halo occupation distributions. This has the effect of increasing the effective weight of our lensing signal to drive the halo occupation constraints. As mentioned before we will explore how the use of abundance changes the constraints of the stellar mass-halo mass relation we obtain.

\section{Weak gravitational lensing} \label{methodology}
Weak gravitational lensing induces statistically coherent, tangential distortions in the shapes of background galaxies due to the intervening matter distribution along the line-of-sight towards the background galaxies. The tangential component of the shear  $\gamma_t$ imparted by an intervening matter distribution is related to its excess surface density (ESD) such that
\begin{align}
    \Delta \Sigma(R) = \overline{\Sigma}(<R) - \langle\Sigma \rangle (R) = \langle\gamma_t\rangle(R) \  \Sigma_{\rm crit}(\zl,\zs)\,. \label{esd}
\end{align}
Here $\Sigma(R)$ is the lens surface mass density at a projected separation $R$ from the lens centre at redshift $\zl$, $\overline{\Sigma}(<R)$ denotes the surface density averaged within a circular aperture $R$ from the lens centre, and $\langle\Sigma \rangle (R)$ is the surface density averaged azimuthally at a distance $R$. The quantity $\Sigma_{\rm crit}(\zl,\zs)$, is a geometrical factor dependent upon the physical angular diameter distances between us (observer) and the lens $D_{\rm a}(\zl)$, us and the source $D_{\rm a}(\zs)$ and between the lens and source, $D_{\rm a}(\zl,\zs)$, and is given by, 
\begin{align}
    \Sigma_{\rm crit}(\zl,\zs) = \frac{c^2}{4\pi G} \frac{D_{\rm a}(\zs)}{D_{\rm a}(\zl) D_{\rm a}(\zl,\zs) (1+\zl)^2} \equiv \Sigma_{\rm crit, ls}\ . \label{sigcrit}
\end{align}
The factor  of $(1+z_{\rm l})^2$ in the denominator corresponds to our use of comoving coordinates. The intrinsic shapes of galaxies contribute to the noise in the determination of this shear from the measured ellipticity of galaxies. Therefore the signal has to be measured statistically by averaging the tangential ellipticity over a large sample of galaxies using weights that yield a minimal variance estimator for $\Delta\Sigma$. For every lens-source pair, we use the weight $\wls = \ws \langle\Sigma^{-1}_{\rm crit, ls}\rangle^{2}$ while performing this average, where $\ws$ is the weight due to error in the shape measurement defined in equation (\ref{eq:shape_weight}). The weight $w_{\rm ls}$ defined above automatically down-weights lens-source pairs which are separated by a small distance from each other.

We will use full PDF of the redshift ($z$-PDF) of each source galaxy and $z_{\rm best}$ estimate of the redshift of each lens galaxy as provided by the photo-$z$ estimation code, and compute the average of the inverse critical surface mass density for each lens-source pair $\langle\Sigma^{-1}_{\rm crit, ls}\rangle$ given by,
\begin{align}
    \langle\Sigma^{-1}_{\rm crit, ls}\rangle = \frac{4\pi G (1+\zl)^2}{c^2} \int_{\zl}^\infty \frac{D_{\rm a}(\zl) D_{\rm a}(\zl,\zs)}{D_{\rm a}(\zs)} p(\zs) d\zs\,. \label{avsigcrit}
\end{align}
The minimum variance estimator for $\Delta \Sigma$ is given by
\begin{align}
        \Delta \Sigma(R) &= \frac{1}{1+\hat{m}} \left( \frac{ \sum_{\rm ls} \wls e_{\rm t,ls} \, \langle\Sigma^{-1}_{\rm crit, ls}\rangle^{-1} }{2\mathcal{R} \sum_{\rm ls} \wls}\right. \nonumber \\
         &\, \qquad\qquad\qquad\qquad\qquad  -\left.\frac{ \sum_{\rm ls} \wls c_{\rm t,ls} \, \langle\Sigma^{-1}_{\rm crit, ls}\rangle^{-1} }{\sum_{\rm ls} \wls} \right)\,, \label{esd_estimator}
\end{align}
where $e_{\rm t,ls}$ and $c_{\rm t,ls}$ are the tangential components of ellipticity and the additive bias for the source galaxy in a lens-source pair, respectively. The quantity $\hat{m}$ is the sample-averaged multiplicative bias and is given by  
\begin{align}
    \hat{m} = \frac{\sum_{\rm ls} \wls m_{\rm ls}} {\sum_{\rm ls} \wls}\,. \label{emsemble_m}
\end{align}
The symbol $\mathcal{R}$ denotes the ensemble responsivity of the measured distortions to a small shear \citep[][]{2018MNRAS.481.3170M} and can be computed using the RMS intrinsic shape distortions $e_{\rm rms}$ provided in the catalog as,
\begin{align}
   \mathcal{R} &= 1 - \frac{\sum_{\rm ls} \wls e^2_{\rm rms,ls}} {\sum_{\rm ls} \wls}\, . \label{responsivity} 
\end{align}

In addition, to minimize effects of the uncertainty in the photometric redshifts, we use only those source galaxies which satisfy 
\begin{align}
    \int_{z_{\rm l, max} + z_{\rm diff}}^\infty p(\zs) d\zs > 0.99 \ , \label{fullpofzcut}
\end{align}
where $z_{\rm l, max}$ is the maximum redshift in a lens galaxy sample, and we use $z_{\rm diff}=0.1$ in our work. This selection implies that based on the posterior of the redshift from the photometry, the source galaxies we use have a $>99\%$ probability of having a redshift greater than the farthest galaxy in the lens sample. Thus they are more likely to be true background galaxies. Even after applying this photo-$z$ filter, source galaxies can still be contaminated by structures correlated to lenses if the posteriors $p(\zs)$ are biased. Therefore, we will quantify any such contamination by looking for source galaxies clustered with our lens galaxies.

The shape noise of galaxies constitutes a dominant component of the error budget on small separation scales between the lens and the source, as the number of lens-source pairs at such separations are small in number. The error on the weak lensing signal measured around a sample of galaxies at various projected radial bins can be expected to be correlated as the same source galaxy may be used for the lensing signal around different lens galaxies in the sample. Such covariance between the measurements which arises due to shape noise can be quantified by randomly rotating the source galaxies and measuring the weak lensing signal around the lens galaxies. This preserves the number of pairs but presents a random realization of the source population ellipticities. However, on large scales we also expect the covariance due to the large scale structure. The large scale over-densities in which the lens galaxies reside can coherently shift the measurements up or down, leading to a larger covariance on such scales than that expected from just the shape noise.

\begin{figure*}
        \includegraphics[width=0.99\textwidth]{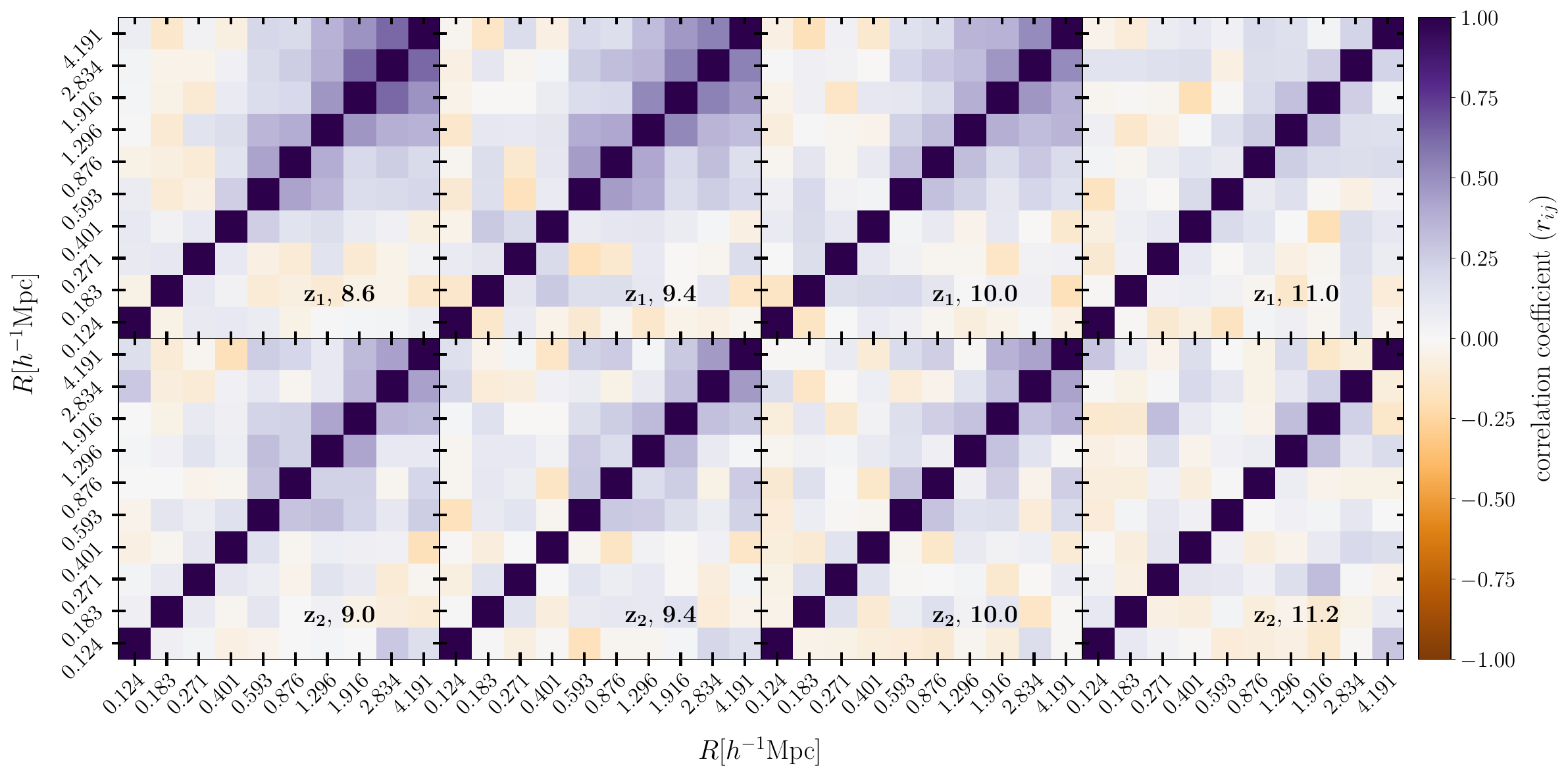}
        \caption{The cross-correlation coefficient matrices obtained for redshift bins $z_1$ (top row) and $z_2$ (bottom row) using the jack-knife technique as a function of logarithmically spaced projected radial bins in the range $[0.1-5.0]\,h^{-1}{\rm Mpc}$. We show signal correlations at the lowest, two intermediate and highest stellar mass threshold sub-samples, indicated by the text placed at bottom-right corners of each panel. The projected radial bin values are quoted as the labels of x- and y-axes.}
        \label{fig:correlation_matrices}
\end{figure*}

We account for the above sources of noise together using the jackknife technique, where we divide the full survey area of the lens catalog in to 103 rectangular jackknife regions, each having an approximately equal area $\sim 1.22\, {\rm deg^2}$, distributed contiguously in each survey field. We utilize the random catalog of points provided by the HSC survey, which have a uniform density of 100 random object per square arc minute, and where we can apply the same exact mask that we applied to our lens samples. Throughout this work, the jackknife sub-division of area remains identical for all the subsamples in each redshift bin. We then measure the lensing signals by excluding each region from the entire data at a time. We use these measurements to compute the covariance matrix $\mathcal{C}$,
\begin{align}
    C_{ij} =  \frac{N-1}{N} \sum _{k=1}^{N} \left[ \Delta\Sigma(R_{i,k}) - \overline{\Delta\Sigma}(R_{i}) \right] \left[\Delta\Sigma(R_{j,k})- \overline{\Delta\Sigma}(R_{j}) \right] \label{jackcov}\,.
\end{align}
Here the indices $i,j$ both vary from 1 to 10 for the 10 projected radial bins, $\Delta\Sigma(R_{i,k})$ is signal computed at $i^{\rm th}$ projected radial bin with removal of $k^{\rm th}$ jackknife chunk, the quantity with bar on top is an average of the jackknife measurements at a particular radial bin. 
We also define the cross-correlation matrix of the measurements between the $i^{\rm th}$ and the $j^{\rm th}$ projected radial bins to be given by
\begin{align}
    r_{ij} = \frac{\mathcal{C}_{ij}}{\sqrt{\mathcal{C}_{ii} \mathcal{C}_{jj}}}.
\end{align}
The cross-correlation matrix of the measurement for a representative set of stellar mass threshold samples in each of the redshift bins is shown in the different rows of Fig.~\ref{fig:correlation_matrices}. As expected we see that on small scales the off-diagonal components of this matrix are close to zero, however as we approach larger scales, neighbouring radial bins show enhancement in the cross-correlation of their errors. 

\begin{figure*}
        \includegraphics[width=0.99\textwidth]{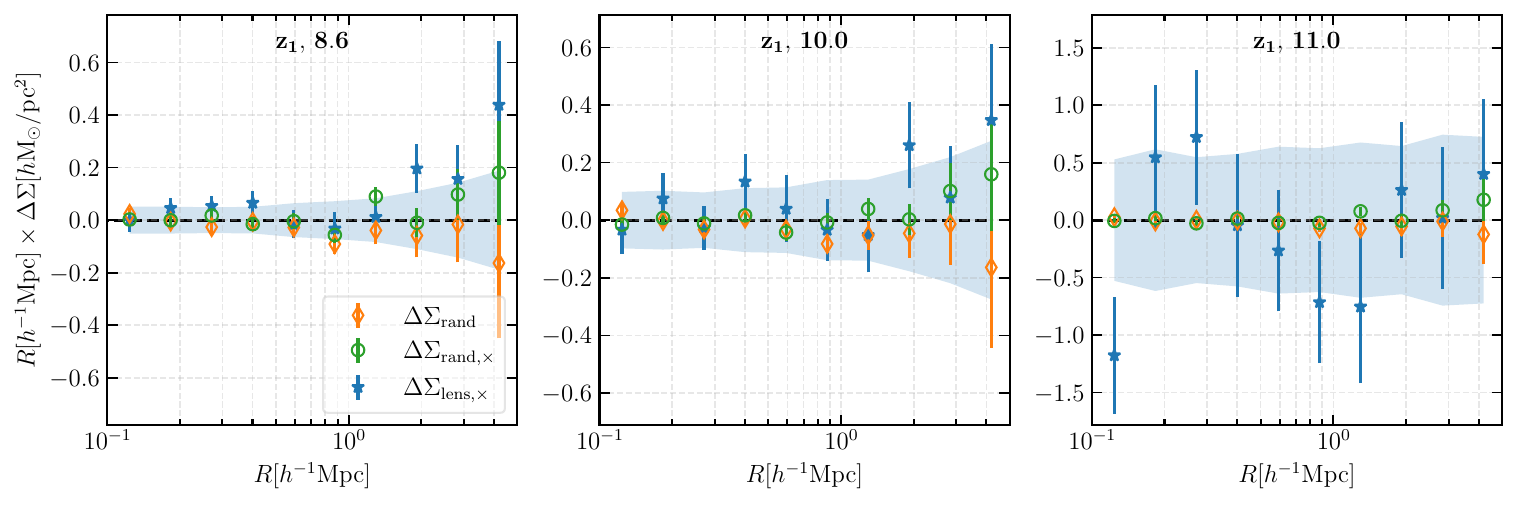} \\
        \includegraphics[width=0.99\textwidth]{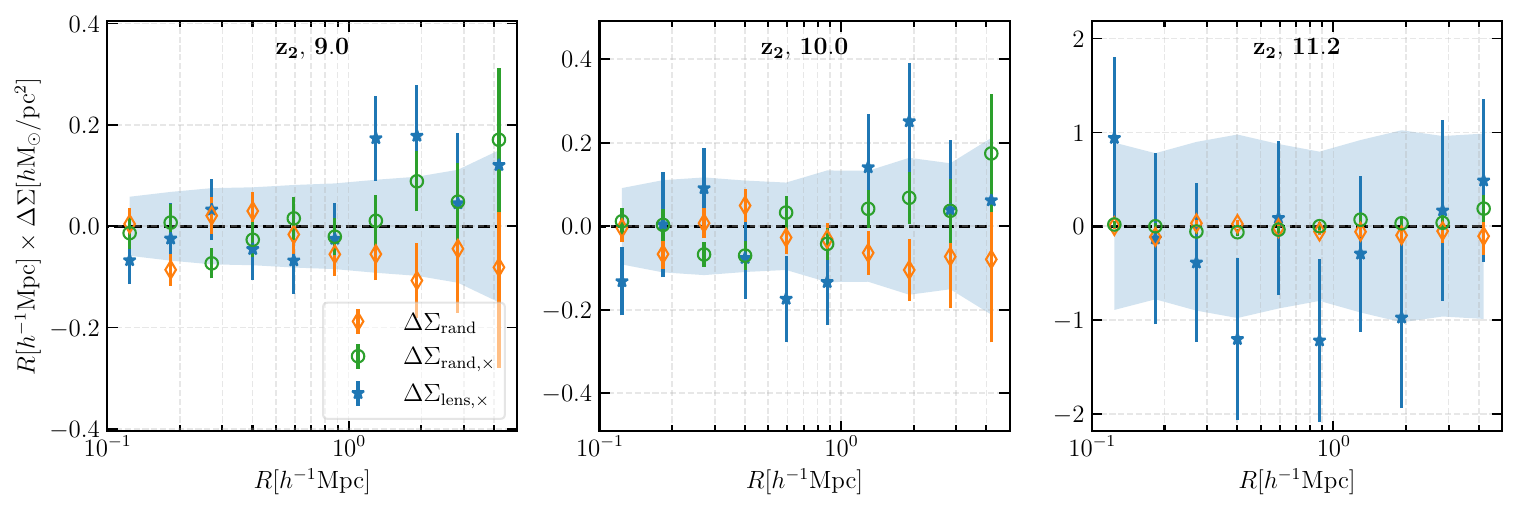}
        \caption{Systematic tests for the two redshift bins $z_1$ (top row) and $z_2$ (bottom row): The markers - diamonds and circles with errors represent 1-$\sigma$ uncertainties in plus $(\Delta\Sigma)$ and cross $(\Delta\Sigma_{\times})$ components of the measured signals around random points. We display the cross component of signals around the actual lens subsamples by stars with 1-$\sigma$ error estimates, while the shaded regions in each panel present a comparison of all three systematic probes with the 1-$\sigma$ uncertainties in the signals $(\Delta\Sigma)$ measured with high SNR (shown in each panel of Figs.~\ref{fig:z1_rpesdfit} and \ref{fig:z2_rpesdfit}) around the lens subsamples after applying all the corrections as mentioned in \ref{methodology}. All the errorbars are computed from jackknife technique as described by eq(\ref{jackcov}). On each row - the left, center and right panels show lowest, an intermediate and the highest stellar mass subsamples respectively, as is indicated by the text placed at the top center of each panel.}
        \label{fig:systematicsplots}
\end{figure*}

Next, we present the results of two null-tests of survey systematics. First, we present the measurement of the weak lensing signal $(\Delta\Sigma_{\rm rand})$ around random points which are distributed in the HSC footprint in the same manner as our lens galaxies. Second, we present the cross-component $(\Delta\Sigma_{\rm rand, \times})$ around the random points and the lens galaxies $(\Delta\Sigma_{\rm lens, \times})$; where $\Delta\Sigma_{\times}$ averages the cross-component of the shear which is the ellipticity induced on circular objects with major/minor axes at $45^\circ$ compared to the line joining the two galaxies. In the absence of systematics, both these measurements should be consistent with zero within the statistical uncertainty. 
In order to measure the signal around random galaxies, $\Delta\Sigma_{\rm rand}$, for a given threshold subsample, we resample the photometric reshifts, $z_{\rm best}$, with replacement from the overall redshift distribution of galaxies in that subsample and assign them to the objects in the random catalog. We follow the procedure described by equations~(\ref{esd}) - (\ref{fullpofzcut}) and compute the tangential component of the weak lensing signal $\Delta\Sigma_{\rm rand}$. We subtract this measured signal from the weak lensing signal around lenses from the true subsamples. Our tests indicate, however, that the measurements around random points for each of our subsamples is consistent with zero given the statistical fluctuations. The measurements $\Delta\Sigma_{\rm rand}$ and cross-components $\Delta\Sigma_{\rm rand, \times}$ around random points, as well as the cross-components $\Delta\Sigma_{\rm lens, \times}$ around lens galaxies along with their jackknife errors are shown in Fig.~\ref{fig:systematicsplots} for the lowest, a middle and the highest stellar mass threshold, respectively. The p-values to exceed $\chi^2$ for all of our subsamples for both the systematics tests are presented in Table~\ref{tab:pvalues}.

In spite of our conservative sample selection cuts and quality filters (Section \ref{lens_sampling} and  equation \ref{fullpofzcut}) in lens and source galaxies, source galaxies can still be contaminated by structures correlated to the lens distribution. These source galaxies may not be even down weighted by the lensing weights if their $p(\zs)$ is biased to high redshifts. This effectively dilutes the lensing signal as a function of projected radius. However, the overall dilution can be estimated and adjusted for by multiplying a boost factor to the  signal (see e.g., \citet{2019MNRAS.489.2511V}). Boost factor, $B(R_i)$ is defined as the ratio of weighted number of $\rm l-s$ pairs per lens galaxy to random-source $\rm (r-s)$ pairs per random point, notationally,
\begin{align}
    B(R_i) = \frac{N_r \sum_{\rm ls} \wls}{N_l \sum_{\rm rs} w_{\rm rs}}.
\end{align}
We adjust the randoms corrected signals by their corresponding boost factors, in each of the ready-to-model signals and their jackknife covariances. The estimated boost factors for few of the threshold bins are described in Fig.~\ref{fig:boostplots}. The errorbars on $B(R)$ values are computed by the jackknife technique outlined by equation~(\ref{jackcov}). Apart from few smallest projected scales in most massive galaxy samples that we probe, redshift bin $z_1$ shows boost factors consistent with unity, indicating presence of a non-zero but small amount of source contamination close to the inner-most radial bin; while the redshift bin $z_2$ shows a consistent contamination of source galaxies at all scales with $B(R)$ ranging from $\sim 4\%$ at inner most radial bin to $\sim 1\%$ around outer most radii. The application of boost factor scales the signal as a function of $R$ and may affect the covariances, however the relative error in the signal remains the same. The relative errors of the signals in bins $z_1$ and $z_2$ evolve slowly from $\sim 5\%$ to $\sim 10\%$ in subsamples of increasing threshold stellar mass within $\log M_{\rm *, limit}=$ $(8.6 - 10.8)$ and $(9.0 - 11.0)$ respectively. The most massive threshold subsamples in each redshift bin have $\sim 15\%$ relative error. Given this level of statistical tolerance, we confirmed that skipping application of boost factors don't change our parameter constraints and thereby the resulting inferences, however, to maintain uniformity throughout our current and future analyses, we include boost factors on weak lensing signal measurements for all subsamples.

Also the photometric redshifts of the galaxies may have both statistical uncertainties and systematic biases. Such uncertainties could cause galaxies that are physically correlated with the lens samples to be included in our source samples, or could cause source galaxies to be wrongly classified as lens galaxies, or could result in background galaxies getting assigned wrong redshifts. The first of these errors are accounted for using boost factors as described in the paragraph above.
We mitigate the second error by using stringent cuts on the choice of source galaxies in this analysis, such that the fraction of source galaxies getting identified as lenses are small. Thus the bias in lensing signals will come mostly from source galaxy photometric redshifts being inconsistent with their true redshifts. We examine this effect using the methodology outlined by \citet[][]{2012MNRAS.420.3240N}\footnote{See appendix.}. We find that the source photo-$z$ biases in bins $z_1$ and $z_2$ are $\sim 1\%$ and $\sim 4\%$ respectively and we have confirmed that such level of biases do not change our results or any of our conclusions in a statistically significant manner. Consequently, we have ignored the photo-$z$ bias correction in our measurements and modelling of the weak lensing signals in this paper.

The weak lensing signals as measured using the above techniques can be seen in Figs.~\ref{fig:z1_rpesdfit} and \ref{fig:z2_rpesdfit} for the two redshift bins we consider in this paper, respectively. The errors on the data points are based on the square root of the diagonal elements of the covariance matrix as defined in equation~(\ref{jackcov}).

\begin{table*}
    \centering
    \begin{tabularx}{0.85\textwidth}{
    >{\centering\arraybackslash}X
    >{\hsize=0.5\hsize\centering\arraybackslash}X
    >{\hsize=0.5\hsize\centering\arraybackslash}X
    >{\hsize=0.5\hsize\centering\arraybackslash}X
    >{\hsize=0.0\hsize\centering\arraybackslash}X
    >{\hsize=0.5\hsize\centering\arraybackslash}X
    >{\hsize=0.5\hsize\centering\arraybackslash}X
    >{\hsize=0.5\hsize\centering\arraybackslash}X
    }
        \hline\hline \\[-1.5ex]
        \multicolumn{1}{c}{}& \multicolumn{3}{c}{$ z_1\, \in [0.30, 0.55)$} &  & \multicolumn{3}{c}{$z_2\, \in [0.55, 0.80)$} \\[1ex]
        \cline{2-4} \cline{6-8}\\[-2ex]
        \multirow{2}{*}{\Large $\log \left[ \frac{M_{*,\rm lim}}{\hiimsun} \right]$} & \multicolumn{3}{c}{\textbf{p-values of null detection hypothesis for}} & & \multicolumn{3}{c}{\textbf{p-values of null detection hypothesis for}} \\ [1ex]
        \multicolumn{1}{c}{} & $\Delta\Sigma_{\rm rand}$ & $\Delta\Sigma_{\rm rand,\times}$ & $\Delta\Sigma_{\rm lens, \times}$ & & $\Delta\Sigma_{\rm rand}$ & $\Delta\Sigma_{\rm rand, \times}$ & $\Delta\Sigma_{\rm lens, \times}$\\[0.5ex]
        \hline
        8.6  & 0.468 & 0.006 & 0.185 & & --- & --- & ---\\
        \hline
        8.8  & 0.485 & 0.304 & 0.164 & & --- & --- & ---\\
        \hline
        9.0  & 0.726 & 0.027 & 0.223 & & 0.110 & 0.106 & 0.313\\
        \hline
        9.2  & 0.249 & 0.111 & 0.330 & & 0.018 & 0.338 & 0.030\\
        \hline
        9.4  & 0.105 & 0.375 & 0.229 & & 0.053 & 0.530 & 0.019\\
        \hline
        9.6  & 0.342 & 0.751 & 0.333 & & 0.322 & 0.395 & 0.069\\
        \hline
        9.8  & 0.317 & 0.628 & 0.379 & & 0.222 & 0.671 & 0.026\\
        \hline
        10.0 & 0.250 & 0.710 & 0.331 & & 0.500 & 0.052 & 0.132\\
        \hline
        10.2 & 0.386 & 0.582 & 0.082 & & 0.337 & 0.620 & 0.671\\
        \hline
        10.4 & 0.022 & 0.986 & 0.310 & & 0.357 & 0.377 & 0.202\\
        \hline
        10.6 & 0.500 & 0.601 & 0.075 & & 0.094 & 0.494 & 0.927\\
        \hline
        10.8 & 0.422 & 0.583 & 0.711 & & 0.084 & 0.422 & 0.857\\
        \hline
        11.0 & 0.476 & 0.236 & 0.395 & & 0.035 & 0.403 & 0.228\\
        \hline
        11.2 & --- & --- & --- & & 0.036 & 0.255 & 0.623\\        
        \hline
    \end{tabularx}
    \caption{The results of the null tests for each of our stellar mass threshold subsamples : The quoted p-values represent the probability to exceed $\chi^2$ for the measured systematic signals in Fig.~\ref{fig:systematicsplots} away from zero given the statistical fluctuations. We have accounted for the full jackknife covariance matrix while computing the observed $\chi^2$ in each systematic test with 10 degrees of freedom and the signal-to-noise of these systematic signals vary in the range $\sim$ 2-5.}
    \label{tab:pvalues}
\end{table*}

\section{Theoretical modelling} \label{theory}
We use a halo occupation distribution (HOD) framework in order to model abundance and weak lensing signal. The HOD framework allows us to relate the theoretical predictions of the abundance of dark matter halos, their clustering and the dark matter distribution around these halos to the observed abundance and lensing of galaxies. The various parameters of the HOD of galaxies describe the average number of galaxies, $\avg{N}(>M_{*, \rm limit}|M)$, in a particular sample of stellar mass threshold $M_{*, \rm limit}$ that reside in halos of mass $M$. In this work we only work with galaxy samples in thresholds of stellar masses, hence for ease of notation we denote it simply as $\avg{N}(M)$. We separate the total HOD of galaxies into a separate term for central galaxies and satellite galaxies, denoted by $\avg{N_{\rm c}}(M)$ and $\avg{N_{\rm s}}(M)$, respectively, such that,
\begin{eqnarray}
\avg{N}(M) = \avg{\ncen}(M) + \avg{\nsat}(M)
\end{eqnarray}
We use a 5-parameter model to describe these separate terms \citep{2005ApJ...630....1Z, 2005ApJ...633..791Z},
\begin{eqnarray}
    \avg{\ncen}(M) &=& \frac{1}{2} \left[1+{\rm erf}\left(\frac{\log {M - \log  \Mmin}} {\sigmalogM} \right)\right] \label{cenHOD}, \\
    \avg{\nsat}(M) &=& \avg{\ncen}(M) \left( \frac{M-M_0}{M_1} \right)^\alpha, \label{satHOD}
\end{eqnarray}
where $\Mmin, \sigmalogM, M_0, M_1, \alpha$ are free parameters which are allowed to vary freely for each threshold subsample. Given that apart from an unknown intrinsic scatter, the relation between central galaxies and their halos is also obscured by uncertainties in the measured signals, we include the total scatter in the host halo masses of the central galaxies by a stochastic model expressed as equation~(\ref{cenHOD}). Assuming that each central halo hosts a single galaxy, the first equation denotes the probability that a halo of mass $M$ hosts a galaxy belonging to threshold subsample. According to the functional form, $\Mmin$ denotes the mass at which half of the halos are occupied by galaxies above the stellar mass threshold of subsample under consideration. Asymptotically, the halo occupation of central galaxies tends to unity. The satellite galaxy halo occupation number is a power law in $M-M_0$, where $M_0$ is the mass scale below which there are no satellite galaxies. $M_1$ can be seen as a typical halo mass to host a satellite galaxy, the exponent $\alpha$ as an indicator of the accumulated star formation history for galaxies of the given mass threshold. The $\avg{\ncen}(M)$ in front of the satellite halo occupation number down weights the galaxies in halos with low $\ncen$. Formally, we treat the two halo occupations to be independent, given that there are cases in which central galaxies are not necessarily the brightest galaxies in their halos.

Further, we also need to specify the position of the central galaxies within the dark matter halos. We assume that the central galaxy resides at the center of the dark matter halo. In our fiducial model we assume that satellite galaxies are distributed according to the NFW profile, 
\begin{align}
    n(r) \propto \left( \frac{r}{r_{\rm s}} \right)^{-1} \left( 1 + \frac{r}{r_{\rm s}}\right)^{-2}
\end{align}
where $r_{\rm s}$ is the scale radius of the halo and is defined as $r_{\rm s} = r_{\rm 200m}/c_{\rm 200m}$. Here $c_{\rm 200m}$ is the concentration of the dark matter within that halo and halo masses are defined to be the masses enclosed within an overdensity of 200 times the background matter density, denoted by $M_{200m}$.

The abundance of galaxies in the threshold sample can be computed from the HOD using
\begin{eqnarray}
n_{\rm gal} = \int \rmd M\,n(M) \left[\avg{\ncen(M)} + \avg{\nsat(M)}\right].
\end{eqnarray}
We use the analytical framework presented in \citep{2013MNRAS.430..725V, 2015ApJ...806....2M} in order to predict the weak lensing signal from the HOD. Here we briefly repeat the formalism for the sake of completeness. 
The ESD profile, equation~(\ref{esd}), depends on the correlated surface density of matter which is a line-of-sight projection of the galaxy-matter cross-correlation function $\xi_{\rm gm}$ at a halo-centric distance R such that
\begin{align}
    \Sigma(R) = \bar\rho \int_0^{\infty} \rmd z \,\xi_{\rm gm}\left([ R^2 + z^2]^{1/2}\right)\,.
\end{align}
Here, we have ignored the uniform density component in the computation of the surface density as it does not impact the weak lensing observables. We have also ignored any possible off-centering of central galaxies. Current modelling assumes that each halo hosts exactly one galaxy at its center 
and that the dark matter contributions from subhalos of the satellite galaxies can be safely ignored. The cross-correlation is a Fourier transform of the cross power spectrum between galaxies and dark matter and can be computed using the analytical framework developed in \citet{2013MNRAS.430..725V}.

The total cross power spectrum between galaxies and dark matter can be divided in to 4 different terms, the one halo central and satellite terms, and the two halo central and satellite terms, such that,
\begin{align}
    P^{\rm gm}(k) = P^{\rm 1h}_{\rm cm}(k) + P^{\rm 1h}_{\rm sm}(k)  + P^{\rm 2h}_{\rm cm}(k)  + P^{\rm 2h}_{\rm sm}(k)\,.
\end{align}
Each of these terms can be expressed as 
\begin{eqnarray}
P^{\rm 1h}_{\rm xm} &=& \int n(M) \rmd M \, {\cal H}_{\rm x}(k, M, z) \frac{M}{\bar\rho} u_{\rm h}(k| M, z)\,, \\
P^{\rm 2h}_{\rm xm} &=& \int n(M') \rmd M' \, {\cal H}_{\rm x}(k, M', z) \times \nonumber\\
&& \int n(\tilde M) d\tilde M \, Q(k|M',\tilde M,z) \frac{\tilde M}{\bar\rho} u_{\rm h}(k| \tilde M, z) \,,
\end{eqnarray}
and `x' stands for either central `c' or satellite `s', $Q(k|M',\tilde{M}, z)$ describes the cross-power spectrum of halos of mass $M'$ and $\tilde M$ at redshift $z$, and we use
\begin{eqnarray}
    {\cal H}_{\rm c}(k|M, z) &=& \frac{\avg{\ncen}(M)}{\bar{n}_{\rm gal}} \\
    {\cal H}_{\rm s}(k|M, z) &=& \frac{\avg{\nsat}(M)}{\bar{n}_{\rm gal}} u_{\rm s}(k|M,z)\,.
\end{eqnarray}
In the equations above, $u_{\rm s/h}(k|M,z)$ denotes the Fourier transform of the number density profile of the satellite galaxy (dark matter) distribution within the halo. As indicated previously we assume this to be given by the NFW profile. We allow the satellite and dark matter concentration to vary from the form given by \citet{2007MNRAS.378...55M} to allow for systematic uncertainties due to baryonic effects, as well as effects of averaging the dark matter profiles of halo of the same mass but varying concentrations \citep[see discussion in][]{2013MNRAS.430..725V}. We implement this with a multiplicative parameter $c_{\rm fac}$ which alters the fiducial concentration-mass relation that we adopt in this paper. We include a Gaussian prior with unit mean and a variance of $0.2$ for this parameter.

The baryonic component within the galaxy is expected to dominate the weak lensing signal at small projected separations. We model this component as a point mass contribution similar to how it has been modelled in previous studies \citep[see e.g.,][]{2015ApJ...806....2M},
\begin{align}
   \Delta \Sigma_{b}(R) = \frac{\bar{M}_{\rm bary}}{\pi R^2} \label{esd_stelm}\,,
\end{align}
where, $\bar{M}_{\rm bary}$ represents average baryonic mass of all the galaxies in a given threshold subsample. We restrict our measurement of the lensing signal to scales above $100 \chikp$, thus our measurements are not very sensitive to the baryonic component (10 percent of the signal at the innermost point for the largest stellar mass bin). Given this relative insensitivity of our results to the baryonic contribution, we simply model this term as the average of the stellar mass contribution of all galaxies within the bin of interest. The total modelled signal is then the sum of ESD due to dark matter-halos and the central baryonic component.

\subsection{HOD model fitting specifications}
We carry out a Bayesian analysis to infer the posterior distribution of model parameters given the data, $P(\Theta|\cal{D, I})$, such that
\begin{align}
P(\Theta|{\cal D, I}) \propto P({\cal D}|\Theta, {\cal I}) P(\Theta|{\cal I}) \,,
\end{align}
where ${\cal I}$ represents the choice of our model, the quantity $P({\cal D}|\Theta, {\cal I})$ is the likelihood of the data given the model parameters. and $P(\Theta|{\cal I})$ the priors on our model parameters. We assume the likelihood to be a multi-variate Gaussian, such that 
\begin{align}
    & \ln P({\cal D}|\Theta, {\cal I}) \propto \frac{\chi^2(\Theta;\mathcal{D},{\cal I})}{2} \, , \nonumber \\
    & \chi^2 = \sum_{\rm i,j} [ \widetilde{\Delta\Sigma} - \Delta\Sigma ]_{\rm i}^{\rm T} [\mathcal{C}^{-1}]_{\rm ij} [ \widetilde{\Delta\Sigma} - \Delta\Sigma ]_{\rm j} + \frac{( \widetilde{n}_{\rm gal} - n_{\rm gal})^2}{\sigma^2_{\rm gal}}\,,
\end{align}
where, the terms with tilde on top are modelled while those without tilde are observed quantities, subscripts $i,j$ stand for the $i^{\rm th}$ and $j^{\rm th}$ radial bins respectively, and the covariance matrix, $\mathcal{C}$, is obtained from jackknife technique discussed in Section \ref{methodology} (equation \ref{jackcov}). We assume uniform priors on most of our parameters (see Table~\ref{tab:priors}), unless mentioned otherwise.

\begin{table}
    \centering 
    \renewcommand{\arraystretch}{1.2}
    \begin{tabular}{cl}
    \hline
         Parameters &  Priors \\
         \hline
         $\log \left[ M_{\rm min}/ (\himsun) \right]$ & flat $(10.0, 16.0]$  \\
         $\log \left[ M_1/ (\himsun) \right]$ & flat $(10.0, 16.0]$ \\ 
         $\log \left[ M_0/ (\himsun) \right]$ & flat $(6.0, 16.0]$ \\
         $\alpha$ & flat (0.001, 5.0] \\
         $\sigmalogM$ & flat (0.001, 5.0] \\
         $c_{\rm fac}$ & $G(\mu=1,\sigma=0.2), >0$ \\ 
         \hline
    \end{tabular}
    \caption{Priors for our model parameters: The same priors have been used for all the stellar mass threshold samples in both redshift bins. All the flat prior ranges are wide enough to keep them uninformative.}
    \label{tab:priors}
\end{table}

We use the analytical HOD modelling framework from \citet{2013MNRAS.430..725V} as implemented by the software {\sc aum} \citep{2021ascl.soft08002M} in order to predict the abundance and galaxy-galaxy lensing predictions given the HOD parameters. We sample the posterior distribution of our parameters given the measurements using the affine invariant MCMC ensemble sampler of \citet{2010CAMCS...5...65G} as implemented in the publicly available package {\sc emcee v3.1.1} \citep{2013PASP..125..306F}. We use 256 walkers for a total of 10000 steps. We remove the first 2000 steps from each walker as a burn-in phase and verify the stationarity of our parameters of interest to confirm convergence.

\subsection{Model predictions} \label{predictions}
In addition to modelling the observables, the $\Delta \Sigma$ and the abundances, we also compute predictions of satellite fractions, 
\begin{eqnarray}
\fsat = \frac{\int \rmd M\,n(M) \avg{\nsat(M)}}{\avg{N}} \label{satfrac}
\end{eqnarray}
and average central halo masses,
\begin{eqnarray}
\avg{M_{\rm cen}} = \frac{\int \rmd M\,M\,n(M) \avg{\ncen(M)}}{\avg{N_{\rm c}}} 
\end{eqnarray}
for each threshold subsample accounting for the full sampled posterior distributions. Where $\avg{N}=\avg{N_{\rm c}}+\avg{N_{\rm s}}$ is the total number of galaxies computed for a given subsample and $\avg{N_{\rm x}}=\int \rmd M\,n(M) \avg{N_{\rm x}}$; `x' stands for either `c' or `s'.

\section{Results and Discussion} \label{results}

We measure the weak gravitational lensing signal for stellar masses from $\log \left[M_*/(\hiimsun) \right] \geq 8.6$ in $z \in [0.3, 0.55]$ and $\log \left[M_*/(\hiimsun) \right] \geq 9.0$ in $z \in [0.55, 0.80]$. Our measurements for the different threshold bins at the two different epochs are shown as black circles in Figs.~\ref{fig:z1_rpesdfit} and \ref{fig:z2_rpesdfit}, respectively. The errors on the points are the square root of the diagonal of the error covariance matrix for each measurement. The figures show $R \Delta\Sigma$ as a function of the projected separation from the lens galaxies and we list the SNR of the measurments in the lower right boxes in each of the subpanels. The weak lensing measurements in each of the redshift bins clearly show that the weak lensing signal increases in strength for lens galaxies with a higher threshold in stellar mass. The lensing signal also show deviation from $\Delta \Sigma \propto R$ as would be expected for a simple isothermal profile.

\subsection{HOD modelling of the abundance and lensing signal}

\begin{figure*}
    \includegraphics[width=0.99\textwidth]{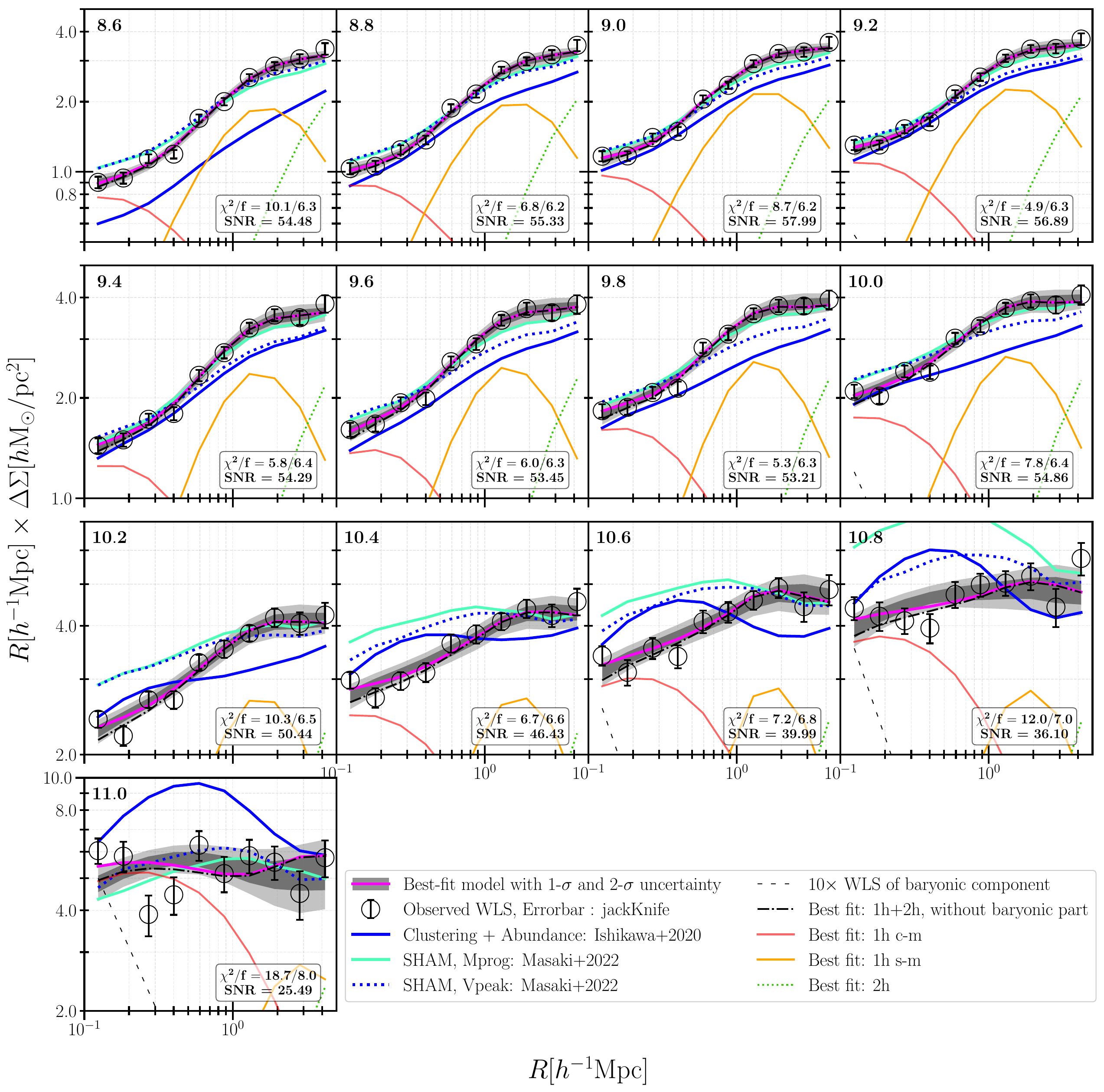}
    \caption{Model fits to the observed WLS as a function of projected radial distance from lens galaxies for all the subsamples of $z_1$ bin: The threshold stellar-mass bin corresponding to each panel is indicated with a number in the top-left corner. The measured WLS is shown as black open circles with errors estimated from the Jackknife technique given by eq(\ref{jackcov}). A text-box at bottom-right of each subsample panel shows the SNR of the measured WLS and $\chi^2$/d.o.f. for the best fitting model, both, after accounting for the full data-covariance matrix. The best-fit model WLS (magenta curve) obtained by fitting both, the observed - WLS and I20 abundance in each subsample, is the sum of: baryonic (loosely-dashed black line due to a central, fixed, point-mass) and the best fit - $1{\rm h}+2{\rm h}$ dark matter contributions (long dashed,dot black-curve). The WLS shaded regions show the 68 and 95 percent credible intervals (labelled as 1-$\sigma$ and  2-$\sigma$) of the posterior signal distribution at each radial bin. The red, orange and dotted green curves represent the best fit WLS contributions from $1h$ central-matter, $1h$ satellite-matter and 2-halo terms. Best fit predictions from other studies are also plotted for the sake of comparison - blue-dotted and light-green lines are predictions of $\Vpeak$ and $\Mprog$ SHAM models of M22 obtained using mini-Uchuu simulations respectively; blue lines are predictions of clustering and abundance based HOD analysis of I20.}
    \label{fig:z1_rpesdfit}
\end{figure*}

\begin{figure*}
    \includegraphics[width=0.99\textwidth]{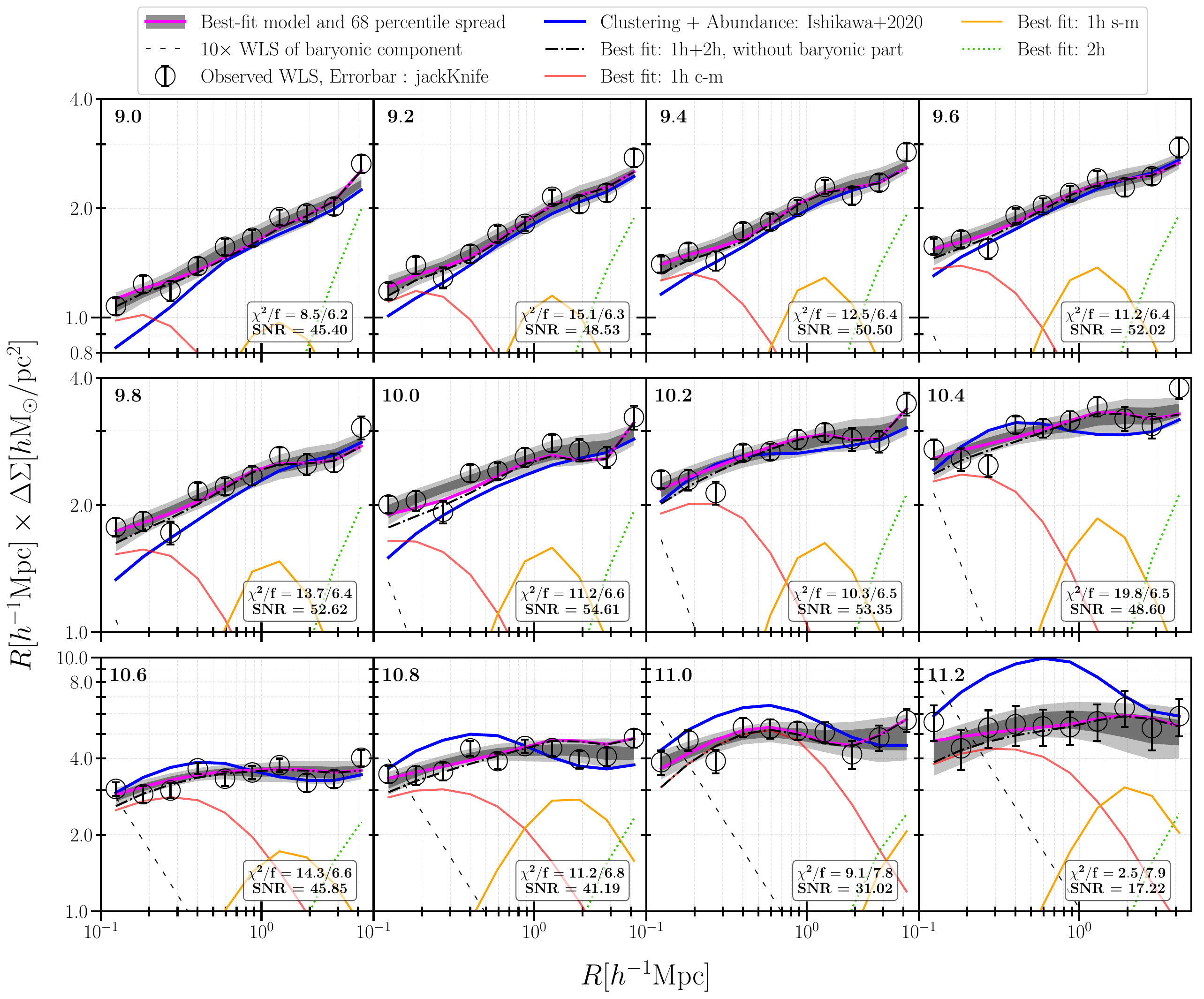}
    \caption{Same as Fig.~\ref{fig:z1_rpesdfit}, but for redshift bin $z_2$.}
    \label{fig:z2_rpesdfit}
\end{figure*}

We fit the analytical HOD model to each of the measurements described above and obtain the posterior distribution of the parameters of our model given the measurements. The priors that we use on the parameters for our analysis are listed in Table~\ref{tab:priors}. The solid magenta lines in Figs.~\ref{fig:z1_rpesdfit} and \ref{fig:z2_rpesdfit} and the associated grey shaded regions indicate the best fit model and the 68, 95 percentile credible intervals using the parameters given the joint fit of the lensing and I20 abundance measurements in the two photometric redshift bins $z_1$ and $z_2$, respectively. The best fit $\chi^2$ value obtained from our measurements alongwith the number of degrees of freedom based on the formalism of \citet[][see eq 29]{2019PhRvD..99d3506R} are also indicated in the boxes on the lower right in each of the subpanels.

We decompose the best fit model we obtain into components that correspond to the 1-halo central and 1-halo satellite term, in addition to the 2-halo term indicated by the solid red, solid orange and dotted green lines, respectively. The baryonic contribution to the lensing signals is quite small and we artificially have boosted it ten times its value for clarity and shown it with a dashed line. The 1-halo central component dominates in the innermost regions upto a few hundred kiloparsecs, followed by the rising 1-halo satellite component as we move further out. 

The increasing amplitude of the observed lensing signal can be fit with a consistently rising 1-halo central component. Statistically this indicates that central galaxies with higher stellar masses live in more massive dark matter halos. The satellite component corresponds to halos which are more massive than that of centrals. These measurements and our modelling allow us to infer the stellar mass-halo mass relation for the central galaxies together with the satellite fractions in each of our subsamples, and these are a reflection of the scale dependence of the measured weak lensing signal.

Our results indicate that a simple dark-matter only HOD model in $\Lambda$CDM cosmology is flexible enough to describe the observed lensing and abundance measurements in each of the threshold stellar mass bins. The best fit $\chi^{2}$ values corresponding to joint fits of weak lensing with either I20 or M13 abundances are listed in Table~\ref{tab:chisq_comparison_for_abundance_choices}. We obtain similar values for $\chi^2$ despite large differences in abundances between I20 and M13, which hint towards a potential degeneracy among HOD parameters when fitting weak lensing and abundances. Even though they appear statistically consistent, we see some evidence that I20 is better fit than M13 in low threshold mass subsamples for the $z_1$ bin, while M13 is better fit for high and low mass thresholds in $z_1$ and $z_2$ bins respectively.
The two-dimensional marginalized posterior distributions\footnote{The posterior distributions for two stellar mass thresholds chosen to be representative at each redshift bin have been made available online in the appendix.} 
of free parameters show familiar degeneracies in the central halo occupation parameters $\Mmin$ and $\sigmalogM$, where an increase in one parameter can be compensated by a corresponding increase in the other parameter. We will discuss the dependence of these degeneracy on our different observable in the following subsection. The satellite parameters are often ill-constrained with a wide variety of satellite parameters leading to similar observables. The constraints on the free parameters of the HOD model along with the inferred satellite fraction, abundances and the $\avMcen$ for each of the threshold bin in two redshift bins are listed in Tables~\ref{tab:Parameter_constraints_z1} and \ref{tab:Parameter_constraints_z2}, respectively.

\subsection{Degeneracy among central HOD parameters and abundance} \label{central_HOD_discussion}
\begin{figure*}
    \begin{subfigure}[b]{0.33\textwidth}
        \includegraphics[width=1.\textwidth]{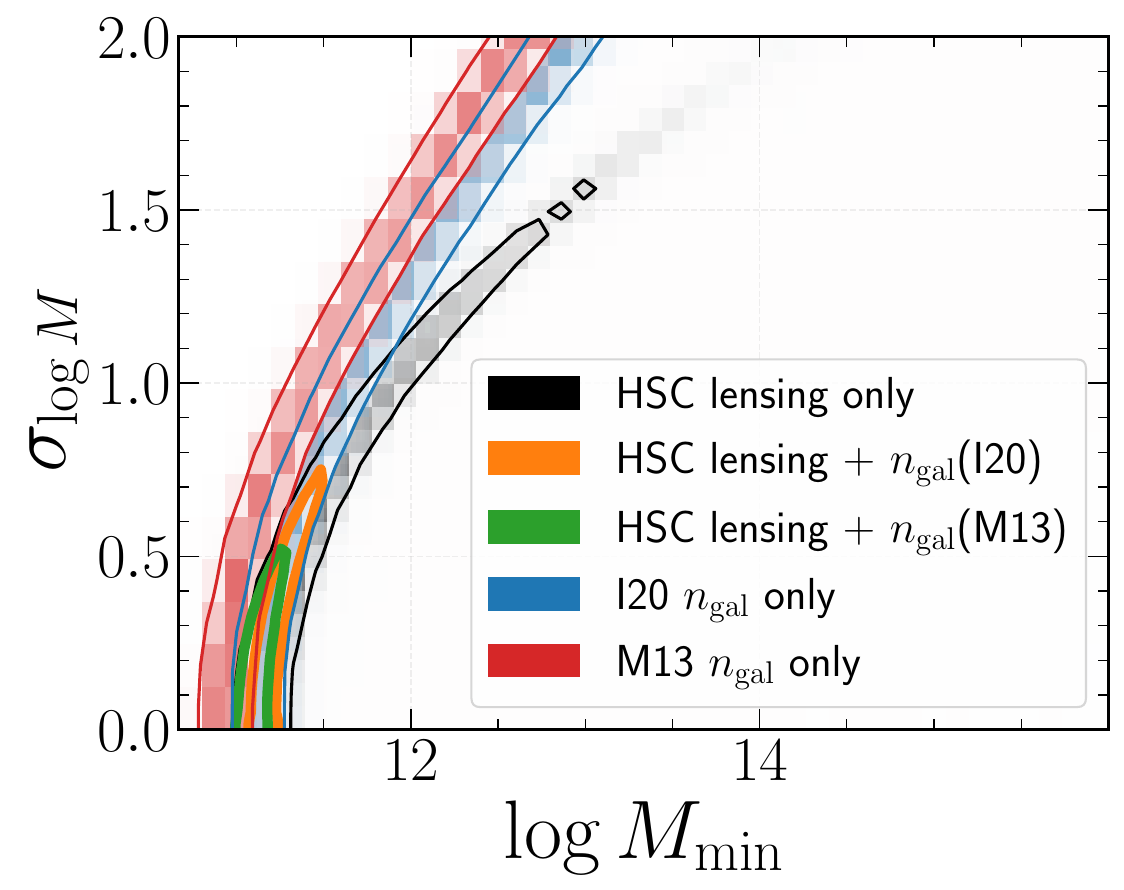}
        \caption{$z_1\, , M_* \geq 10^{8.6} \hiimsun$}
        \label{fig:logMmin_siglogm_degeneracy_z1_8.6}
    \end{subfigure}
    \begin{subfigure}[b]{0.33\textwidth}
        \includegraphics[width=1.\textwidth]{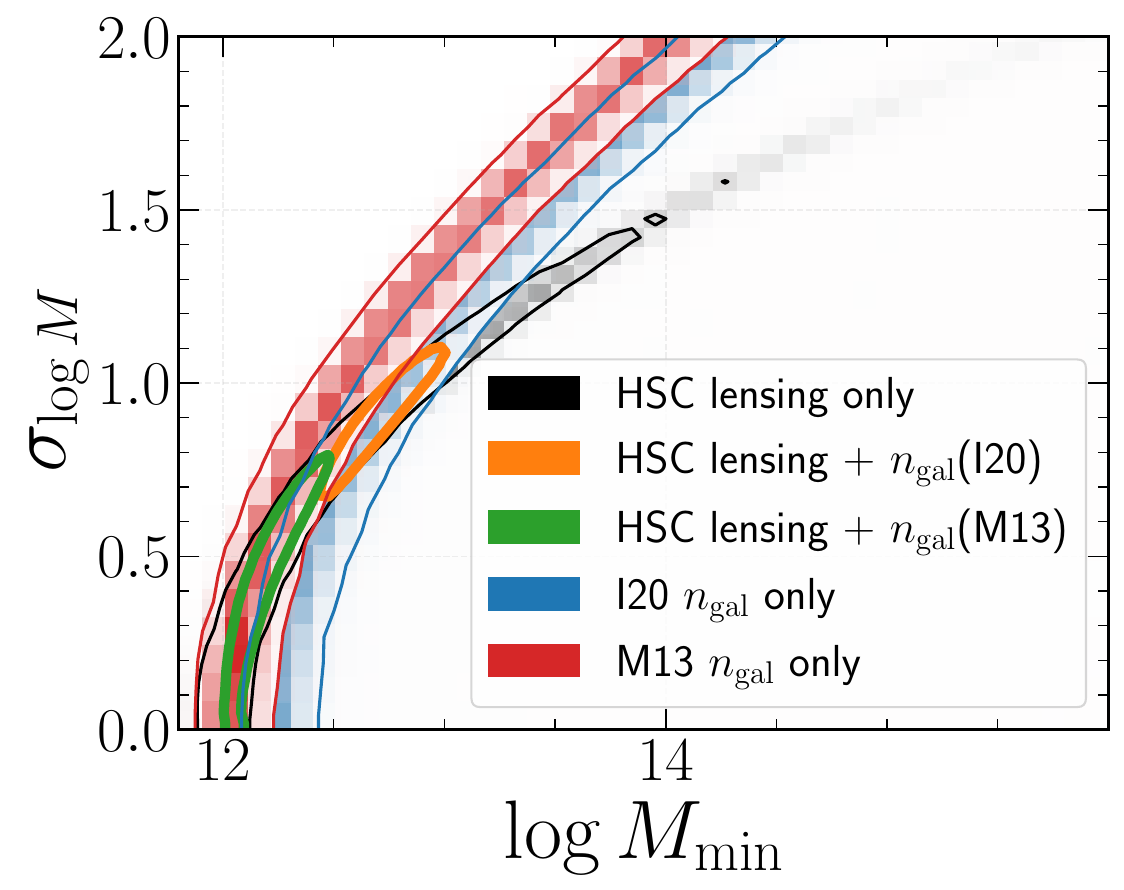}
        \caption{$z_1\, , M_* \geq 10^{10.4} \hiimsun$}
        \label{fig:logMmin_siglogm_degeneracy_z1_10.4}
    \end{subfigure}
    \begin{subfigure}[b]{0.33\textwidth}
        \includegraphics[width=1.\textwidth]{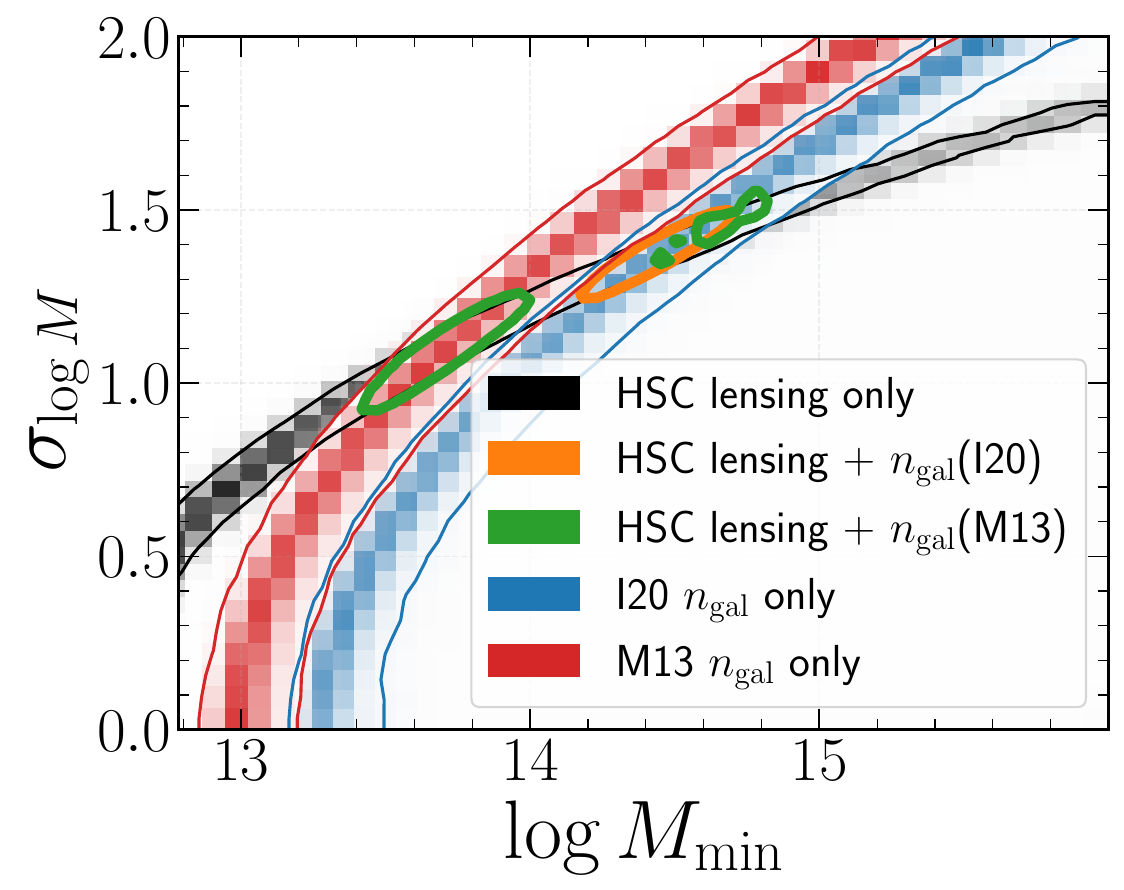}
        \caption{$z_1\, , M_* \geq 10^{11.0} \hiimsun$}
        \label{fig:logMmin_siglogm_degeneracy_z1_11.0}
    \end{subfigure}
    \caption{Various contours represent 2-dimensional marginalised 68 percentile confidence regions of central HOD parameters obtained by performing HOD model fitting to different combinations of weak lensing and abundances plotted in different colors. The HOD fits to only the weak lensing of HSC-MIZUKI galaxies, are shown with black color, joint fits to the HSC lensing and I20 abundances are shown in orange while joint fits to HSC lensing and M13 abundances are shown in green. The HOD fits that use only the I20 abundances are shown in blue, while those with only M13 abundances are shown in red.}
    \label{fig:logMmin_siglogm_degeneracy}
\end{figure*}

Using the posterior distribution of the HOD parameters in our fiducial analysis, we examine the degeneracy between the central HOD parameters and its dependency on the weak lensing and the abundance, separately. The estimates of the abundance of galaxies differ between I20 and M13, and therefore can lead to different constraints on the HOD parameters. Therefore, we fit the HOD model to these observables individually and in combination to demonstrate the impact of using either of these abundance measurements. In Fig.~\ref{fig:logMmin_siglogm_degeneracy}, we present the resulting degeneracy contours between central HOD parameters corresponding to each of these observables. The 68 percent credible regions from the weak lensing only fit, the I20 abundance only fit and the M13 abundance only fit are shown with black, blue and red contours, respectively. The orange and the green correspond to the joint fits between weak lensing and the abundances from I20 and M13, respectively. The different subpanels correspond to lens subsamples with different stellar mass thresholds for bin $z_1$, chosen for illustrative purposes.

The central HOD parameters, $\log \Mmin$ and $\sigmalogM$ for each of the observables individually are degenerate with each other and a positive change in one can be compensated by a positive change in the other parameter. This can be understood as follows. The abundance of halos is a decreasing function of halo mass. Thus increasing $\Mmin$, in general would lead to smaller abundance. However, this can be compensated by increasing the scatter $\sigmalogM$. The scatter allows galaxies to be populated in the more numerous less massive halos, thus satisfying the observed abundance. The relative shift in the degeneracy contours between the contours corresponding to I20 and M13 reflect the smaller abundance of galaxies inferred by I20 compared to M13. The weak lensing signal on the other hand is sensitive to the average halo masses of the lens samples. Thus an increase in $\Mmin$ which would nominally increase the average halo masses, can be compensated by increasing the scatter which brings in lower halo masses, thus compensating the increase. At the highest stellar mass threshold, the degeneracy contours become flatter due to the exponential decrease in the halo abundances at the high mass end. Even though the degeneracies in the $\Mmin-\sigmalogM$ are qualitatively similar, the different dependencies of the abundance of halos and their average halo mass on the HOD parameters implies that the quantitative degeneracies have different directions. The combination of the abundance and weak lensing shown in orange/green contours thus results in tighter constraints on each of the central HOD parameters, which otherwise cannot be constrained by either of the observables on their own.

We also observe that weak lensing prefers somewhat higher values for $\log M_{\rm min}$ than just the abundance information alone until stellar mass thresholds of $10^{10.2}$ ($10^{10.4}) \hiimsun$ in the redshift bin $z_1$ ($z_2$) where the lensing and abundance contours start to cross-over. The exact location of this stellar mass depends upon which study we use the abundance information from. In general, we expect that adding in the abundance information will lead to lower values of $\Mmin$ than just from weak lensing at the low stellar mass threshold end. At the high stellar mass threshold, the inclusion of abundances can have a non-negligible impact on the inferred value of $\Mmin$ as can be seen from the relatively flat degeneracy contours in the $\Mmin-\sigmalogM$ plane in the right hand subpanel of Fig.~\ref{fig:logMmin_siglogm_degeneracy}.

\subsection{Galaxy-halo connection} \label{galaxy_halo_connection}
\begin{figure*}
        \begin{subfigure}[b]{0.99\columnwidth}
        \includegraphics[width=0.99\textwidth]{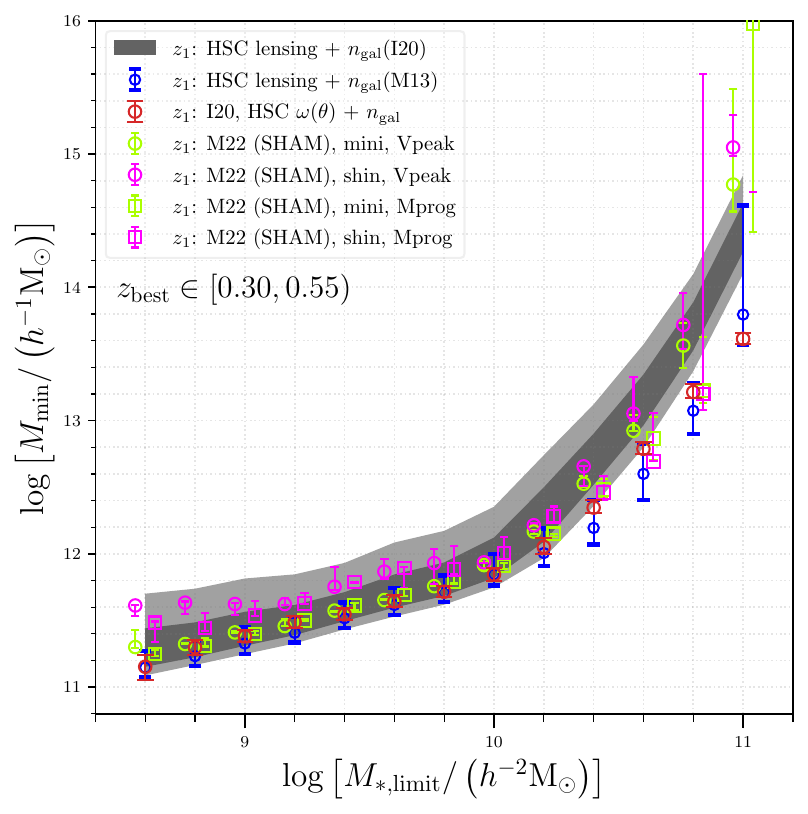}
        \end{subfigure}
        \begin{subfigure}[b]{0.99\columnwidth}
            \includegraphics[width=0.99\textwidth]{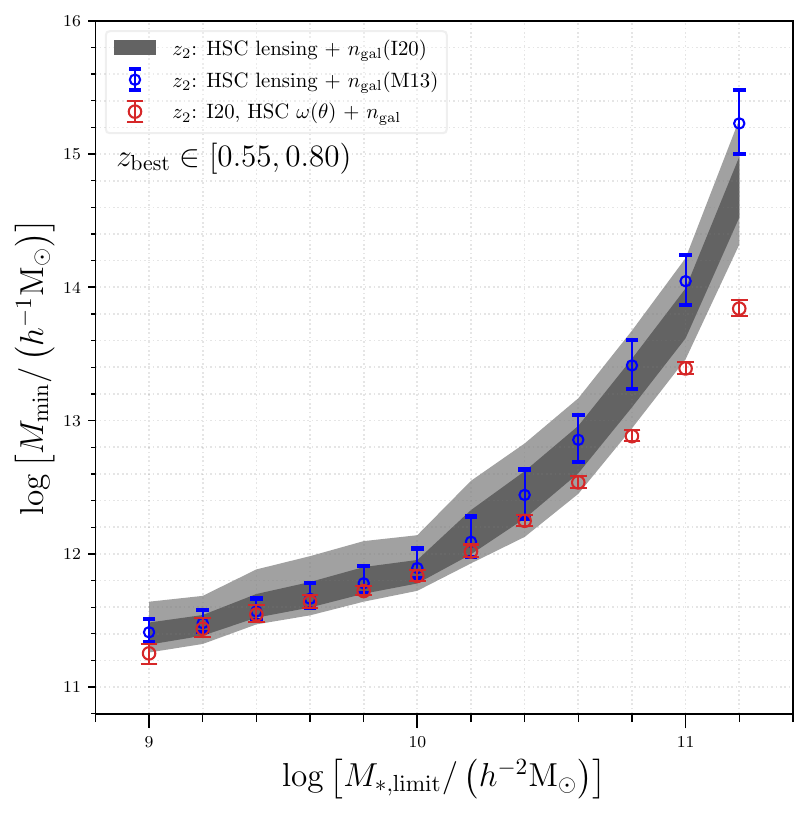}
        \end{subfigure}
        \caption{\textbf{Central halo mass parameter:} The dark and light gray shaded regions represent 68 (1-$\sigma$) and 95 (2-$\sigma$) percent credible intervals of $({\rm \log M_{min} })$ as a function of stellar-mass threshold $(\log \mathrm{M_{*,\rm limit}})$ respectively in redshift bins \textbf{Left:} $z_1$ and \textbf{right:} $z_2$. \textbf{Left panel:} Also compares the best fit results with 1-$\sigma$ errors from M22 based on $\Mprog$ (squares) and $\Vpeak$ (circles) SHAM models for two Uchuu simulations, \textit{mini} (green) and \textit{shin} (magenta). For clarity, we have shifted the $\Mprog$ and $\Vpeak$ model points by 0.04 dex to the right and left respectively. Also given the small volume of shin-Uchuu simulation, we do not compare its results for massive galaxy thresholds beyond $10^{10.4} \hiimsun$. Refer to section \ref{galaxy_halo_connection} for discussion on our results and \ref{comparison} for comparison with other studies.}
        \label{fig:Inferred_Stellar_mass_vs_Central_halo_mass_relation}
\end{figure*}

The galaxy-halo scaling relation that we obtain from our joint analysis of weak lensing and the abundance of galaxies can be summarized with the dependence of $\logMmin$ on the stellar mass threshold $\log M_{*, \rm limit}$. The parameter $\Mmin$ corresponds to the mass at which half of the halos are occupied by galaxies in a given stellar mass threshold sample. This implies that the scaling relation between $\logMmin$ and $\log M_{*, \rm limit}$ can be interpreted as the median of the stellar mass of galaxies at fixed halo mass. We show these constraints for our two redshift bins in different panels of Fig.~\ref{fig:Inferred_Stellar_mass_vs_Central_halo_mass_relation}. Our fiducial constraints are shown as credible intervals using light (95 percent) and dark grey (68 percent) shaded regions that correspond to the use of our weak lensing measurements combined with abundance measurements presented in I20. If instead, we combine with the abundance measurements from M13, we obtain constraints shown in blue points (median) with errors (68 percent credible interval). As discussed in the previous section, the smaller abundance inferred in I20 compared to M13 leads to a higher $\Mmin$ when we use abundance from I20 for redshfit bin $z_1$. In contrast for redshift bin $z_2$, the abundance of galaxies inferred in I20 and M13 both are roughly equivalent (see Fig.~\ref{fig:abundances}) and thus the inferred $\Mmin$ is similar irrespective of which abundances we combine with the weak lensing signal.

\begin{figure*}
    \begin{subfigure}[b]{1.\textwidth}
        \includegraphics[width=1\textwidth]{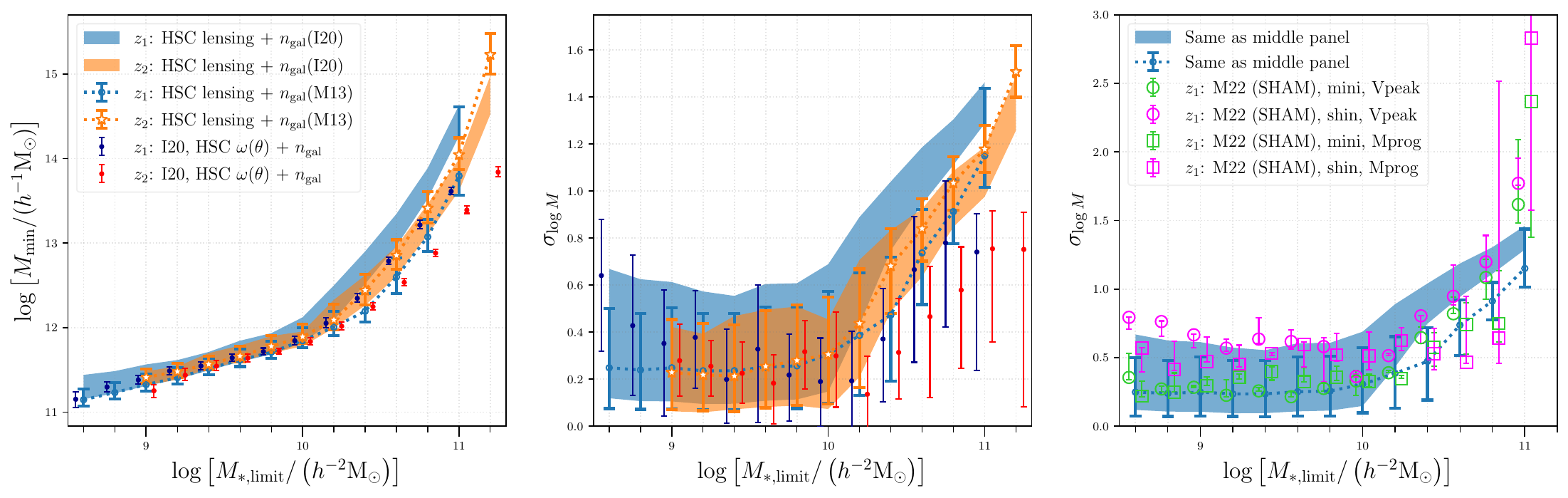}
    \end{subfigure}
    \caption{\textbf{Central HOD parameters $\logMmin$ and $\sigmalogM$, evolution and comparison with other studies:} 
    Left and Middle panels present comparison with I20 study. The blue solid shaded regions and light blue points with errors are results of our joint fits of lensing with I20 and M13 abundances respectively, where the errors represent 68 percent (1-$\sigma$) confidence intervals around median for each parameter. Color coding shows the redshift evolution in each panel, blue (dark blue) for $z_1$ and orange (red) for $z_2$ bin. Dark blue and red points with errorbars in each panel represent the best fit and 1-$\sigma$ constraints from I20. For clarity, we have shifted these blue and red points by 0.04 dex to the right and left respectively. 
    \textbf{Right panel:} compares our 1-$\sigma$ constraints on central halo mass scatter parameter $\sigmalogM$ with those of M22. Blue solid shaded regions and blue points with errors represent the same data as in middle panel. Green/magenta points with errors come from M22 SHAM work for two different simulations \textit{mini/shin} Uchuu. The points with same square/circle shapes refer to same SHAM model Mprog/Vpeak. Again for clarity, we have shifted the $\Mprog$ and $\Vpeak$ model points by 0.04 dex to the right and left respectively. Also given the small volume of shin-Uchuu simulation, we do not compare its results for massive galaxy thresholds beyond $10^{10.4} \hiimsun$. Refer to Section~\ref{galaxy_halo_connection} for discussion on results from this work and \ref{comparison} for comparison with other studies.}
    \label{fig:Central_parameter_evolution}
\end{figure*}

In both redshift bins, we observe a scaling relation which shows that $10^{12} \himsun$ dark matter halos are most efficient at forming stars and become increasingly inefficient as we move away. At lower mass side the inefficiency of star formation manifests in the stellar masses dropping down precipitously to smaller values, while at the high mass end it is seen in a quick rise in halo masses that is required to form more and more massive galaxies. Qualitatively, this picture is consistent with previous studies. We present the comparison of the parameters $\Mmin$ and $\sigmalogM$ obtained from our analysis when combining our weak lensing measurements with the two different abundance estimates in the first two panels of Fig.~\ref{fig:Central_parameter_evolution}. If taken at face value our results in left panel do not indicate a large evolution in the scaling relation within the two redshift bins, especially if we consider the abundance measurements from M13. However, the abundance measurements from I20 indicate that halos of same mass at lower redshift host galaxies with a median stellar mass which is lower by about 0.2 dex.

The scatter $\sigmalogM$ in our HOD parameterization captures the scatter in halo masses of galaxies that have stellar mass at the threshold chosen for our sample. We observe in middle panel of Fig.~\ref{fig:Central_parameter_evolution} that this scatter increases as we increase our threshold to include only massive galaxies. In models which have a fixed scatter in stellar mass of galaxies at fixed halo mass, such behaviour is expected. The slope of the $\avg{\log M_*}-M$ relation is quite shallow at the high mass end, and thus a constant scatter in the stellar masses at fixed halo mass results in a scatter in halo masses that continues to increase with the stellar mass. Our results are therefore qualitatively consistent with studies that indicate such a constant log-normal scatter in stellar masses at fixed halo mass $\sigma_{\log M_*}$ \citep[see e.g.,][]{2013ApJ...770...57B, 2019MNRAS.488.3143B, 2017MNRAS.470..651R}. These trends are consistently observed irrespective of which abundances we use and the redshift bin under consideration.

\begin{figure*}
        \begin{subfigure}[b]{0.99\columnwidth}
        \includegraphics[width=0.99\textwidth]{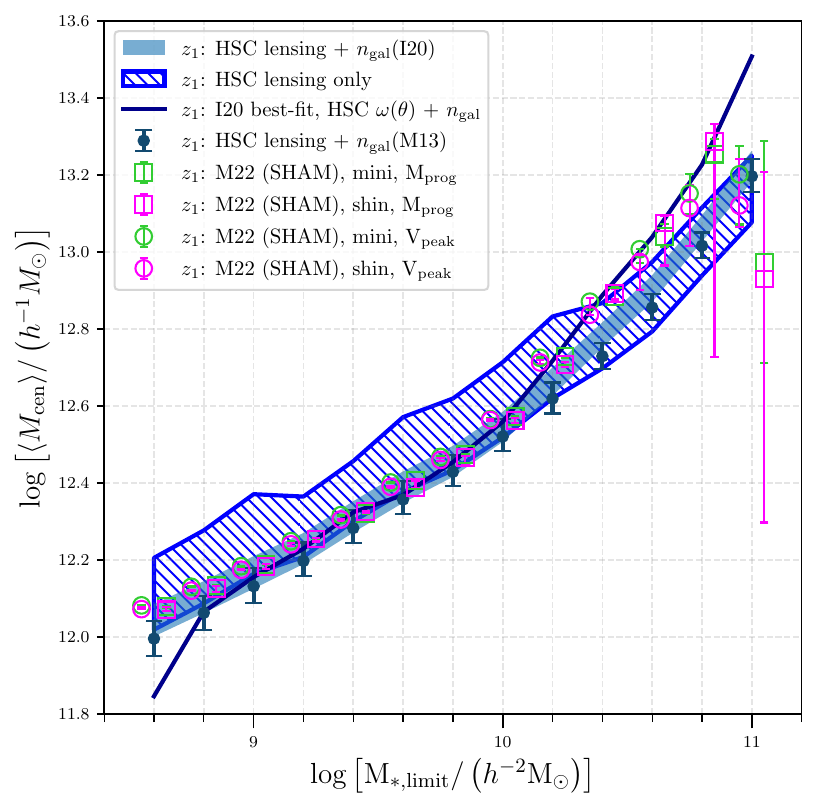}
        \caption{}
        \label{fig:avMcen_z1}
        \end{subfigure}
        \begin{subfigure}[b]{0.99\columnwidth}
            \includegraphics[width=0.99\textwidth]{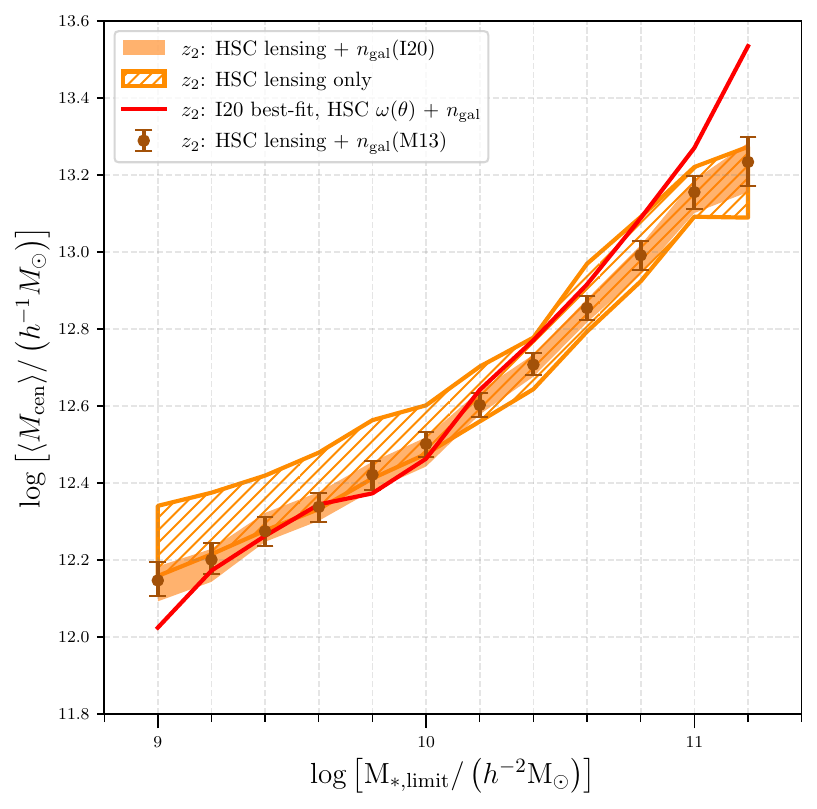}
            \caption{}
            \label{fig:avMcen_z2}
        \end{subfigure}
        \caption{\textbf{Average central halo mass:} The regions shaded with blue (orange) slanted lines, blue (orange) shaded bands and blue (orange) points(median values) with errors represent 68 percent credible intervals obtained using weak lensing only, lensing with abundance from I20 and M13 respectively for redshift bin \textbf{(a)} $z_1$ (\textbf{(b)} $z_2$) respectively. The blue/orange lines in corresponding panels represent the best fit constraints of I20. For clarity in panel \textbf{(a)}, we have shifted the $\Mprog$ and $\Vpeak$ model points by 0.04 dex to the right and left respectively. Also given the small volume of shin-Uchuu simulation, we do not compare its results for massive galaxy thresholds beyond $10^{10.4} \hiimsun$. Refer to \ref{galaxy_halo_connection} for discussion on results of this analysis and \ref{comparison} for cross-study comparisons.}
        \label{fig:avMcen_comparison}
\end{figure*}

Previously, we have shown that the parameters $\Mmin$ is degenerate with $\sigmalogM$ and the posterior constraints on $\Mmin$ are very much dependent on the choice of the abundance measurements, especially in the first redshift bin. The weak lensing signal is expected to be sensitive to the average mass of halos occupied by galaxies in our sample. Given that the small scale weak lensing signal is well measured and is dominated by the 1-halo central term, we expect the average central halo masses $\avMcen$ to be well determined by the lensing signal for every threshold stellar mass bin. In Fig.~\ref{fig:avMcen_comparison}, the blue (orange) shaded region with slanting lines shows our constraints on $\avMcen$ from weak lensing measurements only. The solid blue (orange) shaded region corresponds to the 68 percent credible intervals derived from a joint fit between lensing and abundance from I20 for redshift bin 1 (2), while the blue (orange) solid points with errors correspond to a similar joint fit but using abundance measurements from M13. While both $\Mmin$ and $\avMcen$ have physical meaning, it is clear that $\avg{M_{\rm cen}}$ better reflect the results of our weak lensing measurements and is relatively insensitive to the exact choice of abundance.

We compare the $\avg{M_{\rm cen}}$ obtained for the two redshift bins in Fig.~\ref{fig:avMcen_evolution}. When compared in this manner, we see very small differences in the redshift evolution of the scaling relation. The differences seen in the $\Mmin$ and $\sigmalogM$ compensate to result in a scaling relation between $\avg{M_{\rm cen}}$ as a function of the stellar mass threshold that shows very little evolution over the two redshift bins we use.

\subsection{Satellite fraction} \label{sat_frac}
The weak lensing signal in the innermost regions is dominated by the dark matter halo of the central galaxies in each of our stellar mass threshold subsamples. Some of the galaxies in our subsample are also expected to be satellites. These satellites on average are expected to reside in more massive parent halos than halos hosting centrals of similar stellar mass. However these satellite galaxies do not reside at the centers of their parent dark matter halos, but are distributed within the halo. This signal from the satellite galaxies is thus expected to be a result of convolution of the weak lensing signal expected around the centers of their parent halos with the projected number density of satellite galaxies within the halo. The weak lensing signal at intermediate scales is thus sensitive to the fraction of satellite galaxies within the stellar mass threshold sample as well as the halo occupation distribution of the satellite galaxies in the subsample.

In Fig.~\ref{fig:fsat_evolution}, we show the fraction of satellite galaxies as a function of the stellar mass threshold of our subsamples. The solid blue (orange) colored shaded region shows the 68 percent credible region for the satellite fraction for redshift bin 1 (2) when using the weak lensing measurements along with abundance measurements from I20. The region shaded with slanted line but with the same colors correspond to the case when the lensing measurements alone are used as constraints. To maintain clarity, we do not show results when using M13 abundances here as they are essentially similar within the errors.
Overall, the observations suggest that the satellite fractions decrease as a function of the stellar mass threshold above $10^{10} \hiimsun$ for both redshift bins. There is tentative evidence of a flattening of the satellite fractions at lower stellar mass threshold for redshift bin 1. We do not find significant evidence for the evolution of the satellite fractions with redshift given the large errors in our inference, nor do we find a significant difference depending upon which abundance constraints we use.

\section{Comparison with previous studies}\label{comparison}
As mentioned in Section~\ref{intro}, we examine the results from the two studies, I20 and M22, of clustering and abundance of galaxies from the same samples we use in the paper, against our inferences which are driven by the measured weak lensing signals and abundance estimates from I20. This comparison is well suited even in the photometric observable plane due to use of the same dataset. To briefly summarize the results and approaches of these two studies, I20 modelled the measured projected 2-point correlation functions $\omega (\theta)$, and the measured abundances of galaxies using the same HOD parameterization as used in our modelling scheme.

On the other hand, these same measurements were modelled by M22 using a modified subhalo abundance matching technique using cosmological simulations from the Uchuu suite \citep{2021MNRAS.506.4210I}, namely {\it mini} and {\it shin}. Amongst the two, mini-Uchuu has larger box size of $400 \chimp$ with particle resolution of $3.27\times 10^{8} \himsun$ and shin-Uchuu has larger resolution of $8.97\times 10^{5} \himsun$ with box size of $140 \chimp$. Comparison between the two simulations allows us to test the effect of the resolution. In their paper, M22 explore two different proxies of halo mass which monotonically correlate with the stellar mass of galaxies (albeit with a scatter). The first approach uses the traditional peak maximum circular velocity ($V_{\rm peak}$), while the second one utilizes the halo mass of the progenitor of the subhalo at a prior redshift ($M_{\rm prog}$). The constraints presented by M22 correspond only to the first redshift bin.

We utilize the best fit HOD parameters from I20 and predict the expected weak lensing signal for each of the stellar mass threshold lens samples in redshift bins 1 and 2. We use the framework prescribed in Section~\ref{theory} to predict the lensing signal. These best-fit predictions are shown as blue lines in Figs.~\ref{fig:z1_rpesdfit} and \ref{fig:z2_rpesdfit} which correspond to redshift bins 1 and 2, respectively. In redshift bin 1, we find that the I20 predictions underestimate the measured lensing signal (by about $10-30\%$) around small projected radii corresponding to the 1-halo regime for threshold mass bins up to $10^{10.0} \hiimsun$. For higher threshold bins, the predictions overestimate the measured weak lensing signal by up to $50-60\%$. Although we see qualitatively similar differences in redshift bin 2, the magnitude of these differences is much smaller than in redshift bin 1. For redshift bin 1, we show the best fit predictions for the weak lensing signal from the $\Mprog$ and $\Vpeak$ model for M22 using light green and blue dotted lines, respectively. Both the SHAM models are able to explain the lensing signals relatively well for the galaxies in mass thresholds below $10^{10}\hiimsun$ compared to more massive thresholds, especially when compared to the fits from I20. In this stellar mass range, the $\Mprog$ model seems to fit the measurements better than the $\Vpeak$ model, however, we have checked that this is a resolution dependent statement, and with the higher resolution {\it shin} run, these differences further disappear. For higher threshold stellar masses, both models seem to fare poorly. However, we see evidence that the $\Mprog$ model is at least able to capture the large scale lensing signal beyond $1 \chimp$ well. For these bins, we see appreciable differences between the measurements and the predictions on small scales for both models.

\begin{figure*}
        \begin{subfigure}[b]{0.99\columnwidth}
            \includegraphics[width=\columnwidth]{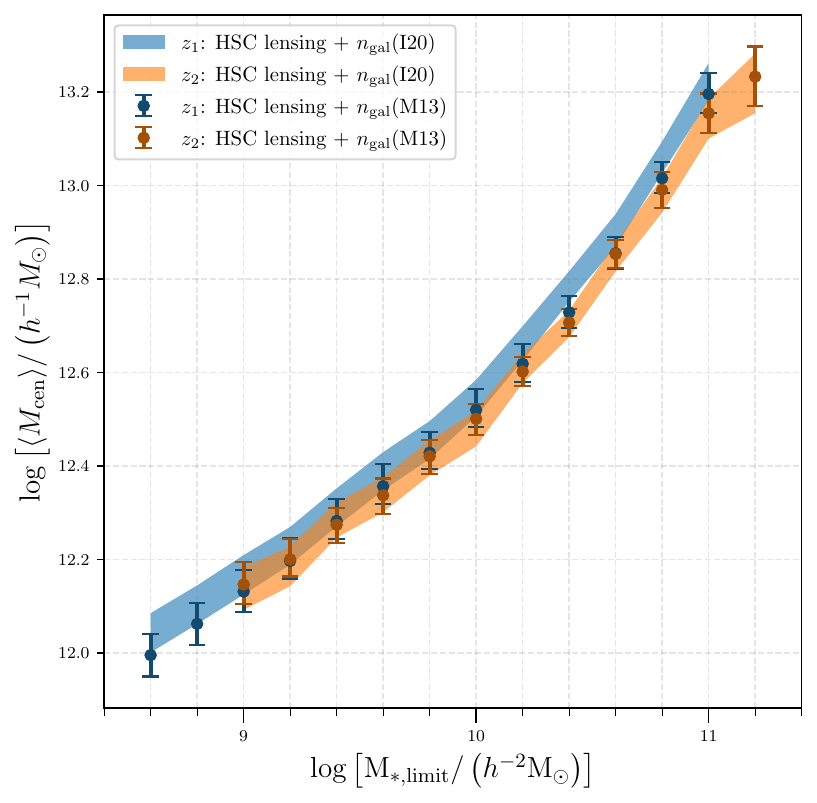}
            \caption{}
            \label{fig:avMcen_evolution}
        \end{subfigure}
        \begin{subfigure}[b]{0.99\columnwidth}
            \includegraphics[width=0.99\columnwidth]{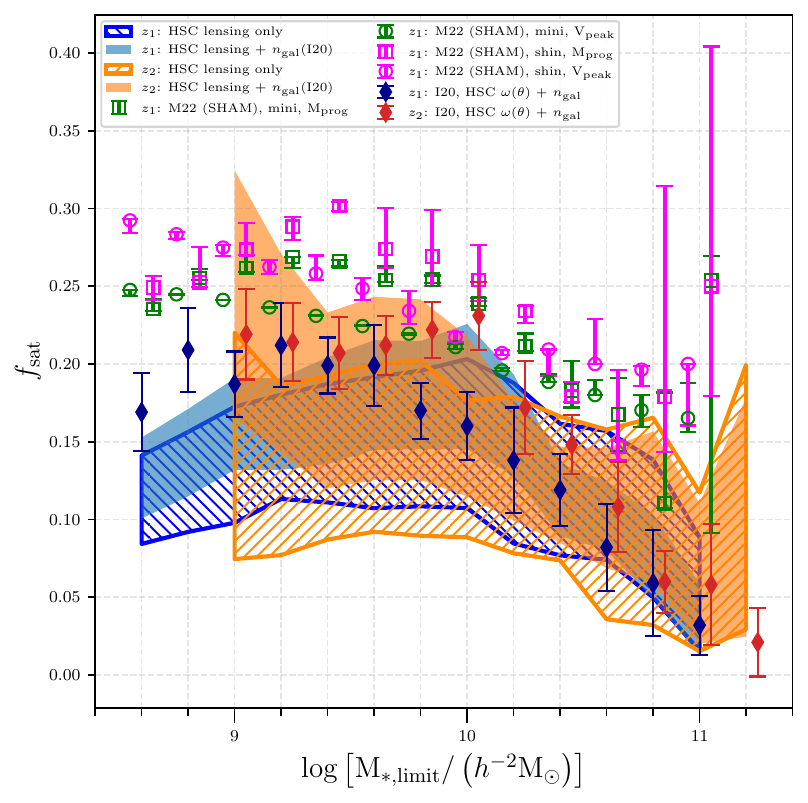}
            \caption{}
            \label{fig:fsat_evolution}
        \end{subfigure}
        \caption{\textbf{(a) Average central halo mass:} shaded bands and points (median values) with errorbars represent 68 percent credible intervals from our joint analyses with I20 and M13 abundances respectively, where color coding shows the redshift evolution - blue and orange for $z_1$ and $z_2$ bins respectively. Refer to \ref{galaxy_halo_connection} for discussion.
        \textbf{(b) satellite fraction:} The meshed bands represent 68 percent confidence intervals from our weak lensing only analysis while filled solid bands are results of joint analysis with I20 abundances. The same color coding as left panel is employed for redshift information. For comparison purposes, we also plot results from other studies in this panel: The 1-$\sigma$ constraints from I20 in both redshift bins, dark-blue and red points with errors for $z_1$ and $z_2$ bins respectively; the best fit and 1-$\sigma$ constraints from M22 for two different models of $\Mprog$ and $\Vpeak$ using two different simulations, mini-Uchuu and shin-Uchuu. Same simulation results are shown in same color, green for mimi and magenta for shin, while same model is shown by same shape, circle for $\Vpeak$ and square for Mprog. Note that, to maintain clarity, we have shifted the $\Mprog$ and $\Vpeak$ model points by 0.04 dex to the right and left respectively. Also given the small volume of shin-Uchuu simulation, we do not compare its results for massive galaxy thresholds beyond $10^{10.4} \hiimsun$. Refer to \ref{sat_frac} and \ref{comparison} for discussion.}
        \label{fig:avMcen_and_fsat}
\end{figure*}

\begin{figure}
\centering
    \begin{subfigure}[b]{1\columnwidth}
        \includegraphics[width=1\textwidth]{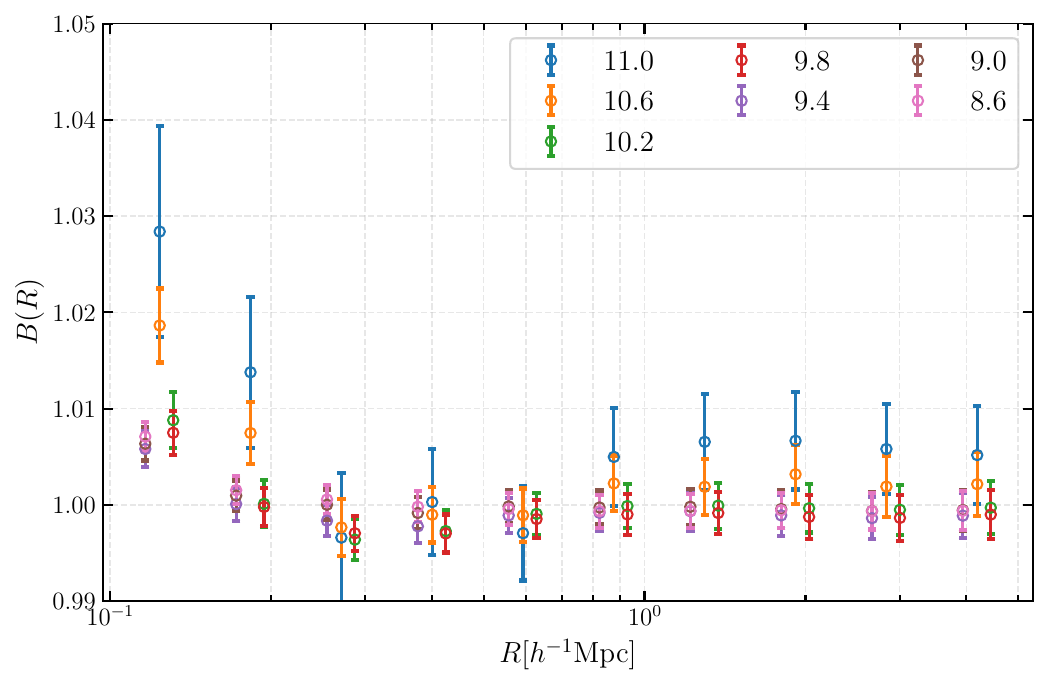}\\
        \includegraphics[width=1\textwidth]{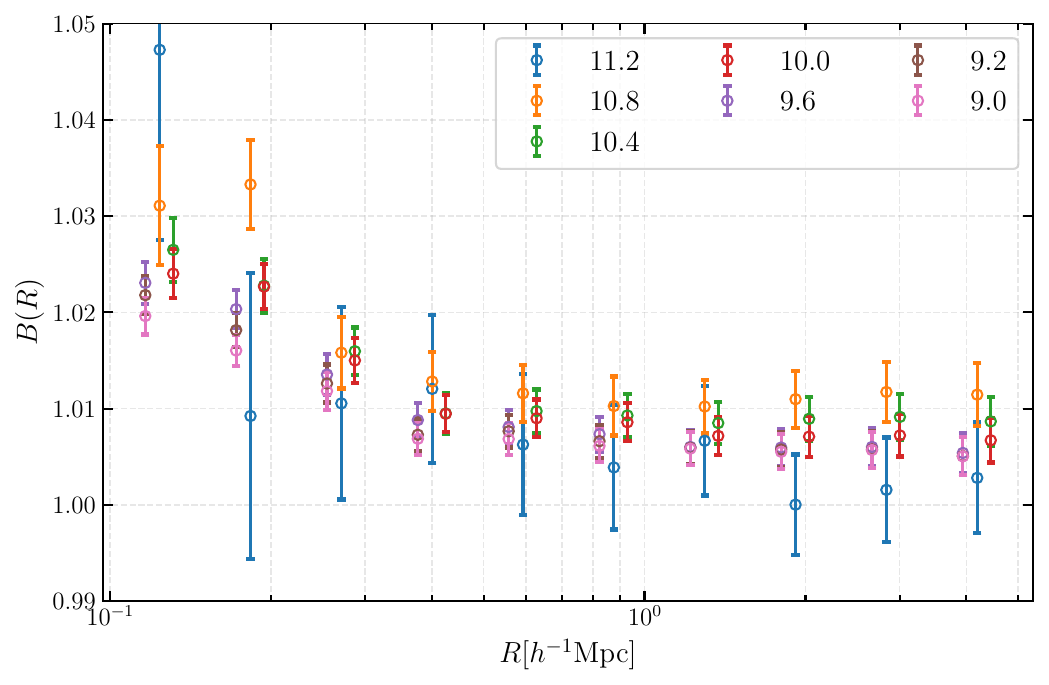}
    \end{subfigure}  
    \caption{The boost factors computed for $z_1$ (upper panel) and $z_2$ (lower panel) bins are shown for a subset of threshold subsamples. The radial bins have been shifted by 6\% about the blue and orange symbols for clarity. The errors are 1-$\sigma$ constraints obtained using jackknife technique.}
    \label{fig:boostplots}
\end{figure}

The weak lensing signal in the 1-halo regime is driven by the average central halo masses $\avMcen$. Therefore, we compare our inference of $\avMcen$ of each of the threshold sample with that inferred from the results of I20 and M22 in Fig.~\ref{fig:avMcen_comparison}. The best-fit predicted average central halo masses from I20 are shown as blue (left panel) and red (right panel) lines for redshift bins 1 and 2 respectively. The comparison shows that the inferences from I20 are statistically larger than ours for $M_{*,\rm limit}>10^{10.0} \hiimsun$, consistent with the expectation based on the comparison of the predicted and measured weak lensing signals. However, for stellar mass thresholds below $10^{10.0} \hiimsun$, the I20 best fit predictions $\avMcen$ appear consistent with our constraints. This implies that such differences in the weak lensing signal are likely absorbed by the difference in satellite fractions in our model compared to that in I20. This is visible in the comparison of the satellite fractions from I20 with our results shown in Fig.~\ref{fig:fsat_evolution}. The comparison shows that when compared with the lensing only results, I20 prefer larger satellite fractions in both redshift bins. 

In the left hand panel of Fig.~\ref{fig:avMcen_comparison}, the results from M22 from the $\Mprog$ model and the $\Vpeak$ model are shown with open squares and open circles with errors, respectively. We distinguish between the results from the two simulations used in M22 with two different colors, green corresponds to the {\it mini} simulation while magenta to the {\it shin} simulation. We see that both the models infer $\avMcen$ results which are consistent with our constraints from the weak lensing and abundance from I20 until a stellar mass threshold of $10^{10}\hiimsun$, consistent with the comparison of the weak lensing signals. At higher stellar mass thresholds the differences seen in the weak lensing signal are a result of the higher average central masses in these models. The results for M22 seem to be much more consistent with the results from I20 at these threshold bins, suggesting that the combination with the clustering is driving the larger halo masses. In the comparison of the satellite fractions we also observe that the models from M22 always prefer higher satellite fractions compared to either I20 or our results, with the exact difference depending upon the resolution of the simulation. While comparing these results, it is worth keeping in mind that the scales over which the lensing measurements are carried out ($< 5 \chimp$) are smaller than the length scales over which the clustering signal was measured by I20 ($\lesssim 25-30 \chimp$ at the median redshifts of the samples). The inferences from clustering are thus expected to be more sensitive to the large scale bias of the dark matter halos or the 2-halo term, while our inferences rely more significantly on the 1-halo term. The signal on large scales can potentially be affected by the presence of galaxy assembly bias \citep[see e.g.][]{2019MNRAS.485.1196Z}, and thus appropriate caution is warranted.

The I20 best fit predictions of SMHM relation for each redshift bin are shown as red circles with errors in the two separate panels of Fig.~\ref{fig:Inferred_Stellar_mass_vs_Central_halo_mass_relation} for comparison with other studies. Despite the overestimate in $\avMcen$ for thresholds greater than $10^{10} \hiimsun$, the halo masses $\Mmin$ are underestimated. As discussed in Section~\ref{central_HOD_discussion}, such a relation between $\Mmin$ and $\avMcen$ can be made possible by choice of small values of halo mass scatters $\sigmalogM$. And we verify in the middle panel of Fig.~\ref{fig:Central_parameter_evolution} that indeed this is the case. Additionally, their $\logMmin$ and $\sigmalogM$ deviation from our constraints increase as we go in the direction of massive galaxy thresholds. Partly this could also be due to clustering information probing different degeneracy direction in degeneracy space of central HOD parameters. We highlight a contrasting feature between lensing and clustering based studies, that the I20 study of galaxy clustering, despite using abundance information which puts strong constraints on the central HOD parameters, is unable to strongly constrain the halo mass scatter parameter $\sigmalogM$ at high stellar mass thresholds whereas lensing is able to unveil the huge ambiguity in scatter parameter. This lack of constraint could be driving the disagreements between the two studies for thresholds beyond $10^{10.0} \hiimsun$. Even though high mass slope of SMHM relation makes the stellar mass a poor tracer of its host halo mass \citep{2012ApJ...744..159L}, lensing is clearly more effective in probing the scatter in SMHM relation than clustering. In the left hand panel for redshift bin 1 of Fig.~\ref{fig:Inferred_Stellar_mass_vs_Central_halo_mass_relation}, we observe that results of M22 (shown with similar color scheme as described before) for either of the $\Vpeak$ and $\Mprog$ model are consistent with our results. We do see a difference between the results depending upon the resolution, and it appears that the two simulation boxes can trade between $\logMmin$ and the scatter $\sigmalogM$ so as to maintain similar value for $\avMcen$. This can be seen in the right panel of Fig.~\ref{fig:Central_parameter_evolution}, where we compare the scatter $\sigmalogM$ from M22 in the two different simulations and compare it to our results.

The best-fit constraint on the halo mass $\logMmin$ and scatter $\sigmalogM$ parameters from I20 are shown as points with 1-$\sigma$ errors in left and middle panels of Fig.~\ref{fig:Central_parameter_evolution}, where blue and red correspond to redshift bins 1 and 2, respectively. The underestimation of the WLS and average central halo mass at the lowest mass threshold of $z_1$ bin (see Figs.~\ref{fig:z1_rpesdfit} and \ref{fig:avMcen_z1}) is caused by the correspondingly larger best fit value of $\sigmalogM$. However, in redshift bin $z_2$ their best fit scatter value is in line with our expectation but then the underestimate of WLS is driven by the lower value of $\Mmin$ preferred by the clustering signal when combined with the abundance.

While we use the same cosmology as I20, we note the fact that differences in their modelling ingredients may have a non-negligible impact on this comparison. To be more specific, I20 uses large scale halo bias function and halo mass function each calibrated from different simulations, that is, bias from \citet{2010ApJ...724..878T} but mass function as given by the Seth \& Tormen halo mass function \citep[]{1999MNRAS.308..119S, 2001MNRAS.323....1S}. Also, I20 uses different halo mass-concentration relation than us, although we have an extra free parameter $c_{\rm fac}$ which can subsume such differences. Similarly M22 use a halo mass definition that contains mass within the virial radius,  $M_{\rm vir}$. We convert $M_{\rm vir}$ to $M_{200m}$ when making a direct comparison of halo masses using \textsc{colossus} \citet{2018ApJS..239...35D}. 

\section{Challenges and future work: photometric data and astrophysical inferences} \label{challenges}
We have inferred the galaxy stellar mass - halo mass scaling relation from a joint analysis of the abundance and weak lensing signal in this paper. The inferred relation assumes the lens galaxy properties given by the photometric redshift and stellar mass estimates from the template fitting method MIZUKI \citep{2018PASJ...70S...9T}. However, it is important to note that the presence of statistical or systematic errors in photometric redshifts can propagate in to the selection of our sample, as well as the measured abundances and the weak lensing signal, in a non-linear manner. As discussed in sec.~\ref{abundance}, the errors in photometric redshifts are expected to positively correlate with those in the stellar masses \citep[see e.g.,][]{2015ApJ...801...20T}. Such correlated errors, even if they are only statistical, can result in a number of lower mass galaxies to scatter into our stellar mass threshold and some of the high mass galaxies to scatter out instead. Similar effects can also be at play at the redshift boundaries of our redshift bins. The stellar mass bin thus does not represent a true stellar mass threshold in the presence of such errors. Moreover, such errors are also expected to affect the true average redshift of the sample, as well as their abundance measurements. The abundance measurements are further complicated due to issues in the determination of the volume associated with the galaxies due to quality cuts on photometric redshift as applied in I20. In case such volume determination uncertainties affect galaxies at all stellar masses equally, then one could correct for such issues by comparing against prior determinations of the abundance in the literature. However in general, the selection effects are often much more nuanced than simple volume misestimates, and are not entirely straightforward to correct.

Even though we explored the effect of photometric redshifts of source galaxies on the weak lensing signal, these measurements can also be affected due to the uncertainties in the photometric redshifts of the lens galaxies. The lens galaxy redshift is used to assign the projected comoving impact parameter at which the light from background galaxies approaches the cluster before it reaches us. The critical surface density estimates used to convert the shear to the excess surface density also depends upon the redshift of the lens galaxies. Thus, the interpretation of the weak lensing signal can also be affected due to the use of photometric redshifts for the lenses. Therefore, each of the above mentioned measurements can impact the inferred HOD parameters in a variety of ways.

Given these uncertainties, we refrain from making direct comparisons between these results and those present in the literature on the stellar mass halo mass relation. We restrict our comparison to those studies which use the same sample of galaxies and have similar assumptions in order to make a fair comparison between the results of these studies with the results we obtain. In order to enable comparison with the broader literature, in a future study, we will use a forward modelling approach and ascertain the level of systematic bias by making use of mock galaxy catalogs that have the errors in photometric redshifts as expected from the photometry from the HSC survey.

The Subaru HSC survey can map out galaxies out to even higher redshifts than those considered in this study. However beyond the median redshift of the survey we become exceedingly sensitive to potential systematic biases due to the use of photometric redshift estimates of the source galaxies. We also expect that magnification bias to start to play a role by correlating the lens and the source number densities, especially for galaxies that lie at the steep end of the luminosity function \citep[see e.g.,][]{2020A&A...638A..96U, 2020A&ARv..28....7U}. Eventually, once we have a control over all the above systematics, it would become interesting to model the clustering, the lensing and the abundance of galaxies as a function of stellar mass and at multiple redshift bins.

\section{Summary and Conclusions} \label{summary}
We have investigated the galaxy-dark matter connection and its evolution using samples of photometric galaxies from the HSC survey with varying thresholds of stellar masses from $8.6 \leq \log \left[ M_*/(\hiimsun) \right] \leq 11.2$ in the redshift ranges $[0.30,0.55)$ and $[0.55,0.80)$. Our results are based on the weak lensing signal measured for these samples using the Year $1$ catalog of source galaxy shapes from the HSC survey, and the measurements of the abundance of galaxies. We carry out a Bayesian analysis to infer the posterior distribution of parameters that describe the halo occupation distribution of these galaxies. The key results and findings of our study are summarized as follows.

\begin{itemize}
    \item We present high signal-to-noise ratio measurements (SNR ranging from $30-50$) of the lensing signals in both redshift bins for all of our samples. We show the robustness of the measured lensing lensing signals with multiple null tests, such as the tangential and cross components of the lensing signal around random points and the cross component around lens galaxies. We also find that the boost factors for our signals are statistically insignificant and the biases due to use of photometric redshifts for the source galaxies are $\sim 1\%$ and $\sim 4\%$ for the redshift bins 1 and 2, respectively. These tests of systematics indicate that our measurements are not heavily affected by contamination of either the foreground or the background galaxies. 
    \item We fit these weak lensing measurements together with the abundances of galaxies with a simple 5 parameter HOD model per sample in the context of the Planck cosmological model and show that the model provides a reasonable description of the data. We infer the posterior distribution of these parameters given the measurements.
    \item We show that the weak lensing measurements and the abundances on their own constrain the central HOD parameters $\logMmin$ and the scatter $\sigmalogM$ in a degenerate manner. However these degeneracy directions are different for each of the observables and hence a combination of the two helps break the degeneracy. We also show the impact of using different abundances in the literature. We show that the average halo masses of central galaxies are well constrained irrespective of the use of either abundances.
    \item We find that the average halo masses of central galaxies increases with the threshold for the stellar mass subsample for both redshift bin 1 and 2. Comparison between these scaling relations at the two different redshifts shows very mild evolution between these two redshifts, if any.
    \item We also compare our results with the constraints obtained by the study of I20 who jointly model the abundance and clustering of the same sample of galaxies. We show that the best fit model of I20 underestimates the observed lensing signals by varying amounts of $10\%-30\%$ in the 1-halo central term regime and $50\%-60\%$ at larger radii for mass thresholds up to $10^{10} \hiimsun$ and overestimate the lensing signal for more massive threshold samples. Nevertheless, we find excellent agreement between the constraints on the average halo masses of central galaxies for these samples for thresholds until $10^{10} \hiimsun$, and the results from I20 overestimate these average halo masses for higher threshold samples.
    \item We also compare our results with the subhalo abundance matching method of M22 which uses the abundance and clustering measurements of I20 as constraints. We find that their models which use a monotonic relation between $\Mprog$ or $\Vpeak$ of the subhalos and the stellar mass of galaxies, is able to predict lensing signal consistent with our measurements for stellar mass thresholds up to $10^{10} \hiimsun$. Both models fail to explain the lensing signal especially within the 1-halo regime for higher stellar mass threshold samples.
    \item Finally, we find that the satellite fractions predicted by our fiducal analysis are consistent with the clustering study of I20 given the statistical errors. However, we find that the models from M22 based on subhalo abundance matching predict an additional satellite fraction of up to $15\%$ over our constraints. 
\end{itemize}

The paper demonstrates the great potential of large imaging surveys such as the HSC to infer the galaxy-dark matter connection over a large range of redshifts using multiple observational probes such as the abundance of galaxies, their clustering and their galaxy-galaxy lensing signal. An accurate inference of the true underlying scaling relations between stellar mass and halo mass, however, will depend upon quantitative estimates of how the photometric redshift errors in the lens galaxy population affect the underlying stellar mass threshold samples. Assessment of the extent of such biases will be subject of our work in the near future.

\section*{Acknowledgements}
We thank Divya Rana, Amit Kumar, Preetish K. Mishra, Susmita Adhikari, Arka Banerjee, Supranta S. Boruah and Priyanka Gawade for useful discussions and their comments on the draft version of the paper. We also thank our research advisory committee members Aseem Paranjape, Masamune Oguri, Anupreeta More for useful discussions on the current project along with comments on the draft version of this paper. NC is thankful for the financial support provided by the University Grants Commission (UGC) of India. He is also thankful to IUCAA for its the amicable environment and hospitality offered to students.  
We acknowledge the use of Pegasus, the high performance computing facility at IUCAA. The calculations in part were carried out on Cray XC50 at Center for Computational Astrophysics, National Astronomical Observatory of Japan. Data analysis was in part carried out on the Multi-wavelength Data Analysis System operated by the Astronomy Data Center (ADC), National Astronomical Observatory of Japan.
This work was supported in part by JSPS KAKENHI Grant Numbers JP23K13145 (SI), JP19H00677, JP21H05465, JP22K03644 (S. Masaki) and JP21K13956 (DK).
TO acknowledges support from the Ministry of Science and Technology of Taiwan under Grant Nos. MOST 111-2112-M-001-061- and NSTC 112-2112-M-001-034- and the Career Development Award, Academia Sinica (AS-CDA-108-M02) for the period of 2019 to 2023.

The Hyper Suprime-Cam (HSC) collaboration includes the astronomical communities of Japan and Taiwan, and Princeton University. The HSC instrumentation and software were developed by the National Astronomical Observatory of Japan (NAOJ), the Kavli Institute for the Physics and Mathematics of the Universe (Kavli IPMU), the University of Tokyo, the High Energy Accelerator Research Organization (KEK), the Academia Sinica Institute for Astronomy and Astrophysics in Taiwan (ASIAA), and Princeton University. Funding was contributed by the FIRST program from the Japanese Cabinet Office, the Ministry of Education, Culture, Sports, Science and Technology (MEXT), the Japan Society for the Promotion of Science (JSPS), Japan Science and Technology Agency (JST), the Toray Science Foundation, NAOJ, Kavli IPMU, KEK, ASIAA, and Princeton University.

We also thank Instituto de Astrofisica de Andalucia (IAA-CSIC), Centro de Supercomputacion de Galicia (CESGA) and the Spanish academic and research network (RedIRIS) in Spain for hosting Uchuu DR1, DR2 and DR3 in the Skies \& Universes site for cosmological simulations. The Uchuu simulations were carried out on Aterui II supercomputer at Center for Computational Astrophysics, CfCA, of National Astronomical Observatory of Japan, and the K computer at the RIKEN Advanced Institute for Computational Science. The Uchuu Data Releases efforts have made use of the skun@IAA\_RedIRIS and skun6@IAA computer facilities managed by the IAA-CSIC in Spain (MICINN EU-Feder grant EQC2018-004366-P).

We have used \texttt{corner.py} \citep{2016JOSS....1...24F} to create degeneracy plots and \texttt{PYGTC} \citep{Bocquet2016} to create triangle/corner plots.

\section*{Data Availability}
The weak lensing signal measurements after applying all correction as mentioned in Section~\ref{methodology} for our stellar mass threshold lens samples along with the measured covariance matrices and abundances are made available in a public github repository, \url{https://github.com/0Navin0/galaxy_halo_connection_in_HSC}. This repository also contains our modelling constraints from Tables~\ref{tab:Parameter_constraints_z1} and \ref{tab:Parameter_constraints_z2} along with additional relevant plots for interested readers. 

\bibliographystyle{mnras}
\bibliography{example} 

\begin{thebibliography}{}
\makeatletter
\relax
\def\mn@urlcharsother{\let\do\@makeother \do\$\do\&\do\#\do\^\do\_\do\%\do\~}
\def\mn@doi{\begingroup\mn@urlcharsother \@ifnextchar [ {\mn@doi@}
  {\mn@doi@[]}}
\def\mn@doi@[#1]#2{\def\@tempa{#1}\ifx\@tempa\@empty \href
  {http://dx.doi.org/#2} {doi:#2}\else \href {http://dx.doi.org/#2} {#1}\fi
  \endgroup}
\def\mn@eprint#1#2{\mn@eprint@#1:#2::\@nil}
\def\mn@eprint@arXiv#1{\href {http://arxiv.org/abs/#1} {{\tt arXiv:#1}}}
\def\mn@eprint@dblp#1{\href {http://dblp.uni-trier.de/rec/bibtex/#1.xml}
  {dblp:#1}}
\def\mn@eprint@#1:#2:#3:#4\@nil{\def\@tempa {#1}\def\@tempb {#2}\def\@tempc
  {#3}\ifx \@tempc \@empty \let \@tempc \@tempb \let \@tempb \@tempa \fi \ifx
  \@tempb \@empty \def\@tempb {arXiv}\fi \@ifundefined
  {mn@eprint@\@tempb}{\@tempb:\@tempc}{\expandafter \expandafter \csname
  mn@eprint@\@tempb\endcsname \expandafter{\@tempc}}}

\bibitem[\protect\citeauthoryear{{Aihara} et~al.,}{{Aihara}
  et~al.}{2018a}]{2018PASJ...70S...4A}
{Aihara} H.,  et~al., 2018a, \mn@doi [\pasj] {10.1093/pasj/psx066}, \href
  {https://ui.adsabs.harvard.edu/abs/2018PASJ...70S...4A} {70, S4}

\bibitem[\protect\citeauthoryear{{Aihara} et~al.,}{{Aihara}
  et~al.}{2018b}]{2018PASJ...70S...8A}
{Aihara} H.,  et~al., 2018b, \mn@doi [\pasj] {10.1093/pasj/psx081}, \href
  {https://ui.adsabs.harvard.edu/abs/2018PASJ...70S...8A} {70, S8}

\bibitem[\protect\citeauthoryear{{Aihara} et~al.,}{{Aihara}
  et~al.}{2019}]{2019PASJ...71..114A}
{Aihara} H.,  et~al., 2019, \mn@doi [\pasj] {10.1093/pasj/psz103}, \href
  {https://ui.adsabs.harvard.edu/abs/2019PASJ...71..114A} {71, 114}

\bibitem[\protect\citeauthoryear{{Aihara} et~al.,}{{Aihara}
  et~al.}{2022}]{2022PASJ...74..247A}
{Aihara} H.,  et~al., 2022, \mn@doi [\pasj] {10.1093/pasj/psab122}, \href
  {https://ui.adsabs.harvard.edu/abs/2022PASJ...74..247A} {74, 247}

\bibitem[\protect\citeauthoryear{{Behroozi}, {Conroy}  \&
  {Wechsler}}{{Behroozi} et~al.}{2010}]{2010ApJ...717..379B}
{Behroozi} P.~S.,  {Conroy} C.,   {Wechsler} R.~H.,  2010, \mn@doi [\apj]
  {10.1088/0004-637X/717/1/379}, \href
  {https://ui.adsabs.harvard.edu/abs/2010ApJ...717..379B} {717, 379}

\bibitem[\protect\citeauthoryear{{Behroozi}, {Wechsler}  \&
  {Conroy}}{{Behroozi} et~al.}{2013}]{2013ApJ...770...57B}
{Behroozi} P.~S.,  {Wechsler} R.~H.,   {Conroy} C.,  2013, \mn@doi [\apj]
  {10.1088/0004-637X/770/1/57}, \href
  {https://ui.adsabs.harvard.edu/abs/2013ApJ...770...57B} {770, 57}

\bibitem[\protect\citeauthoryear{{Behroozi}, {Wechsler}, {Hearin}  \&
  {Conroy}}{{Behroozi} et~al.}{2019}]{2019MNRAS.488.3143B}
{Behroozi} P.,  {Wechsler} R.~H.,  {Hearin} A.~P.,   {Conroy} C.,  2019,
  \mn@doi [\mnras] {10.1093/mnras/stz1182}, \href
  {https://ui.adsabs.harvard.edu/abs/2019MNRAS.488.3143B} {488, 3143}

\bibitem[\protect\citeauthoryear{{Bhattacharya}, {Heitmann}, {White},
  {Luki{\'c}}, {Wagner}  \& {Habib}}{{Bhattacharya}
  et~al.}{2011}]{2011ApJ...732..122B}
{Bhattacharya} S.,  {Heitmann} K.,  {White} M.,  {Luki{\'c}} Z.,  {Wagner} C.,
   {Habib} S.,  2011, \mn@doi [\apj] {10.1088/0004-637X/732/2/122}, \href
  {https://ui.adsabs.harvard.edu/abs/2011ApJ...732..122B} {732, 122}

\bibitem[\protect\citeauthoryear{Bocquet \& Carter}{Bocquet \&
  Carter}{2016}]{Bocquet2016}
Bocquet S.,  Carter F.~W.,  2016, \mn@doi [The Journal of Open Source Software]
  {10.21105/joss.00046}, 1

\bibitem[\protect\citeauthoryear{{Bosch} et~al.,}{{Bosch}
  et~al.}{2018}]{2018PASJ...70S...5B}
{Bosch} J.,  et~al., 2018, \mn@doi [\pasj] {10.1093/pasj/psx080}, \href
  {https://ui.adsabs.harvard.edu/abs/2018PASJ...70S...5B} {70, S5}

\bibitem[\protect\citeauthoryear{{Bose}, {Eisenstein}, {Hernquist},
  {Pillepich}, {Nelson}, {Marinacci}, {Springel}  \& {Vogelsberger}}{{Bose}
  et~al.}{2019}]{2019MNRAS.490.5693B}
{Bose} S.,  {Eisenstein} D.~J.,  {Hernquist} L.,  {Pillepich} A.,  {Nelson} D.,
   {Marinacci} F.,  {Springel} V.,   {Vogelsberger} M.,  2019, \mn@doi [\mnras]
  {10.1093/mnras/stz2546}, \href
  {https://ui.adsabs.harvard.edu/abs/2019MNRAS.490.5693B} {490, 5693}

\bibitem[\protect\citeauthoryear{{Bower}, {Benson}, {Malbon}, {Helly}, {Frenk},
  {Baugh}, {Cole}  \& {Lacey}}{{Bower} et~al.}{2006}]{2006MNRAS.370..645B}
{Bower} R.~G.,  {Benson} A.~J.,  {Malbon} R.,  {Helly} J.~C.,  {Frenk} C.~S.,
  {Baugh} C.~M.,  {Cole} S.,   {Lacey} C.~G.,  2006, \mn@doi [\mnras]
  {10.1111/j.1365-2966.2006.10519.x}, \href
  {https://ui.adsabs.harvard.edu/abs/2006MNRAS.370..645B} {370, 645}

\bibitem[\protect\citeauthoryear{{Cacciato}, {van den Bosch}, {More}, {Li},
  {Mo}  \& {Yang}}{{Cacciato} et~al.}{2009}]{2009MNRAS.394..929C}
{Cacciato} M.,  {van den Bosch} F.~C.,  {More} S.,  {Li} R.,  {Mo} H.~J.,
  {Yang} X.,  2009, \mn@doi [\mnras] {10.1111/j.1365-2966.2008.14362.x}, \href
  {https://ui.adsabs.harvard.edu/abs/2009MNRAS.394..929C} {394, 929}

\bibitem[\protect\citeauthoryear{Cacciato, van~den Bosch, More, Mo  \&
  Yang}{Cacciato et~al.}{2013}]{10.1093/mnras/sts525}
Cacciato M.,  van~den Bosch F.~C.,  More S.,  Mo H.,   Yang X.,  2013, \mn@doi
  [Monthly Notices of the Royal Astronomical Society] {10.1093/mnras/sts525},
  430, 767

\bibitem[\protect\citeauthoryear{{Calzetti}, {Armus}, {Bohlin}, {Kinney},
  {Koornneef}  \& {Storchi-Bergmann}}{{Calzetti}
  et~al.}{2000}]{2000ApJ...533..682C}
{Calzetti} D.,  {Armus} L.,  {Bohlin} R.~C.,  {Kinney} A.~L.,  {Koornneef} J.,
   {Storchi-Bergmann} T.,  2000, \mn@doi [\apj] {10.1086/308692}, \href
  {https://ui.adsabs.harvard.edu/abs/2000ApJ...533..682C} {533, 682}

\bibitem[\protect\citeauthoryear{{Chabrier}}{{Chabrier}}{2003}]{2003PASP..115..763C}
{Chabrier} G.,  2003, \mn@doi [\pasp] {10.1086/376392}, \href
  {https://ui.adsabs.harvard.edu/abs/2003PASP..115..763C} {115, 763}

\bibitem[\protect\citeauthoryear{{Conroy}, {Wechsler}  \& {Kravtsov}}{{Conroy}
  et~al.}{2006}]{2006ApJ...647..201C}
{Conroy} C.,  {Wechsler} R.~H.,   {Kravtsov} A.~V.,  2006, \mn@doi [\apj]
  {10.1086/503602}, \href
  {https://ui.adsabs.harvard.edu/abs/2006ApJ...647..201C} {647, 201}

\bibitem[\protect\citeauthoryear{{Conselice}}{{Conselice}}{2014}]{2014ARA&A..52..291C}
{Conselice} C.~J.,  2014, \mn@doi [\araa]
  {10.1146/annurev-astro-081913-040037}, \href
  {https://ui.adsabs.harvard.edu/abs/2014ARA&A..52..291C} {52, 291}

\bibitem[\protect\citeauthoryear{{Coupon} et~al.,}{{Coupon}
  et~al.}{2015}]{2015MNRAS.449.1352C}
{Coupon} J.,  et~al., 2015, \mn@doi [\mnras] {10.1093/mnras/stv276}, \href
  {https://ui.adsabs.harvard.edu/abs/2015MNRAS.449.1352C} {449, 1352}

\bibitem[\protect\citeauthoryear{{Coupon}, {Czakon}, {Bosch}, {Komiyama},
  {Medezinski}, {Miyazaki}  \& {Oguri}}{{Coupon}
  et~al.}{2018}]{2018PASJ...70S...7C}
{Coupon} J.,  {Czakon} N.,  {Bosch} J.,  {Komiyama} Y.,  {Medezinski} E.,
  {Miyazaki} S.,   {Oguri} M.,  2018, \mn@doi [\pasj] {10.1093/pasj/psx047},
  \href {https://ui.adsabs.harvard.edu/abs/2018PASJ...70S...7C} {70, S7}

\bibitem[\protect\citeauthoryear{{Courtin}, {Rasera}, {Alimi}, {Corasaniti},
  {Boucher}  \& {F{\"u}zfa}}{{Courtin} et~al.}{2011}]{2011MNRAS.410.1911C}
{Courtin} J.,  {Rasera} Y.,  {Alimi} J.~M.,  {Corasaniti} P.~S.,  {Boucher} V.,
    {F{\"u}zfa} A.,  2011, \mn@doi [\mnras] {10.1111/j.1365-2966.2010.17573.x},
  \href {https://ui.adsabs.harvard.edu/abs/2011MNRAS.410.1911C} {410, 1911}

\bibitem[\protect\citeauthoryear{{Cui}, {Baldi}  \& {Borgani}}{{Cui}
  et~al.}{2012}]{2012MNRAS.424..993C}
{Cui} W.,  {Baldi} M.,   {Borgani} S.,  2012, \mn@doi [\mnras]
  {10.1111/j.1365-2966.2012.21267.x}, \href
  {https://ui.adsabs.harvard.edu/abs/2012MNRAS.424..993C} {424, 993}

\bibitem[\protect\citeauthoryear{{Diemer}}{{Diemer}}{2018}]{2018ApJS..239...35D}
{Diemer} B.,  2018, \mn@doi [\apjs] {10.3847/1538-4365/aaee8c}, \href
  {https://ui.adsabs.harvard.edu/abs/2018ApJS..239...35D} {239, 35}

\bibitem[\protect\citeauthoryear{{Dvornik} et~al.,}{{Dvornik}
  et~al.}{2020}]{2020A&A...642A..83D}
{Dvornik} A.,  et~al., 2020, \mn@doi [\aap] {10.1051/0004-6361/202038693},
  \href {https://ui.adsabs.harvard.edu/abs/2020A&A...642A..83D} {642, A83}

\bibitem[\protect\citeauthoryear{{Dvornik} et~al.,}{{Dvornik}
  et~al.}{2022}]{2022arXiv221003110D}
{Dvornik} A.,  et~al., 2022, \mn@doi [arXiv e-prints]
  {10.48550/arXiv.2210.03110}, \href
  {https://ui.adsabs.harvard.edu/abs/2022arXiv221003110D} {p. arXiv:2210.03110}

\bibitem[\protect\citeauthoryear{{Foreman-Mackey}}{{Foreman-Mackey}}{2016}]{2016JOSS....1...24F}
{Foreman-Mackey} D.,  2016, \mn@doi [The Journal of Open Source Software]
  {10.21105/joss.00024}, \href
  {https://ui.adsabs.harvard.edu/abs/2016JOSS....1...24F} {1, 24}

\bibitem[\protect\citeauthoryear{{Foreman-Mackey}, {Hogg}, {Lang}  \&
  {Goodman}}{{Foreman-Mackey} et~al.}{2013}]{2013PASP..125..306F}
{Foreman-Mackey} D.,  {Hogg} D.~W.,  {Lang} D.,   {Goodman} J.,  2013, \mn@doi
  [\pasp] {10.1086/670067}, \href
  {https://ui.adsabs.harvard.edu/abs/2013PASP..125..306F} {125, 306}

\bibitem[\protect\citeauthoryear{{Genel} et~al.,}{{Genel}
  et~al.}{2014}]{2014MNRAS.445..175G}
{Genel} S.,  et~al., 2014, \mn@doi [\mnras] {10.1093/mnras/stu1654}, \href
  {https://ui.adsabs.harvard.edu/abs/2014MNRAS.445..175G} {445, 175}

\bibitem[\protect\citeauthoryear{{Goodman} \& {Weare}}{{Goodman} \&
  {Weare}}{2010}]{2010CAMCS...5...65G}
{Goodman} J.,  {Weare} J.,  2010, \mn@doi [Communications in Applied
  Mathematics and Computational Science] {10.2140/camcos.2010.5.65}, \href
  {https://ui.adsabs.harvard.edu/abs/2010CAMCS...5...65G} {5, 65}

\bibitem[\protect\citeauthoryear{{Guo}, {White}, {Li}  \&
  {Boylan-Kolchin}}{{Guo} et~al.}{2010}]{2010MNRAS.404.1111G}
{Guo} Q.,  {White} S.,  {Li} C.,   {Boylan-Kolchin} M.,  2010, \mn@doi [\mnras]
  {10.1111/j.1365-2966.2010.16341.x}, \href
  {https://ui.adsabs.harvard.edu/abs/2010MNRAS.404.1111G} {404, 1111}

\bibitem[\protect\citeauthoryear{{Hirata} \& {Seljak}}{{Hirata} \&
  {Seljak}}{2003}]{2003MNRAS.343..459H}
{Hirata} C.,  {Seljak} U.,  2003, \mn@doi [\mnras]
  {10.1046/j.1365-8711.2003.06683.x}, \href
  {https://ui.adsabs.harvard.edu/abs/2003MNRAS.343..459H} {343, 459}

\bibitem[\protect\citeauthoryear{{Hoekstra}, {Hsieh}, {Yee}, {Lin}  \&
  {Gladders}}{{Hoekstra} et~al.}{2005}]{2005ApJ...635...73H}
{Hoekstra} H.,  {Hsieh} B.~C.,  {Yee} H.~K.~C.,  {Lin} H.,   {Gladders} M.~D.,
  2005, \mn@doi [\apj] {10.1086/496913}, \href
  {https://ui.adsabs.harvard.edu/abs/2005ApJ...635...73H} {635, 73}

\bibitem[\protect\citeauthoryear{{Huang} et~al.,}{{Huang}
  et~al.}{2020}]{2020MNRAS.492.3685H}
{Huang} S.,  et~al., 2020, \mn@doi [\mnras] {10.1093/mnras/stz3314}, \href
  {https://ui.adsabs.harvard.edu/abs/2020MNRAS.492.3685H} {492, 3685}

\bibitem[\protect\citeauthoryear{{Ishikawa} et~al.,}{{Ishikawa}
  et~al.}{2020}]{2020ApJ...904..128I}
{Ishikawa} S.,  et~al., 2020, \mn@doi [\apj] {10.3847/1538-4357/abbd95}, \href
  {https://ui.adsabs.harvard.edu/abs/2020ApJ...904..128I} {904, 128}

\bibitem[\protect\citeauthoryear{{Ishiyama} et~al.,}{{Ishiyama}
  et~al.}{2021}]{2021MNRAS.506.4210I}
{Ishiyama} T.,  et~al., 2021, \mn@doi [\mnras] {10.1093/mnras/stab1755}, \href
  {https://ui.adsabs.harvard.edu/abs/2021MNRAS.506.4210I} {506, 4210}

\bibitem[\protect\citeauthoryear{{Juri{\'c}} et~al.,}{{Juri{\'c}}
  et~al.}{2017}]{2017ASPC..512..279J}
{Juri{\'c}} M.,  et~al., 2017, in {Lorente} N.~P.~F.,  {Shortridge} K.,
  {Wayth} R.,  eds,  Astronomical Society of the Pacific Conference Series Vol.
  512, Astronomical Data Analysis Software and Systems XXV. p.~279 (\mn@eprint
  {arXiv} {1512.07914}), \mn@doi{10.48550/arXiv.1512.07914}

\bibitem[\protect\citeauthoryear{{Kere{\v{s}}}, {Katz}, {Weinberg}  \&
  {Dav{\'e}}}{{Kere{\v{s}}} et~al.}{2005}]{2005MNRAS.363....2K}
{Kere{\v{s}}} D.,  {Katz} N.,  {Weinberg} D.~H.,   {Dav{\'e}} R.,  2005,
  \mn@doi [\mnras] {10.1111/j.1365-2966.2005.09451.x}, \href
  {https://ui.adsabs.harvard.edu/abs/2005MNRAS.363....2K} {363, 2}

\bibitem[\protect\citeauthoryear{{Komiyama} et~al.,}{{Komiyama}
  et~al.}{2018}]{2018PASJ...70S...2K}
{Komiyama} Y.,  et~al., 2018, \mn@doi [\pasj] {10.1093/pasj/psx069}, \href
  {https://ui.adsabs.harvard.edu/abs/2018PASJ...70S...2K} {70, S2}

\bibitem[\protect\citeauthoryear{{Kravtsov}, {Berlind}, {Wechsler}, {Klypin},
  {Gottl{\"o}ber}, {Allgood}  \& {Primack}}{{Kravtsov}
  et~al.}{2004}]{2004ApJ...609...35K}
{Kravtsov} A.~V.,  {Berlind} A.~A.,  {Wechsler} R.~H.,  {Klypin} A.~A.,
  {Gottl{\"o}ber} S.,  {Allgood} B.,   {Primack} J.~R.,  2004, \mn@doi [\apj]
  {10.1086/420959}, \href
  {https://ui.adsabs.harvard.edu/abs/2004ApJ...609...35K} {609, 35}

\bibitem[\protect\citeauthoryear{{Leauthaud}, {Tinker}, {Behroozi}, {Busha}  \&
  {Wechsler}}{{Leauthaud} et~al.}{2011}]{2011ApJ...738...45L}
{Leauthaud} A.,  {Tinker} J.,  {Behroozi} P.~S.,  {Busha} M.~T.,   {Wechsler}
  R.~H.,  2011, \mn@doi [\apj] {10.1088/0004-637X/738/1/45}, \href
  {https://ui.adsabs.harvard.edu/abs/2011ApJ...738...45L} {738, 45}

\bibitem[\protect\citeauthoryear{{Leauthaud} et~al.,}{{Leauthaud}
  et~al.}{2012}]{2012ApJ...744..159L}
{Leauthaud} A.,  et~al., 2012, \mn@doi [\apj] {10.1088/0004-637X/744/2/159},
  \href {https://ui.adsabs.harvard.edu/abs/2012ApJ...744..159L} {744, 159}

\bibitem[\protect\citeauthoryear{{Macci{\`o}}, {Dutton}, {van den Bosch},
  {Moore}, {Potter}  \& {Stadel}}{{Macci{\`o}}
  et~al.}{2007}]{2007MNRAS.378...55M}
{Macci{\`o}} A.~V.,  {Dutton} A.~A.,  {van den Bosch} F.~C.,  {Moore} B.,
  {Potter} D.,   {Stadel} J.,  2007, \mn@doi [\mnras]
  {10.1111/j.1365-2966.2007.11720.x}, \href
  {https://ui.adsabs.harvard.edu/abs/2007MNRAS.378...55M} {378, 55}

\bibitem[\protect\citeauthoryear{{Macci{\`o}}, {Dutton}  \& {van den
  Bosch}}{{Macci{\`o}} et~al.}{2008}]{2008MNRAS.391.1940M}
{Macci{\`o}} A.~V.,  {Dutton} A.~A.,   {van den Bosch} F.~C.,  2008, \mn@doi
  [\mnras] {10.1111/j.1365-2966.2008.14029.x}, \href
  {https://ui.adsabs.harvard.edu/abs/2008MNRAS.391.1940M} {391, 1940}

\bibitem[\protect\citeauthoryear{{Mandelbaum}, {Seljak}, {Kauffmann}, {Hirata}
  \& {Brinkmann}}{{Mandelbaum} et~al.}{2006}]{2006MNRAS.368..715M}
{Mandelbaum} R.,  {Seljak} U.,  {Kauffmann} G.,  {Hirata} C.~M.,   {Brinkmann}
  J.,  2006, \mn@doi [\mnras] {10.1111/j.1365-2966.2006.10156.x}, \href
  {https://ui.adsabs.harvard.edu/abs/2006MNRAS.368..715M} {368, 715}

\bibitem[\protect\citeauthoryear{{Mandelbaum} et~al.,}{{Mandelbaum}
  et~al.}{2014}]{2014ApJS..212....5M}
{Mandelbaum} R.,  et~al., 2014, \mn@doi [\apjs] {10.1088/0067-0049/212/1/5},
  \href {https://ui.adsabs.harvard.edu/abs/2014ApJS..212....5M} {212, 5}

\bibitem[\protect\citeauthoryear{{Mandelbaum} et~al.,}{{Mandelbaum}
  et~al.}{2015}]{2015MNRAS.450.2963M}
{Mandelbaum} R.,  et~al., 2015, \mn@doi [\mnras] {10.1093/mnras/stv781}, \href
  {https://ui.adsabs.harvard.edu/abs/2015MNRAS.450.2963M} {450, 2963}

\bibitem[\protect\citeauthoryear{{Mandelbaum} et~al.,}{{Mandelbaum}
  et~al.}{2018a}]{2018PASJ...70S..25M}
{Mandelbaum} R.,  et~al., 2018a, \mn@doi [\pasj] {10.1093/pasj/psx130}, \href
  {https://ui.adsabs.harvard.edu/abs/2018PASJ...70S..25M} {70, S25}

\bibitem[\protect\citeauthoryear{{Mandelbaum} et~al.,}{{Mandelbaum}
  et~al.}{2018b}]{2018MNRAS.481.3170M}
{Mandelbaum} R.,  et~al., 2018b, \mn@doi [\mnras] {10.1093/mnras/sty2420},
  \href {https://ui.adsabs.harvard.edu/abs/2018MNRAS.481.3170M} {481, 3170}

\bibitem[\protect\citeauthoryear{{Masaki}, {Kashino}  \& {Lin}}{{Masaki}
  et~al.}{2022}]{2022arXiv221011713M}
{Masaki} S.,  {Kashino} D.,   {Lin} Y.-T.,  2022, \mn@doi [arXiv e-prints]
  {10.48550/arXiv.2210.11713}, \href
  {https://ui.adsabs.harvard.edu/abs/2022arXiv221011713M} {p. arXiv:2210.11713}

\bibitem[\protect\citeauthoryear{{Mishra}, {Rana}  \& {More}}{{Mishra}
  et~al.}{2023}]{2023arXiv230104664M}
{Mishra} P.~K.,  {Rana} D.,   {More} S.,  2023, \mn@doi [arXiv e-prints]
  {10.48550/arXiv.2301.04664}, \href
  {https://ui.adsabs.harvard.edu/abs/2023arXiv230104664M} {p. arXiv:2301.04664}

\bibitem[\protect\citeauthoryear{{Miyazaki} et~al.,}{{Miyazaki}
  et~al.}{2018}]{2018PASJ...70S...1M}
{Miyazaki} S.,  et~al., 2018, \mn@doi [\pasj] {10.1093/pasj/psx063}, \href
  {https://ui.adsabs.harvard.edu/abs/2018PASJ...70S...1M} {70, S1}

\bibitem[\protect\citeauthoryear{{More}}{{More}}{2021}]{2021ascl.soft08002M}
{More} S.,  2021, {AUM: A Unified Modeling scheme for galaxy abundance, galaxy
  clustering and galaxy-galaxy lensing}, Astrophysics Source Code Library,
  record ascl:2108.002 (\mn@eprint {ascl} {2108.002})

\bibitem[\protect\citeauthoryear{{More}, {van den Bosch}  \& {Cacciato}}{{More}
  et~al.}{2009}]{2009MNRAS.392..917M}
{More} S.,  {van den Bosch} F.~C.,   {Cacciato} M.,  2009, \mn@doi [\mnras]
  {10.1111/j.1365-2966.2008.14114.x}, \href
  {https://ui.adsabs.harvard.edu/abs/2009MNRAS.392..917M} {392, 917}

\bibitem[\protect\citeauthoryear{{More}, {van den Bosch}, {Cacciato}, {Skibba},
  {Mo}  \& {Yang}}{{More} et~al.}{2011}]{2011MNRAS.410..210M}
{More} S.,  {van den Bosch} F.~C.,  {Cacciato} M.,  {Skibba} R.,  {Mo} H.~J.,
  {Yang} X.,  2011, \mn@doi [\mnras] {10.1111/j.1365-2966.2010.17436.x}, \href
  {https://ui.adsabs.harvard.edu/abs/2011MNRAS.410..210M} {410, 210}

\bibitem[\protect\citeauthoryear{{More}, {Miyatake}, {Mandelbaum}, {Takada},
  {Spergel}, {Brownstein}  \& {Schneider}}{{More}
  et~al.}{2015}]{2015ApJ...806....2M}
{More} S.,  {Miyatake} H.,  {Mandelbaum} R.,  {Takada} M.,  {Spergel} D.~N.,
  {Brownstein} J.~R.,   {Schneider} D.~P.,  2015, \mn@doi [\apj]
  {10.1088/0004-637X/806/1/2}, \href
  {https://ui.adsabs.harvard.edu/abs/2015ApJ...806....2M} {806, 2}

\bibitem[\protect\citeauthoryear{{Murray}, {Power}  \& {Robotham}}{{Murray}
  et~al.}{2013}]{2013MNRAS.434L..61M}
{Murray} S.~G.,  {Power} C.,   {Robotham} A.~S.~G.,  2013, \mn@doi [\mnras]
  {10.1093/mnrasl/slt079}, \href
  {https://ui.adsabs.harvard.edu/abs/2013MNRAS.434L..61M} {434, L61}

\bibitem[\protect\citeauthoryear{{Muzzin} et~al.,}{{Muzzin}
  et~al.}{2013}]{2013ApJ...777...18M}
{Muzzin} A.,  et~al., 2013, \mn@doi [\apj]
  {10.1088/0004-637X/777/1/1810.48550/arXiv.1303.4409}, \href
  {https://ui.adsabs.harvard.edu/abs/2013ApJ...777...18M} {777, 18}

\bibitem[\protect\citeauthoryear{{Nakajima}, {Mandelbaum}, {Seljak}, {Cohn},
  {Reyes}  \& {Cool}}{{Nakajima} et~al.}{2012}]{2012MNRAS.420.3240N}
{Nakajima} R.,  {Mandelbaum} R.,  {Seljak} U.,  {Cohn} J.~D.,  {Reyes} R.,
  {Cool} R.,  2012, \mn@doi [\mnras] {10.1111/j.1365-2966.2011.20249.x}, \href
  {https://ui.adsabs.harvard.edu/abs/2012MNRAS.420.3240N} {420, 3240}

\bibitem[\protect\citeauthoryear{{Planck Collaboration} et~al.,}{{Planck
  Collaboration} et~al.}{2016}]{2016A&A...594A..13P}
{Planck Collaboration} et~al., 2016, \mn@doi [\aap]
  {10.1051/0004-6361/201525830}, \href
  {https://ui.adsabs.harvard.edu/abs/2016A&A...594A..13P} {594, A13}

\bibitem[\protect\citeauthoryear{{Rana}, {More}, {Miyatake}, {Nishimichi},
  {Takada}, {Robotham}, {Hopkins}  \& {Holwerda}}{{Rana}
  et~al.}{2022}]{2022MNRAS.510.5408R}
{Rana} D.,  {More} S.,  {Miyatake} H.,  {Nishimichi} T.,  {Takada} M.,
  {Robotham} A. S.~G.,  {Hopkins} A.~M.,   {Holwerda} B.~W.,  2022, \mn@doi
  [\mnras] {10.1093/mnras/stac007}, \href
  {https://ui.adsabs.harvard.edu/abs/2022MNRAS.510.5408R} {510, 5408}

\bibitem[\protect\citeauthoryear{{Raveri} \& {Hu}}{{Raveri} \&
  {Hu}}{2019}]{2019PhRvD..99d3506R}
{Raveri} M.,  {Hu} W.,  2019, \mn@doi [\prd] {10.1103/PhysRevD.99.043506},
  \href {https://ui.adsabs.harvard.edu/abs/2019PhRvD..99d3506R} {99, 043506}

\bibitem[\protect\citeauthoryear{{Reddick}, {Wechsler}, {Tinker}  \&
  {Behroozi}}{{Reddick} et~al.}{2013}]{2013ApJ...771...30R}
{Reddick} R.~M.,  {Wechsler} R.~H.,  {Tinker} J.~L.,   {Behroozi} P.~S.,  2013,
  \mn@doi [\apj] {10.1088/0004-637X/771/1/30}, \href
  {https://ui.adsabs.harvard.edu/abs/2013ApJ...771...30R} {771, 30}

\bibitem[\protect\citeauthoryear{{Reddick}, {Tinker}, {Wechsler}  \&
  {Lu}}{{Reddick} et~al.}{2014}]{2014ApJ...783..118R}
{Reddick} R.~M.,  {Tinker} J.~L.,  {Wechsler} R.~H.,   {Lu} Y.,  2014, \mn@doi
  [\apj] {10.1088/0004-637X/783/2/118}, \href
  {https://ui.adsabs.harvard.edu/abs/2014ApJ...783..118R} {783, 118}

\bibitem[\protect\citeauthoryear{{Rodr{\'\i}guez-Puebla}, {Primack},
  {Avila-Reese}  \& {Faber}}{{Rodr{\'\i}guez-Puebla}
  et~al.}{2017}]{2017MNRAS.470..651R}
{Rodr{\'\i}guez-Puebla} A.,  {Primack} J.~R.,  {Avila-Reese} V.,   {Faber}
  S.~M.,  2017, \mn@doi [\mnras] {10.1093/mnras/stx1172}, \href
  {https://ui.adsabs.harvard.edu/abs/2017MNRAS.470..651R} {470, 651}

\bibitem[\protect\citeauthoryear{{Rowe} et~al.,}{{Rowe}
  et~al.}{2015}]{2015A&C....10..121R}
{Rowe} B.~T.~P.,  et~al., 2015, \mn@doi [Astronomy and Computing]
  {10.1016/j.ascom.2015.02.002}, \href
  {https://ui.adsabs.harvard.edu/abs/2015A&C....10..121R} {10, 121}

\bibitem[\protect\citeauthoryear{{Sheth} \& {Tormen}}{{Sheth} \&
  {Tormen}}{1999}]{1999MNRAS.308..119S}
{Sheth} R.~K.,  {Tormen} G.,  1999, \mn@doi [\mnras]
  {10.1046/j.1365-8711.1999.02692.x}, \href
  {https://ui.adsabs.harvard.edu/abs/1999MNRAS.308..119S} {308, 119}

\bibitem[\protect\citeauthoryear{{Sheth}, {Mo}  \& {Tormen}}{{Sheth}
  et~al.}{2001}]{2001MNRAS.323....1S}
{Sheth} R.~K.,  {Mo} H.~J.,   {Tormen} G.,  2001, \mn@doi [\mnras]
  {10.1046/j.1365-8711.2001.04006.x}, \href
  {https://ui.adsabs.harvard.edu/abs/2001MNRAS.323....1S} {323, 1}

\bibitem[\protect\citeauthoryear{{Springel} et~al.,}{{Springel}
  et~al.}{2018}]{2018MNRAS.475..676S}
{Springel} V.,  et~al., 2018, \mn@doi [\mnras] {10.1093/mnras/stx3304}, \href
  {https://ui.adsabs.harvard.edu/abs/2018MNRAS.475..676S} {475, 676}

\bibitem[\protect\citeauthoryear{{Tanaka}}{{Tanaka}}{2015}]{2015ApJ...801...20T}
{Tanaka} M.,  2015, \mn@doi [\apj] {10.1088/0004-637X/801/1/20}, \href
  {https://ui.adsabs.harvard.edu/abs/2015ApJ...801...20T} {801, 20}

\bibitem[\protect\citeauthoryear{{Tanaka} et~al.,}{{Tanaka}
  et~al.}{2018}]{2018PASJ...70S...9T}
{Tanaka} M.,  et~al., 2018, \mn@doi [\pasj] {10.1093/pasj/psx077}, \href
  {https://ui.adsabs.harvard.edu/abs/2018PASJ...70S...9T} {70, S9}

\bibitem[\protect\citeauthoryear{{Taylor} et~al.,}{{Taylor}
  et~al.}{2020}]{2020MNRAS.499.2896T}
{Taylor} E.~N.,  et~al., 2020, \mn@doi [\mnras] {10.1093/mnras/staa2648}, \href
  {https://ui.adsabs.harvard.edu/abs/2020MNRAS.499.2896T} {499, 2896}

\bibitem[\protect\citeauthoryear{{Tinker}, {Weinberg}, {Zheng}  \&
  {Zehavi}}{{Tinker} et~al.}{2005}]{2005ApJ...631...41T}
{Tinker} J.~L.,  {Weinberg} D.~H.,  {Zheng} Z.,   {Zehavi} I.,  2005, \mn@doi
  [\apj] {10.1086/432084}, \href
  {https://ui.adsabs.harvard.edu/abs/2005ApJ...631...41T} {631, 41}

\bibitem[\protect\citeauthoryear{{Tinker}, {Robertson}, {Kravtsov}, {Klypin},
  {Warren}, {Yepes}  \& {Gottl{\"o}ber}}{{Tinker}
  et~al.}{2010}]{2010ApJ...724..878T}
{Tinker} J.~L.,  {Robertson} B.~E.,  {Kravtsov} A.~V.,  {Klypin} A.,  {Warren}
  M.~S.,  {Yepes} G.,   {Gottl{\"o}ber} S.,  2010, \mn@doi [\apj]
  {10.1088/0004-637X/724/2/878}, \href
  {https://ui.adsabs.harvard.edu/abs/2010ApJ...724..878T} {724, 878}

\bibitem[\protect\citeauthoryear{{Tinker} et~al.,}{{Tinker}
  et~al.}{2012}]{2012ApJ...745...16T}
{Tinker} J.~L.,  et~al., 2012, \mn@doi [\apj] {10.1088/0004-637X/745/1/16},
  \href {https://ui.adsabs.harvard.edu/abs/2012ApJ...745...16T} {745, 16}

\bibitem[\protect\citeauthoryear{{Umetsu}}{{Umetsu}}{2020}]{2020A&ARv..28....7U}
{Umetsu} K.,  2020, \mn@doi [\aapr] {10.1007/s00159-020-00129-w}, \href
  {https://ui.adsabs.harvard.edu/abs/2020A&ARv..28....7U} {28, 7}

\bibitem[\protect\citeauthoryear{{Unruh}, {Schneider}, {Hilbert}, {Simon},
  {Martin}  \& {Puertas}}{{Unruh} et~al.}{2020}]{2020A&A...638A..96U}
{Unruh} S.,  {Schneider} P.,  {Hilbert} S.,  {Simon} P.,  {Martin} S.,
  {Puertas} J.~C.,  2020, \mn@doi [\aap] {10.1051/0004-6361/201936915}, \href
  {https://ui.adsabs.harvard.edu/abs/2020A&A...638A..96U} {638, A96}

\bibitem[\protect\citeauthoryear{{Varga} et~al.,}{{Varga}
  et~al.}{2019}]{2019MNRAS.489.2511V}
{Varga} T.~N.,  et~al., 2019, \mn@doi [\mnras] {10.1093/mnras/stz2185}, \href
  {https://ui.adsabs.harvard.edu/abs/2019MNRAS.489.2511V} {489, 2511}

\bibitem[\protect\citeauthoryear{{Velander} et~al.,}{{Velander}
  et~al.}{2014}]{2014MNRAS.437.2111V}
{Velander} M.,  et~al., 2014, \mn@doi [\mnras] {10.1093/mnras/stt2013}, \href
  {https://ui.adsabs.harvard.edu/abs/2014MNRAS.437.2111V} {437, 2111}

\bibitem[\protect\citeauthoryear{{Viola} et~al.,}{{Viola}
  et~al.}{2015}]{2015MNRAS.452.3529V}
{Viola} M.,  et~al., 2015, \mn@doi [\mnras] {10.1093/mnras/stv1447}, \href
  {https://ui.adsabs.harvard.edu/abs/2015MNRAS.452.3529V} {452, 3529}

\bibitem[\protect\citeauthoryear{{Vogelsberger} et~al.,}{{Vogelsberger}
  et~al.}{2014}]{2014MNRAS.444.1518V}
{Vogelsberger} M.,  et~al., 2014, \mn@doi [\mnras] {10.1093/mnras/stu1536},
  \href {https://ui.adsabs.harvard.edu/abs/2014MNRAS.444.1518V} {444, 1518}

\bibitem[\protect\citeauthoryear{{Vogelsberger}, {Marinacci}, {Torrey}  \&
  {Puchwein}}{{Vogelsberger} et~al.}{2020}]{2020NatRP...2...42V}
{Vogelsberger} M.,  {Marinacci} F.,  {Torrey} P.,   {Puchwein} E.,  2020,
  \mn@doi [Nature Reviews Physics] {10.1038/s42254-019-0127-2}, \href
  {https://ui.adsabs.harvard.edu/abs/2020NatRP...2...42V} {2, 42}

\bibitem[\protect\citeauthoryear{{Zehavi} et~al.,}{{Zehavi}
  et~al.}{2005}]{2005ApJ...630....1Z}
{Zehavi} I.,  et~al., 2005, \mn@doi [\apj] {10.1086/431891}, \href
  {https://ui.adsabs.harvard.edu/abs/2005ApJ...630....1Z} {630, 1}

\bibitem[\protect\citeauthoryear{{Zehavi} et~al.,}{{Zehavi}
  et~al.}{2011}]{2011ApJ...736...59Z}
{Zehavi} I.,  et~al., 2011, \mn@doi [\apj] {10.1088/0004-637X/736/1/59}, \href
  {https://ui.adsabs.harvard.edu/abs/2011ApJ...736...59Z} {736, 59}

\bibitem[\protect\citeauthoryear{{Zentner}, {Hearin}, {van den Bosch}, {Lange}
  \& {Villarreal}}{{Zentner} et~al.}{2019}]{2019MNRAS.485.1196Z}
{Zentner} A.~R.,  {Hearin} A.,  {van den Bosch} F.~C.,  {Lange} J.~U.,
  {Villarreal} A.~S.,  2019, \mn@doi [\mnras] {10.1093/mnras/stz470}, \href
  {https://ui.adsabs.harvard.edu/abs/2019MNRAS.485.1196Z} {485, 1196}

\bibitem[\protect\citeauthoryear{{Zheng} et~al.,}{{Zheng}
  et~al.}{2005}]{2005ApJ...633..791Z}
{Zheng} Z.,  et~al., 2005, \mn@doi [\apj] {10.1086/466510}, \href
  {https://ui.adsabs.harvard.edu/abs/2005ApJ...633..791Z} {633, 791}

\bibitem[\protect\citeauthoryear{{Zheng}, {Coil}  \& {Zehavi}}{{Zheng}
  et~al.}{2007}]{2007ApJ...667..760Z}
{Zheng} Z.,  {Coil} A.~L.,   {Zehavi} I.,  2007, \mn@doi [\apj]
  {10.1086/521074}, \href
  {https://ui.adsabs.harvard.edu/abs/2007ApJ...667..760Z} {667, 760}

\bibitem[\protect\citeauthoryear{{Zu} \& {Mandelbaum}}{{Zu} \&
  {Mandelbaum}}{2015}]{2015MNRAS.454.1161Z}
{Zu} Y.,  {Mandelbaum} R.,  2015, \mn@doi [\mnras] {10.1093/mnras/stv2062},
  \href {https://ui.adsabs.harvard.edu/abs/2015MNRAS.454.1161Z} {454, 1161}

\bibitem[\protect\citeauthoryear{{Zu} \& {Mandelbaum}}{{Zu} \&
  {Mandelbaum}}{2016}]{2016MNRAS.457.4360Z}
{Zu} Y.,  {Mandelbaum} R.,  2016, \mn@doi [\mnras] {10.1093/mnras/stw221},
  \href {https://ui.adsabs.harvard.edu/abs/2016MNRAS.457.4360Z} {457, 4360}

\bibitem[\protect\citeauthoryear{{van Uitert}, {Hoekstra}, {Velander},
  {Gilbank}, {Gladders}  \& {Yee}}{{van Uitert}
  et~al.}{2011}]{2011A&A...534A..14V}
{van Uitert} E.,  {Hoekstra} H.,  {Velander} M.,  {Gilbank} D.~G.,  {Gladders}
  M.~D.,   {Yee} H.~K.~C.,  2011, \mn@doi [\aap] {10.1051/0004-6361/201117308},
  \href {https://ui.adsabs.harvard.edu/abs/2011A&A...534A..14V} {534, A14}

\bibitem[\protect\citeauthoryear{{van Uitert}, {Cacciato}, {Hoekstra}  \&
  {Herbonnet}}{{van Uitert} et~al.}{2015}]{2015A&A...579A..26V}
{van Uitert} E.,  {Cacciato} M.,  {Hoekstra} H.,   {Herbonnet} R.,  2015,
  \mn@doi [\aap] {10.1051/0004-6361/201525834}, \href
  {https://ui.adsabs.harvard.edu/abs/2015A&A...579A..26V} {579, A26}

\bibitem[\protect\citeauthoryear{{van Uitert} et~al.,}{{van Uitert}
  et~al.}{2016}]{2016MNRAS.459.3251V}
{van Uitert} E.,  et~al., 2016, \mn@doi [\mnras] {10.1093/mnras/stw747}, \href
  {https://ui.adsabs.harvard.edu/abs/2016MNRAS.459.3251V} {459, 3251}

\bibitem[\protect\citeauthoryear{{van den Bosch}, {More}, {Cacciato}, {Mo}  \&
  {Yang}}{{van den Bosch} et~al.}{2013}]{2013MNRAS.430..725V}
{van den Bosch} F.~C.,  {More} S.,  {Cacciato} M.,  {Mo} H.,   {Yang} X.,
  2013, \mn@doi [\mnras] {10.1093/mnras/sts006}, \href
  {https://ui.adsabs.harvard.edu/abs/2013MNRAS.430..725V} {430, 725}

\makeatother
\end{thebibliography}

\begin{table*}
    \centering
    \renewcommand{\arraystretch}{1.5}
    \begin{tabularx}{0.99\textwidth}{
    | c ||
     C |
     >{\centering\arraybackslash}X |
     C |
     >{\centering\arraybackslash}X |
     C |
     >{\centering\arraybackslash}X ||
     >{\centering\arraybackslash}X |
     c |
     c |
    }
        \hline\hline
        \multicolumn{10}{|c|}{$z_1 \in [0.30, 0.55)$} \\
        \hline
        \multirow{2}{*}{$\log \left[ \frac{M_{*,\rm lim}}{\hiimsun} \right]$} & \multirow{2}{*}{$\log \left[ \frac{\Mmin}{\himsun} \right]$} & \multirow{2}{*}{$\sigmalogM$} & \multirow{2}{*}{$\log \left[ \frac{M_1}{\himsun} \right]$} & \multirow{2}{*}{$\alpha$} & \multirow{2}{*}{$\log \left[ \frac{M_0}{\himsun} \right]$} & \multirow{2}{*}{$c_{\rm fac}$}  & \multirow{2}{*}{$\fsat$} & \multirow{2}{*}{{\large $\frac{\ngal}{(h^{-1} {\rm Mpc})^3}$}\,$10^3$} & \multirow{2}{*}{$\log \left[ \frac{\avg{M_{\rm cen}}}{\himsun} \right]$}\\
        &  &  &  &  &  &  &  &  & \\
        \hline
        8.6 & $11.25_{-0.10}^{+0.19}$ & $0.37_{-0.25}^{+0.30}$ & $12.96_{-0.13}^{+0.12}$ & $1.48_{-0.14}^{+0.14}$ & $9.46_{-2.38}^{+2.39}$ & $1.01_{-0.17}^{+0.19}$ & $0.12_{-0.02}^{+0.03}$ & $29.00_{-3.54}^{+3.52}$ & $12.04_{-0.04}^{+0.05}$\\
        \hline
        8.8 & $11.32_{-0.09}^{+0.17}$ & $0.34_{-0.23}^{+0.29}$ & $12.95_{-0.12}^{+0.12}$ & $1.44_{-0.12}^{+0.13}$ & $9.50_{-2.40}^{+2.36}$ & $1.05_{-0.16}^{+0.17}$ & $0.14_{-0.02}^{+0.03}$ & $25.34_{-3.14}^{+2.87}$ & $12.10_{-0.04}^{+0.04}$\\
        \hline
        9.0 & $11.40_{-0.09}^{+0.17}$ & $0.33_{-0.23}^{+0.28}$ & $12.94_{-0.11}^{+0.11}$ & $1.41_{-0.11}^{+0.12}$ & $9.50_{-2.40}^{+2.37}$ & $1.07_{-0.16}^{+0.17}$ & $0.16_{-0.03}^{+0.03}$ & $21.63_{-2.81}^{+2.50}$ & $12.17_{-0.04}^{+0.04}$\\
        \hline
        9.2 & $11.47_{-0.08}^{+0.15}$ & $0.31_{-0.21}^{+0.27}$ & $12.98_{-0.14}^{+0.11}$ & $1.42_{-0.14}^{+0.12}$ & $9.97_{-2.71}^{+2.49}$ & $1.08_{-0.15}^{+0.17}$ & $0.16_{-0.03}^{+0.03}$ & $18.24_{-2.15}^{+2.13}$ & $12.23_{-0.04}^{+0.04}$\\
        \hline
        9.4 & $11.57_{-0.08}^{+0.14}$ & $0.30_{-0.20}^{+0.26}$ & $12.99_{-0.22}^{+0.11}$ & $1.37_{-0.20}^{+0.13}$ & $10.24_{-2.90}^{+2.63}$ & $1.09_{-0.15}^{+0.17}$ & $0.17_{-0.03}^{+0.04}$ & $14.76_{-1.83}^{+1.79}$ & $12.31_{-0.04}^{+0.04}$\\
        \hline
        9.6 & $11.68_{-0.09}^{+0.16}$ & $0.33_{-0.22}^{+0.27}$ & $13.04_{-0.20}^{+0.11}$ & $1.35_{-0.19}^{+0.12}$ & $10.25_{-2.90}^{+2.57}$ & $1.09_{-0.15}^{+0.16}$ & $0.18_{-0.03}^{+0.04}$ & $12.06_{-1.53}^{+1.54}$ & $12.39_{-0.04}^{+0.04}$\\
        \hline
        9.8 & $11.77_{-0.09}^{+0.16}$ & $0.35_{-0.23}^{+0.26}$ & $13.10_{-0.19}^{+0.11}$ & $1.36_{-0.19}^{+0.13}$ & $10.28_{-2.94}^{+2.51}$ & $1.12_{-0.15}^{+0.16}$ & $0.18_{-0.03}^{+0.04}$ & $10.01_{-1.26}^{+1.29}$ & $12.45_{-0.04}^{+0.04}$\\
        \hline
        10.0 & $11.93_{-0.11}^{+0.20}$ & $0.42_{-0.27}^{+0.27}$ & $13.12_{-0.45}^{+0.12}$ & $1.29_{-0.35}^{+0.14}$ & $10.85_{-3.33}^{+2.32}$ & $1.10_{-0.15}^{+0.16}$ & $0.18_{-0.04}^{+0.04}$ & $7.67_{-1.04}^{+1.00}$ & $12.55_{-0.04}^{+0.04}$\\
        \hline
        10.2 & $12.28_{-0.19}^{+0.22}$ & $0.69_{-0.26}^{+0.20}$ & $13.34_{-0.19}^{+0.11}$ & $1.39_{-0.22}^{+0.16}$ & $10.25_{-2.92}^{+2.68}$ & $1.07_{-0.16}^{+0.18}$ & $0.16_{-0.03}^{+0.03}$ & $4.82_{-0.73}^{+0.72}$ & $12.66_{-0.04}^{+0.04}$\\
        \hline
        10.4 & $12.70_{-0.18}^{+0.20}$ & $0.90_{-0.15}^{+0.14}$ & $13.60_{-0.46}^{+0.12}$ & $1.54_{-0.48}^{+0.24}$ & $10.90_{-3.34}^{+2.62}$ & $1.15_{-0.16}^{+0.17}$ & $0.11_{-0.02}^{+0.03}$ & $2.68_{-0.41}^{+0.40}$ & $12.78_{-0.03}^{+0.03}$\\
        \hline
        10.6 & $13.14_{-0.18}^{+0.20}$ & $1.07_{-0.12}^{+0.12}$ & $13.74_{-0.83}^{+0.13}$ & $1.57_{-0.74}^{+0.34}$ & $11.37_{-3.66}^{+2.42}$ & $1.15_{-0.16}^{+0.17}$ & $0.10_{-0.02}^{+0.03}$ & $1.45_{-0.22}^{+0.22}$ & $12.90_{-0.03}^{+0.04}$\\
        \hline
        10.8 & $13.70_{-0.17}^{+0.19}$ & $1.21_{-0.10}^{+0.10}$ & $13.99_{-0.64}^{+0.17}$ & $1.62_{-0.82}^{+0.66}$ & $11.06_{-3.47}^{+2.76}$ & $1.20_{-0.15}^{+0.17}$ & $0.07_{-0.02}^{+0.03}$ & $0.59_{-0.09}^{+0.09}$ & $13.06_{-0.03}^{+0.03}$\\
        \hline
        11.0 & $14.44_{-0.17}^{+0.19}$ & $1.37_{-0.09}^{+0.09}$ & $14.36_{-1.05}^{+0.25}$ & $2.41_{-1.45}^{+1.62}$ & $11.90_{-4.05}^{+2.79}$ & $1.17_{-0.17}^{+0.18}$ & $0.03_{-0.01}^{+0.03}$ & $0.17_{-0.03}^{+0.03}$ & $13.22_{-0.04}^{+0.04}$\\
        \hline
    \end{tabularx}
    \caption{Constraints on HOD parameters, satellite fraction, abundance and the average halo mass from joint analysis of weak lensing and I20 abundance for redshift bin $z_1$.}
    \label{tab:Parameter_constraints_z1}
\end{table*}

\begin{table*}
    \centering
    \renewcommand{\arraystretch}{1.5}
    \begin{tabularx}{0.99\textwidth}{
    | c ||
     C |
     >{\centering\arraybackslash}X |
     C |
     >{\centering\arraybackslash}X |
     C |
     >{\centering\arraybackslash}X ||
     >{\centering\arraybackslash}X |
     c |
     c |
    }
        \hline\hline
        \multicolumn{10}{|c|}{$z_2 \in [0.55, 0.80)$} \\
        \hline
        \multirow{2}{*}{$\log \left[ \frac{M_{*,\rm lim}}{\hiimsun} \right]$} & \multirow{2}{*}{$\log \left[ \frac{\Mmin}{\himsun} \right]$} & \multirow{2}{*}{$\sigmalogM$} & \multirow{2}{*}{$\log \left[ \frac{M_1}{\himsun} \right]$} & \multirow{2}{*}{$\alpha$} & \multirow{2}{*}{$\log \left[ \frac{M_0}{\himsun} \right]$} & \multirow{2}{*}{$c_{\rm fac}$}  & \multirow{2}{*}{$\fsat$} & \multirow{2}{*}{{\large $\frac{\ngal}{(h^{-1} {\rm Mpc})^3}$}\,$10^3$} & \multirow{2}{*}{$\log \left[ \frac{\avg{M_{\rm cen}}}{\himsun} \right]$}\\
        &  &  &  &  &  &  &  &  & \\
        \hline
        9.0 & $11.39_{-0.07}^{+0.10}$ & $0.21_{-0.15}^{+0.21}$ & $12.58_{-0.21}^{+0.18}$ & $0.99_{-0.14}^{+0.14}$ & $9.24_{-2.21}^{+2.23}$ & $1.13_{-0.16}^{+0.16}$ & $0.24_{-0.07}^{+0.09}$ & $22.66_{-3.22}^{+3.45}$ & $12.13_{-0.04}^{+0.05}$\\
        \hline
        9.2 & $11.45_{-0.07}^{+0.09}$ & $0.19_{-0.13}^{+0.20}$ & $12.74_{-0.19}^{+0.16}$ & $1.12_{-0.15}^{+0.15}$ & $9.33_{-2.28}^{+2.33}$ & $1.14_{-0.15}^{+0.16}$ & $0.20_{-0.06}^{+0.07}$ & $18.54_{-2.50}^{+2.85}$ & $12.19_{-0.04}^{+0.04}$\\
        \hline
        9.4 & $11.59_{-0.07}^{+0.11}$ & $0.24_{-0.17}^{+0.23}$ & $12.83_{-0.47}^{+0.16}$ & $1.13_{-0.31}^{+0.17}$ & $10.48_{-3.03}^{+2.41}$ & $1.11_{-0.15}^{+0.16}$ & $0.17_{-0.05}^{+0.06}$ & $13.50_{-1.69}^{+1.69}$ & $12.28_{-0.03}^{+0.04}$\\
        \hline
        9.6 & $11.67_{-0.07}^{+0.12}$ & $0.26_{-0.18}^{+0.24}$ & $12.82_{-0.76}^{+0.17}$ & $1.08_{-0.40}^{+0.17}$ & $11.14_{-3.52}^{+1.87}$ & $1.16_{-0.14}^{+0.15}$ & $0.18_{-0.05}^{+0.07}$ & $11.56_{-1.41}^{+1.50}$ & $12.34_{-0.04}^{+0.04}$\\
        \hline
        9.8 & $11.78_{-0.07}^{+0.13}$ & $0.28_{-0.19}^{+0.23}$ & $12.88_{-1.01}^{+0.18}$ & $1.07_{-0.48}^{+0.19}$ & $11.44_{-3.73}^{+1.68}$ & $1.15_{-0.15}^{+0.15}$ & $0.18_{-0.05}^{+0.07}$ & $9.19_{-1.14}^{+1.19}$ & $12.42_{-0.04}^{+0.04}$\\
        \hline
        10.0 & $11.84_{-0.07}^{+0.11}$ & $0.24_{-0.16}^{+0.22}$ & $12.81_{-1.59}^{+0.30}$ & $0.98_{-0.53}^{+0.30}$ & $12.65_{-4.23}^{+0.66}$ & $1.22_{-0.14}^{+0.15}$ & $0.15_{-0.04}^{+0.07}$ & $7.49_{-0.93}^{+0.93}$ & $12.48_{-0.04}^{+0.04}$\\
        \hline
        10.2 & $12.13_{-0.14}^{+0.19}$ & $0.49_{-0.28}^{+0.22}$ & $13.19_{-1.21}^{+0.18}$ & $1.19_{-0.63}^{+0.28}$ & $11.95_{-4.04}^{+1.43}$ & $1.10_{-0.15}^{+0.16}$ & $0.13_{-0.03}^{+0.05}$ & $4.62_{-0.67}^{+0.60}$ & $12.61_{-0.03}^{+0.03}$\\
        \hline
        10.4 & $12.43_{-0.17}^{+0.19}$ & $0.68_{-0.19}^{+0.16}$ & $13.38_{-1.23}^{+0.16}$ & $1.34_{-0.74}^{+0.33}$ & $11.87_{-3.99}^{+1.68}$ & $1.15_{-0.15}^{+0.16}$ & $0.11_{-0.03}^{+0.04}$ & $2.97_{-0.44}^{+0.44}$ & $12.70_{-0.03}^{+0.03}$\\
        \hline
        10.6 & $12.77_{-0.17}^{+0.19}$ & $0.79_{-0.15}^{+0.14}$ & $13.59_{-0.71}^{+0.17}$ & $1.39_{-0.66}^{+0.45}$ & $10.99_{-3.41}^{+2.52}$ & $1.05_{-0.17}^{+0.17}$ & $0.10_{-0.03}^{+0.05}$ & $1.63_{-0.25}^{+0.25}$ & $12.85_{-0.03}^{+0.03}$\\
        \hline
        10.8 & $13.27_{-0.17}^{+0.19}$ & $0.97_{-0.12}^{+0.12}$ & $13.76_{-0.35}^{+0.21}$ & $1.57_{-0.66}^{+0.75}$ & $10.56_{-3.09}^{+2.82}$ & $1.03_{-0.17}^{+0.18}$ & $0.10_{-0.04}^{+0.06}$ & $0.75_{-0.12}^{+0.12}$ & $12.98_{-0.04}^{+0.04}$\\
        \hline
        11.0 & $13.79_{-0.17}^{+0.20}$ & $1.08_{-0.11}^{+0.11}$ & $14.09_{-0.85}^{+0.29}$ & $2.36_{-1.47}^{+1.68}$ & $11.38_{-3.68}^{+3.06}$ & $0.94_{-0.17}^{+0.19}$ & $0.04_{-0.02}^{+0.07}$ & $0.27_{-0.04}^{+0.04}$ & $13.14_{-0.04}^{+0.04}$\\
        \hline
        11.2 & $14.74_{-0.21}^{+0.24}$ & $1.37_{-0.11}^{+0.12}$ & $14.13_{-0.75}^{+0.32}$ & $1.74_{-1.23}^{+1.77}$ & $11.01_{-3.41}^{+2.98}$ & $1.03_{-0.19}^{+0.19}$ & $0.06_{-0.04}^{+0.12}$ & $0.08_{-0.01}^{+0.01}$ & $13.22_{-0.06}^{+0.06}$\\
        \hline
    \end{tabularx}
    \caption{Constraints on HOD parameters, satellite fraction, abundance and the average halo mass from joint analysis of weak lensing and I20 abundance for redshift bin $z_2$.}
    \label{tab:Parameter_constraints_z2}    
\end{table*}

\begin{table*}
    \centering
    \begin{tabularx}{0.99\textwidth}{
    | c |
    >{\centering\arraybackslash}X |
    >{\centering\arraybackslash}X |
    >{\centering\arraybackslash}X |
    >{\centering\arraybackslash}X |
    >{\centering\arraybackslash}X |
    >{\centering\arraybackslash}X |
    >{\centering\arraybackslash}X |
    >{\centering\arraybackslash}X |
    >{\centering\arraybackslash}X |
    >{\centering\arraybackslash}X |
    >{\centering\arraybackslash}X |
    >{\centering\arraybackslash}X |
    >{\centering\arraybackslash}X |
    >{\centering\arraybackslash}X |
    }
        \hline
        \multirow{2}{*}{best fit $\chi^{2}$ for} & \multirow{2}{*}{8.6 }& \multirow{2}{*}{8.8 }& \multirow{2}{*}{9.0 }& \multirow{2}{*}{9.2 }& \multirow{2}{*}{9.4 }& \multirow{2}{*}{9.6 }& \multirow{2}{*}{9.8 }& \multirow{2}{*}{10.0} & \multirow{2}{*}{10.2} & \multirow{2}{*}{10.4} & \multirow{2}{*}{10.6} & \multirow{2}{*}{10.8} & \multirow{2}{*}{11.0} & \multirow{2}{*}{11.2} \\
        & & & & & & & & & & & & & & \\
        \hline
        {\bf I20}, $z_{1}$ & 10.13 & 6.84 & 8.70 & 4.88 & 5.77 & 5.99 & 5.28 & 7.79 & 10.29 & 6.73 & 7.22 & 11.96 & 18.68 & ---\\
        \hline
        {\bf M13}, $z_{1}$ & 12.20 & 8.01 & 9.25 & 5.15 & 6.49 & 6.23 & 5.07 & 7.71 & 10.29 & 6.60 & 6.73 & 11.75 & 18.06 & ---\\
        \hline
        {\bf I20}, $z_{2}$ & --- & --- & 14.01 & 18.10 & 13.35 & 10.93 & 13.51 & 9.29 & 15.58 & 19.57 & 14.23 & 11.90 & 9.12 & 2.51\\
        \hline
        {\bf M13}, $z_{2}$ & --- & --- & 13.34 & 16.82 & 13.78 & 10.97 & 13.52 & 9.79 & 15.49 & 19.74 & 14.26 & 12.08 & 8.96 & 2.51\\
        \hline
    \end{tabularx}
    \caption{$\chi^{2}$ comparison between joint fits of lensing with abundances from I20 and M13. The top row is $\log_{10}$ of stellar mass thresholds in units of $\hiimsun$ and the left most column indicates the choice of redshift bin and abundance used in joint fits. Each threshold stellar mass subsample of a given redshift bin probes almost the same degrees of freedom for both choices of abundance.}
    \label{tab:chisq_comparison_for_abundance_choices}    
\end{table*}

\appendix

\section{Photo-$z$ bias computation due to errors in $\zs$ alone} \label{Nakajima_way}
Our full lens sample had no overlap with the HST COSMOS survey footprint. To perform this photo-$z$ bias analysis, we required a large enough sample of galaxies which can act as a representative population of our lens subsamples, a calibration sample. One has to make sure that the number densities and color-magnitude distributions of the calibration sample should be at least similar, if not the same, to the original source galaxy catalog (in our case, S16A catalog) used for the ESD signal computation. HSC-SSP survey, in its PDR2, processed data from COSMOS field as part of S17A shape catalog release; and provides `self organizing map' (SOM) weights ($w_{\rm som}$) and full PDF ($p(z)$) along with point estimates of photo-$z$s of the galaxies in the shape catalog. The point photo-$z$ estimates are available from both COSMOS photo-$z$ catalog and from the HSC photometric catalogs. The SOM weights are designed to match the color-magnitude distributions of galaxies in COSMOS field with those selected for weak lensing in S16A. Therefore we redefine the source weights $w_{s}$ by multiplying $w_{\rm som,s}$ to its previous definition eq(2), resuting $\wls$ is still computed by the expression defined in section 4 with redefined $\ws$ (strictly for this sub-section only). 

Therefore we take all the galaxies in the \texttt{wide\_s17a\_[9812,9813]} tracts of HSC, observed in the COSMOS field (henceforth: calibration catalog, CC). We depend on the goodness of COSMOS photo-$z$s and treat them as true redshifts of the galaxies. We use the CC catalog as both lenses and sources for the photo-$z$ bias analysis. Along with other quality filters as chosen in I20 work for the lens subsamples, we also apply the same {\sc photoz\_risk\_best} cuts of $< 0.5$ and $\leq 0.1$ for the sources and the lenses respectively in order to match the selection criteria used in the actual ESD signal computation. We obtain calibration lens subsamples defined with the same redshift trimming and binning and stellar mass cuts on CC as our actual lenses described in Table . The average bias in ESD signal due to errors in photometric redshifts of source galaxies $\zs$, computed at a lens redshift $\zl$ is written as,
\begin{align}
    b_{z}\left(\zl \right)+1 \equiv \frac{\Delta \Sigma}{\widetilde{\Delta \Sigma}}(\zl)= \frac{\sum_{\rm s} \wls \langle \Sigma^{-1}_{\rm crit, ls}\rangle^{-1} \widetilde{\Sigma}_{\rm crit, ls}^{-1}}{\sum_{\rm s} \wls} \label{nakajima_avbias_for_one_lens} 
\end{align}
where, quantities with/without tilde are true/observed estimates of corresponding quantities. $\widetilde{\Sigma}_{\rm crit, ls}$ is obtained using eq(4) with true redshifts of the lens and the source in each $\rm l-s$ pair. $\wls$ is computed as explained in the above paragraph and $\langle\Sigma^{-1}_{\rm crit, ls}\rangle$ from eq(5), wherein we perform the same source selection treatment eq(9) to get $\langle\Sigma^{-1}_{\rm crit, ls}\rangle$, as was done while computing the actual ESD signals for each subsample of lens galaxies. Eq(\ref{nakajima_avbias_for_one_lens}) performs summation over all the sources behind the given lens photo-$z$, $\zl$.

Thus, for a selected lens galaxy subsample, we compute an estimate of the photo-$z$ bias averaged over the redshift distribution of all the lens galaxies in the sample by following \citep[eq(22), ][]{2012MNRAS.420.3240N}, which we mention here for completeness -
\begin{align}
    \left\langle b_{z}\right\rangle=\frac{\int \mathrm{d} \zl \  p\left(\zl \right) w_{\rm l} \left(\zl \right) b_{z}\left(\zl \right)}{\int \mathrm{d} \zl \  p\left(\zl \right) w_{\rm l}\left(\zl \right)}. \label{nakajima_avbias}
\end{align}
Here we compute the probability for a galaxy to have HSC photo-$z$ within $\zl \text{ and } \zl+{\rm d}\zl$ i.e. $p(\zl){\rm d}\zl$ by binning the actual lens galaxies in bins of photo-$z$s and counting the fraction of galaxies falling in each bin. $\wl = D^{-2}_c(\zl)\ \sum_{\rm s} \wls$ where $D_c(\zl)$ denotes the comoving angular diameter distance to the lens and summation is performed over all the sources defined for a fixed lens redshift. The factor $\wl \left(\zl \right)$ is supposed to weight the bias $b_z$ depending on the effectiveness of a lens at a redshift $\zl$ for a given distribution of source redshifts, in the same fashion that lensing measurements are performed. The integration in eq(\ref{nakajima_avbias}) are done over the range of lens redshifts.\\
Note: We imposed more stringent quality cut on {\sc photoz\_risk\_best} for lenses compared to the S16A catalog of background galaxies in order to avoid galaxies with multiple peaks in the $z$-PDF. But we don't know the true redshifts of these lenses, hence above analysis can not be performed for the actual lenses. However we chose to use the $p(\zl)$ due to actual lenses in above calculation, whereas $\wl$ comes from the HSC-COSMOS matched objects following the similar selection criteria as the actual lenses. This is justified in the expectation that the photo-$z$ bias in the HSC survey will be independent of the choice of LOS. We have checked that the point photo-$z$ distribution $p(\zl)$ of calibration sample after applying the similar cuts as lenses, are similar to $p(\zl)$ of lenses. 

\begin{figure}
    \centering
    \begin{subfigure}[b]{0.99\columnwidth}
        \includegraphics[width=1\textwidth]{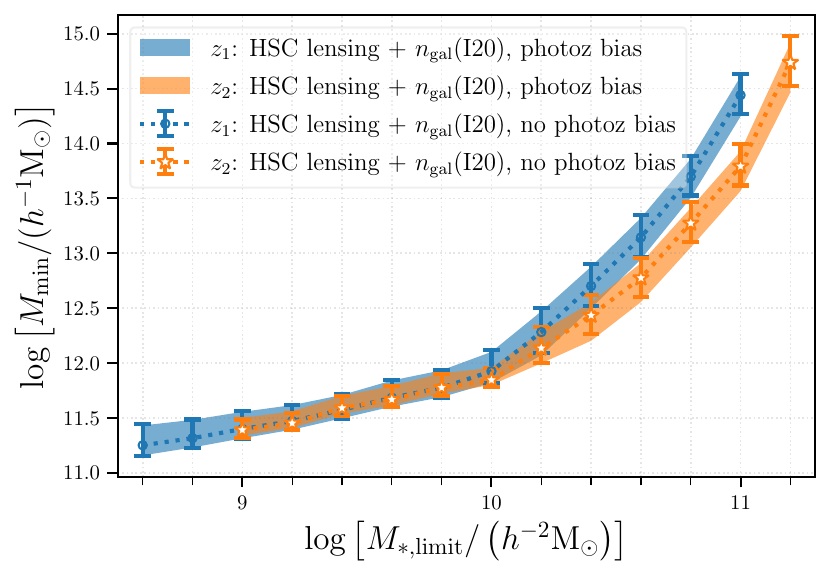}\\
        \includegraphics[width=1\textwidth]{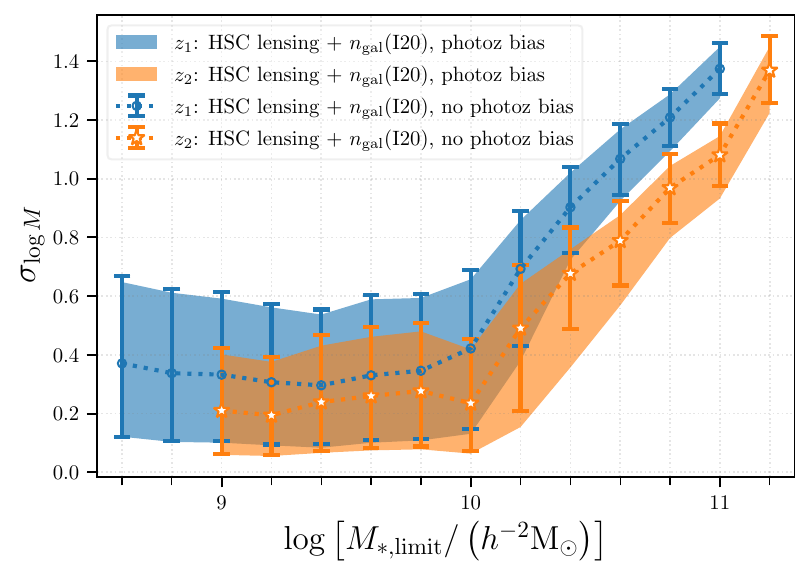}
    \end{subfigure}
    \caption{Effect of applying photo-$z$ bias on central HOD parameters in our fiducal analysis.}
    \label{fig:photozbias_effect_centralHOD}
\end{figure}

Fig.~\ref{fig:photozbias_effect_centralHOD} demonstrates the effect of applying photo-$z$ bias of the background galaxies in our fiducial joint analysis. We find that these biases can be safely ignored without affecting any of our inferences. 

\subsection{Halo occupation distributions of stellar mass threshold bins}
\begin{figure*}
    \begin{subfigure}[b]{0.99\columnwidth}
    \includegraphics[width=0.99\textwidth]{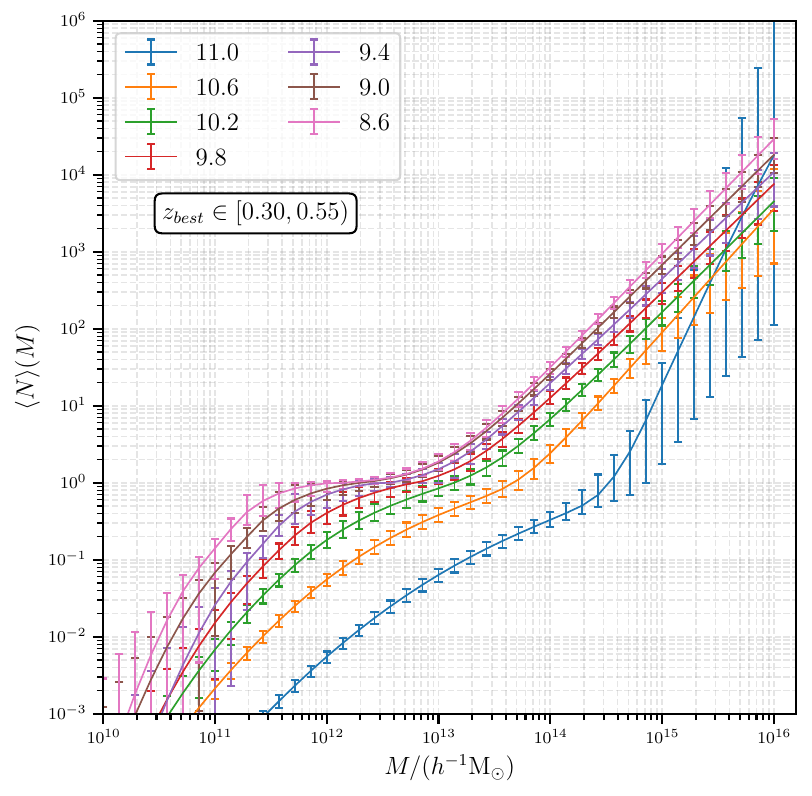}
    \end{subfigure}
    \begin{subfigure}[b]{0.99\columnwidth}
        \includegraphics[width=0.99\textwidth]{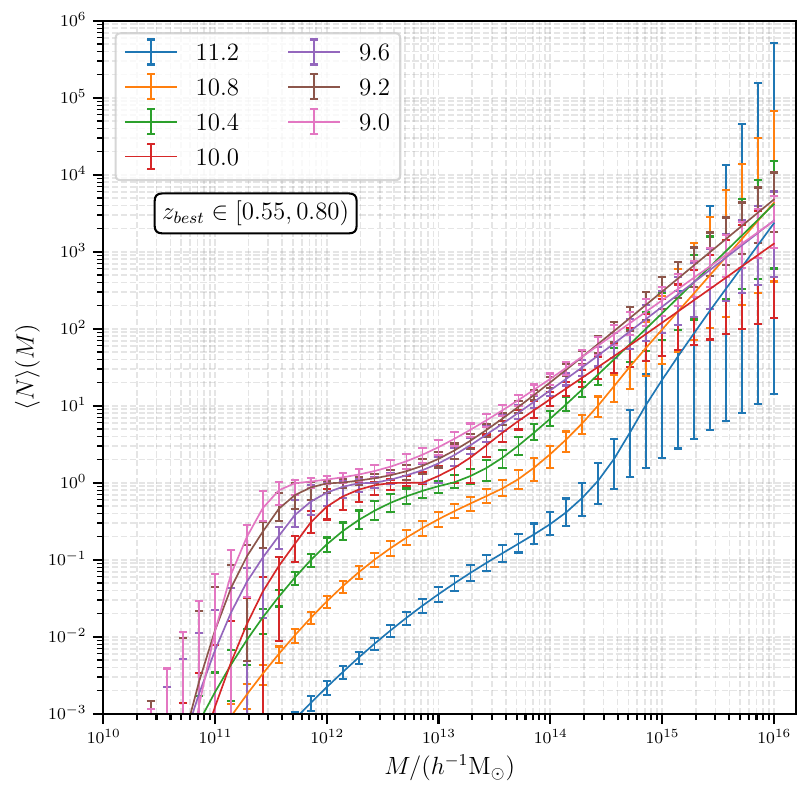}
    \end{subfigure}
    \caption{HOD constraints with 68 percent credible interval around median HODs shown for only a subset of threshold subsamples obtained from our joint analyses of lensing and I20 abundances for redshift bins $z_1$ (Left) and $z_2$ (right) respectively.}
    \label{fig:HODplot}
\end{figure*}

We show in the two panels of Fig.~\ref{fig:HODplot} the HODs constrained from individual threshold bins for the two different redshift bins. Ideally one would expect that the HODs of the individual threshold stellar mass bins should not cross, given that the sample of galaxies in the higher threshold bin are a subsample of the galaxies in the lower threshold bin. Although these conditions apparently are not always satisfied by the constrained HODs, the violations are not significant given the errors. In principle a joint fit to all the threshold bins should get rid of such inconsistencies. However, a joint fit will also require computation of a much larger covariance matrix given the significant overlap in the galaxy samples in each threshold bin. We defer such joint fits of different threshold stellar mass bins to future work. 

\section{Sample Posterior Distributions of free parameters for our fiducial analyses}
\begin{figure*}
        \includegraphics[width=0.99\textwidth]{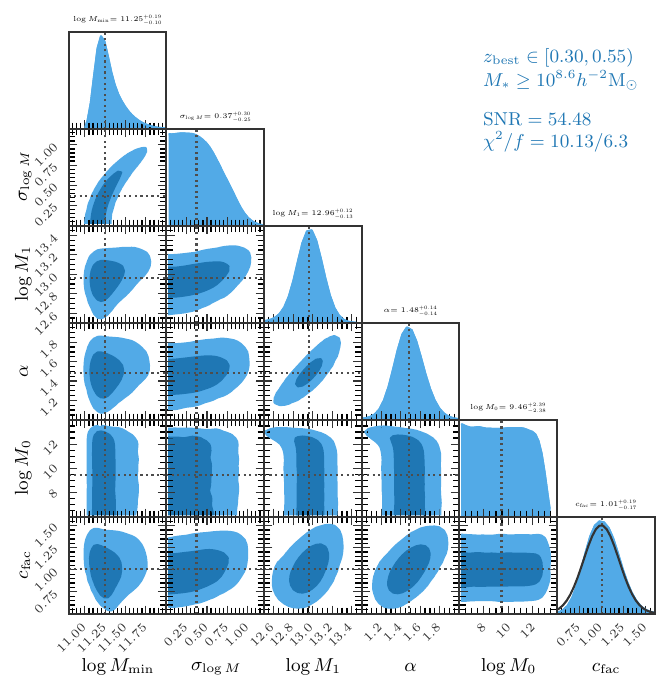} 
        \caption{Parameter constraints and covariances in parameter space for redshift bin $z_1$, $M_* \ge 10^{8.6} \hiimsun$.}
        \label{fig:Parameter_constraints_z1_8.6}
\end{figure*}

\begin{figure*}
    \includegraphics[width=0.99\textwidth]{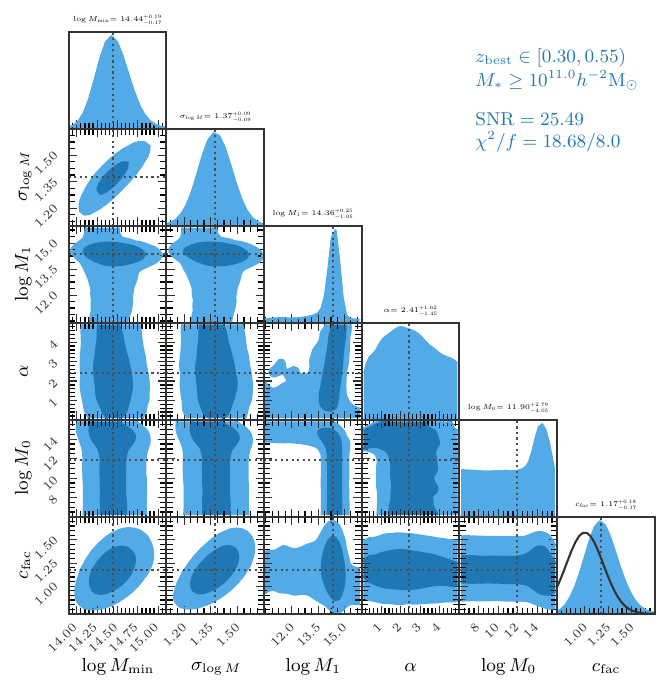}
    \caption{Parameter constraints and covariances in parameter space for redshift bin $z_1$, $M_* \ge 10^{11.0} \hiimsun$.}
    \label{fig:Parameter_constraints_z1_11.0}    
\end{figure*}

\begin{figure*}
    \includegraphics[width=0.99\textwidth]{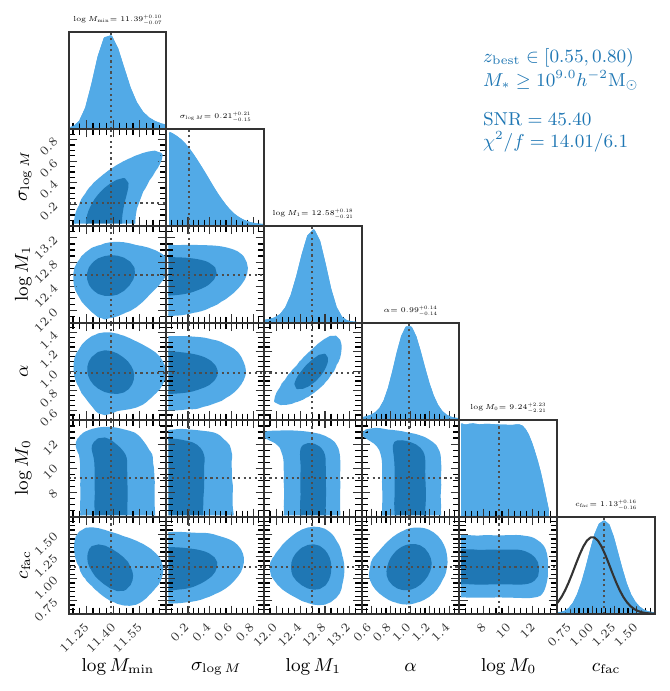}
    \caption{Parameter constraints and covariances in parameter space for redshift bin $z_2$, $M_* \ge 10^{9.0} \hiimsun$.}
    \label{fig:Parameter_constraints_z2_9.0}
\end{figure*}

\begin{figure*}
            \includegraphics[width=0.99\textwidth]{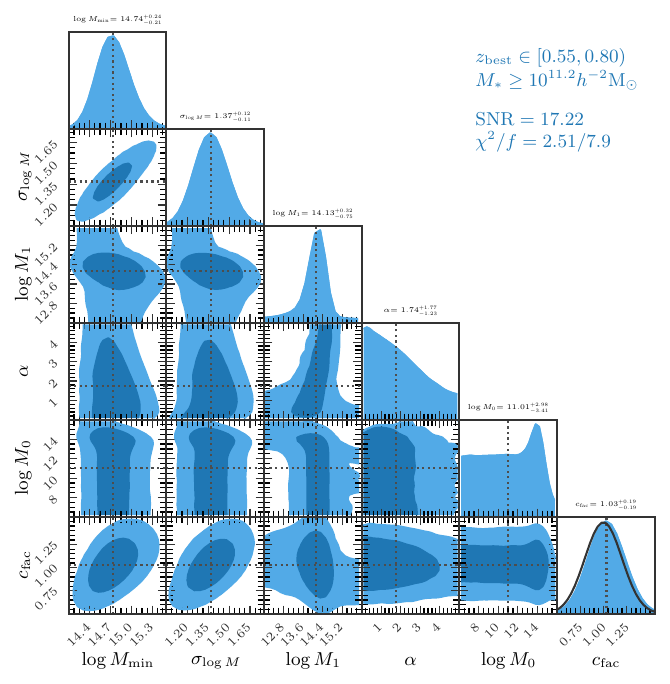}
    \caption{Parameter constraints and covariances in parameter space for redshift bin $z_2$, $M_* \ge 10^{11.2} \hiimsun$.}
    \label{fig:Parameter_constraints_z2_11.2}
\end{figure*}

We present the triangle plots which show the degeneracy of the inferred parameters when fitting the weak lensing signal and the I20 abundance in two threshold stellar mass bins for each redshift bin in Figures~\ref{fig:Parameter_constraints_z1_8.6}-\ref{fig:Parameter_constraints_z2_11.2}.

\bsp
\label{lastpage}
\end{document}